\documentclass[fleqn,usenatbib]{mnras}

\usepackage{newtxtext,newtxmath}
\usepackage{comment}
\usepackage{scalerel}
\usepackage[T1]{fontenc}
\usepackage{pdflscape}

\DeclareRobustCommand{\VAN}[3]{#2}
\let\VANthebibliography\thebibliography
\def\thebibliography{\DeclareRobustCommand{\VAN}[3]{##3}\VANthebibliography}

\usepackage{graphicx}	
\usepackage{amsmath}	
\usepackage{subcaption}

\title[Parameterization of clusters in the LMC and SMC]{Spatio-temporal map of star clusters in the Magellanic Clouds using \textit{Gaia}: Synchronized peaks and radial shrinkage of cluster formation}

\author[Dhanush et al.]{
S.R. Dhanush,$^{1,2}$\thanks{E-mail: dhanush.sr@iiap.res.in, srdhanushsr@gmail.com}
A. Subramaniam,$^{1}$
Prasanta K. Nayak,$^{3,4}$
S. Subramanian$^{1}$
\\
$^{1}$Indian Institute of Astrophysics, 2$^{nd}$ block Koramangala, Bangalore, 560034, India\\
$^{2}$Pondicherry University, R.V. Naga, Kalapet, 605014, Puducherry, India\\
$^{3}$Institute of Astrophysics, Pontificia Universidad Católica de Chile, Av. Vicuña MacKenna 4860, 7820436, Santiago, Chile\\
$^{4}$Department of Astronomy and Astrophysics, Tata Institute of Fundamental Research, 2 Homi Bhabha Road, Old Navy Nagar, Mumbai, 400005, India
}

\date{Accepted XXX. Received YYY; in original form ZZZ}

\pubyear{2023}

\begin{document}
\label{firstpage}
\pagerange{\pageref{firstpage}--\pageref{lastpage}}
\maketitle

\begin{abstract} 
We present a detailed view of cluster formation (CF) to trace the evolution and interaction history of the Magellanic Clouds (MCs) in the last 3.5 Gyr. Using the \textit{Gaia} DR3 data, we parameterized 1710 and 280 star clusters in the Large Magellanic Cloud (LMC) and the Small Magellanic Cloud (SMC), where 847 and 113 clusters are newly characterized in the outer LMC and SMC, respectively. 
We estimated the age-extinction-metallicity-distance parameters using an automated fitting of the color-magnitude diagram (CMD) after field star removal, followed by an MCMC technique. We report a first-time detection of two synchronized CF peaks in the MCs at 1.5$\pm$0.12 Gyr and 800$\pm$60 Myr. We recommend that the choice of the metallicity ($Z$) values of isochrones for clusters with age $\le$ 1 - 2 Gyr are Z$_{\text{LMC}}$ = 0.004 - 0.008 and Z$_{\text{SMC}}$ = 0.0016 - 0.004 for the LMC and SMC, respectively. We found evidence for spiral arms in the LMC, as traced by the cluster count profiles over the last 3.5 Gyr. The density maps provide evidence of ram-pressure stripping in the North-East of the LMC, a severe truncation of CF in the South of the LMC, and a radial shrinkage of CF in the SMC in the last 450 Myr. The last SMC-LMC interaction ($\sim$ 150 Myr) resulted in a substantial CF in the North and Eastern SMC, with a marginal impact on the LMC. This study provides important insights into the CF episodes in the MCs and their connection to the LMC-SMC-MW interactions.
 
\end{abstract}




\section{Introduction}
The Magellanic Clouds (MCs) comprise two satellite irregular galaxies, the Large Magellanic Cloud (LMC) and the Small Magellanic Cloud (SMC). The LMC and the SMC are located at a distance of $\sim$ 50 kpc \citep{lmc_dit2019Natur.567..200P} and $\sim$ 60 kpc \citep{smc_dist2020ApJ...904...13G}, respectively. They are the nearest interacting dwarf galaxies in the local group, and they are considered the laboratory to study similar distant systems in our universe. The Magellanic system is composed of the MCs along with a stream of gas, namely the Leading Arm (LA, \citealt{lu1998AJ....115..162L,put1998Natur.394..752P}), the Magellanic Stream (MS), which is seen as a trail of neutral hydrogen that spans more than 100$^{\circ}$ across the sky \citep{PUT2003ApJ...586..170P}, and the Magellanic Bridge (MB) appears as an elongated structure of HI structure, resulting from the tidal interaction between the MCs \citep{ltgard10.1093/mnras/266.3.567, muller10.1111/j.1745-3933.2007.00356.x}. Notably, the MB is not solely observable in gas form but also evident in stars \citep{mbstar2Jacyszyn-Dobrzeniecka_2020,gaia2021A&A...649A...7G}.

The tidal interaction history of the MCs with our Milky Way (MW) has been studied by several authors \citep{put1998Natur.394..752P,wein2000ApJ...532..922W,diaz2012ApJ...750...36D,Hammer_2015} and we understand that the structure and evolution of MCs depend not only on their mutual interactions but also on the interaction with the MW. The proper motion studies by \cite{kal2006ApJ...638..772K,kal22006ApJ...652.1213K} and \cite{beslakaliay2007ApJ...668..949B} showed that the MCs are likely to be in their first orbital passage towards the MW. However the recent study by \cite{vas10.1093/mnras/stad2612} has shown the possibility of the MCs being on the second infall towards the MW. Also, the LMC and the SMC are bound for the past $\sim$ 6.3 Gyr \citep{besla2016ApJ...825...20B}. The recent relative proper motion study \citep{zivick2018ApJ...864...55Z} between the MCs suggested that a close interaction between them happened at $\sim$ 150 Myr ago. A study of the HI velocity profiles from the Leiden Argentine-Bonn (LAB) all-sky HI survey \citep{krab2005A&A...440..775K} traced the partial origins of LA and MS in the southeast HI overdensity (SEHO) of the LMC. Simulation studies of the Magellanic system \citep{bes2012MNRAS.421.2109B,diaz2012ApJ...750...36D,Lucchini_2021} suggested that the gaseous features in the MCs must have formed because of their mutual interaction. These models suggested that the MS was formed $\sim$ 1.5 Gyr ago from the gas stripped from the Clouds, whereas the MB was formed $\sim$ 100-300 Myr ago due to material stripped mainly from the SMC. The ram pressure is considered to have an effect in shaping the MCs, as their passage through the circumgalactic medium is expected to truncate the disc of the LMC \citep{ramp12015ApJ...815...77S}.

 The evolution of the MCs is a complex phenomenon, the mutual interactions among the MCs and interaction with the MW are understood to have played a major role in triggering as well as truncating star formation in the MCs \citep{bekki10.1111/j.1365-2966.2004.08510.x,Yasuo10.1093/pasj/psaa103}. To trace these, it is necessary to have proxies whose ages can be estimated accurately and that are available throughout the MCs. Star clusters are therefore the ideal candidates to trace the common triggers of star formation as well as truncation due to tidal/ram-pressure effects \citep{Chilingarian_2018, AE10.1093/mnras/sty1048}. Therefore, a spatio-temporal map of cluster formation (CF) over the full coverage of the MCs will help us understand the evolution of MCs. It will also help us understand the interaction history of the MCs with the MW and among themselves.

The global and spatially resolved star formation history (SFH) of the MCs was studied by \cite{zaritsky_1997AJ....114.1002Z} using the Magellanic Cloud Photometric Surveys (MCPS). The study found peaks of star formation (SF) at $\sim$ 2 Gyr, 500 Myr and 100 Myr for the LMC \citep{Harris_2009}, and peaks at $\sim$ 2.5 Gyr, 400 Myr and 60 Myr for the SMC \citep{Harris_2004AJ....127.1531H}. The catalog of  Long Period Variable (LPV) stars  \citep{spano2011A&A...536A..60S,Soszy2011AcA....61....1S} was used to trace the SFH \citep{sara_rez2014MNRAS.445.2214R} of the MCs, and the authors found that the ancient SFH of the LMC ($\sim$ 10 Gyr ago) and the SMC ($\sim$ 3 - 5 Gyr ago) were found to be significantly distinct. A recent study that obtained SFH using the Survey of the Magellanic Stellar History (SMASH,  \citealt{smash2017AJ....154..199N}) found the synchronous and tidally entangled evolution of the LMC and the SMC for the past at ~3.5 Gyr \citep{synchronised2022MNRAS.513L..40M}. The studies of SFH trace the population from very old to very young ages, with a typical resolution of about 1 Gyr. As the MCs interacted with each other and with the MW, in the last 1 - 2 Gyr, age-dating of star clusters in the above age range is required to trace the history of CF. Hence a spatio-temporal tracing of the CF in the MCs will help us to probe the interaction-driven evolution of the MCs.

The extensive catalog of star clusters and associations in the MCs is provided by \citet[hereafter B08]{bica2008MNRAS.389..678B}\label{b2o08}, \citet[hereafter S16]{sitek_lmc2016AcA....66..255S} and \citet[hereafter S17]{sitek2017AcA....67..363S}. A significant number of studies have been performed with several of the cataloged star clusters to obtain their properties, including age. \citet[hereafter G10]{glatt2010} estimated the ages of 1193 LMC and 324 SMC star clusters using the MCPS data. The study estimated peaks of CF at $\sim$ 125 Myr and $\sim$ 800 Myr for the LMC, and, $\sim$ 160 Myr and $\sim$ 630 Myr for the SMC, respectively. \citet[hereafter N16]{nayak2016MNRAS.463.1446N}, \citet[hereafter N18]{smc_nayak_2018A&A...616A.187N} estimated ages of 1072 LMC star clusters and 179 SMC star clusters, respectively, using the data from Optical Gravitational Lensing Experiment III (OGLE III, \citealt{ogle32008AcA....58...69U}). They obtained  CF peaks of $\sim$ 125 Myr for the LMC and  $\sim$ 130 Myr for the SMC, respectively. Several authors have performed similar studies to estimate the ages of star clusters \citep{pusmc1999AcA....49..157P,pu2000AcA....50..337P,chiosi2006yCat..34520179C,palma2016A&A...586A..41P,andres2018MNRAS.473..105P}. Even though these studies traced the age distribution, the MCs' global and spatial age distribution was limited to the inner regions, particularly in the LMC. The clusters in the outskirts of the LMC are not adequately studied yet as it requires a significantly large sky coverage. As interactions between the MCs and the MW would have a larger impact in the outer regions of the LMC/ SMC, age estimations of the outer clusters are very important to trace the effect of interactions. This study is an attempt to produce a wide spatio-temporal CF history in the MCs.

In this study, we aim to achieve the following: 1) To increase the number of characterized clusters in the MCs, especially in the outer regions of LMC; 2) To develop an automated method to estimate age-extinction-metallicity-distance for clusters from their CMD; 3) To trace the Episodic Cluster Formation (ECF) history and obtain the signatures of interaction history between the MCs or with the MW; 4) To trace the density structures traced by clusters (such as bar/ spiral arm) as well as signatures of trigger/ shrinkage/ truncation of CF in the outer parts of the MCs.

The paper is arranged as follows. In Section \ref{sec_2}, we describe the data set and data reduction procedures used in this study. In Section \ref{sec_3}, we describe our automated methods for the cluster parameter estimation. In Section \ref{sec_4}, we present our estimated age-extinction-metallicity-distance values for the clusters we characterized, along with the spatial age maps, ECF history, and radial cluster density profiles. The discussions based on our results are presented in Section \ref{sec_5}, followed by Section \ref{sec_6}, which covers the summary of our work.

\section{Data}\label{sec_2}

We have made use of the \textit{Gaia} data in this study. \textit{Gaia} is an ongoing astronomical space mission launched by the European Space Agency (ESA) in 2013. The \textit{Gaia} Data Releases, DR1 and DR2 \citep{gaiadr12016A&A...595A...2G, gaiadr22018A&A...616A...1G} changed our entire perspective of visualizing the Galaxy and its neighborhood. It provides astrometry, photometry, and spectroscopy of nearly 2 billion stars in the MW and of nearby extragalactic systems \citep{gaia2016A&A...595A...1G,gaia_extra2018A&A...616A..12G}. Recently, the \textit{Gaia} Early Data Release 3 (EDR3) was used to study the structure of the MCs including the outskirts of both galaxies \citep{gaia2021A&A...649A...7G}.
The latest Data Release from \textit{Gaia} is the \textit{Gaia} DR3 \citep{gaia2022arXiv220800211G}, which adds corrections to the G band photometry \citep{corre2021A&A...649A...1G,corr22021A&A...649A...3R} from EDR3. The sky coverage in the \textit{Gaia} Data Releases is an advantage to create spatial age maps of star clusters in the MCs.

This study used the data from the \textit{Gaia} DR3 survey in \textit{G}, \textit{$G_{BP}$} and \textit{$G _{RP}$} passbands. We considered 4041 star clusters taken from B08 and S16\&17 catalogs for this study, among which 3330 and 711 clusters belong to LMC and SMC, respectively. The information on central coordinates of the clusters in Right Ascension and Declination ($\alpha$,$\delta$) and radius (r${_c}$) are adopted from these catalogs. This study aims to characterize a large number of star clusters with a wide range of ages. As manual estimation of parameters is tiresome and liable to subjectivity, the entire process of cluster parameterization was automated. Before the automated parameterization was performed, we first selected clusters suitable for this study based on their position in the sky and the presence of neighboring/overlapping clusters in the line of sight (LOS). The bulk handling of the \textit{Gaia} archival data in our analyses was performed with the \textit{Gaia} Asynchronous query services available in the Astroquery \citep{Astroquery2019AJ....157...98G} python package. The next stage involves cleaning of star cluster data sets from field star contamination, where we used a field star decontamination (FSD) algorithm, which was obtained by modifying and improving the technique used by N16. The following subsections are the detailed data reduction techniques we adopted in our study.

\subsection{Classifications of clusters based on spatial overlap}

The estimation of cluster membership is performed using the statistical subtraction of field stars selected from one or more field regions near the cluster. First and foremost, we need to ensure that the chosen field region is not a part of another cluster, as there may be a number of clusters located in close proximity. Therefore, each cluster is checked to identify star clusters that overlap in their spatial extend along the LOS.

\begin{figure*}
        \centering
        \begin{subfigure}[b]{\textwidth}
            \centering
            \includegraphics[width=\textwidth]{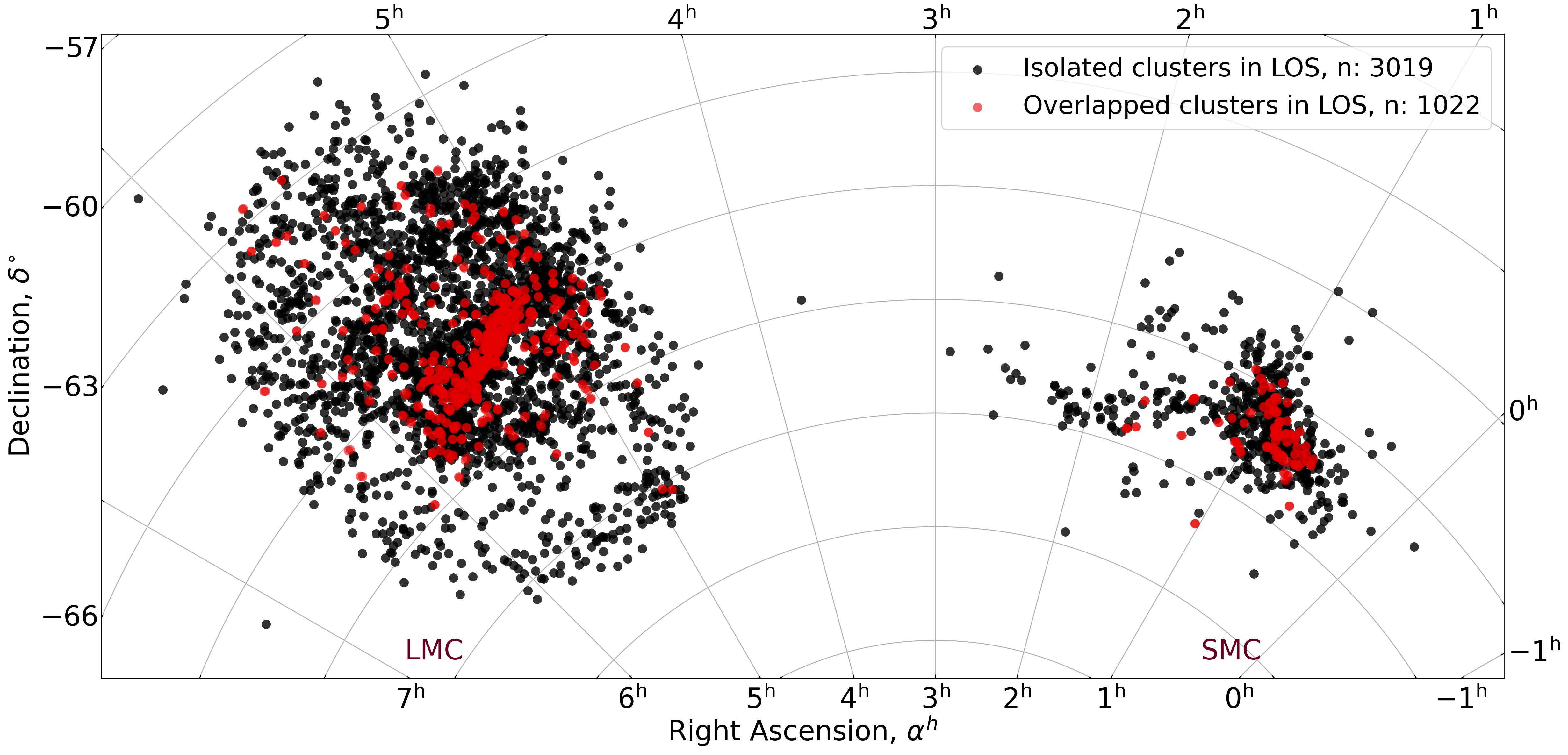}
            \caption[Network2]%
            {{\small }}    
            \label{fig1_a}
        \end{subfigure}
        \vskip\baselineskip
        \begin{subfigure}[b]{\textwidth}   
            \centering 
            \includegraphics[width=\textwidth]{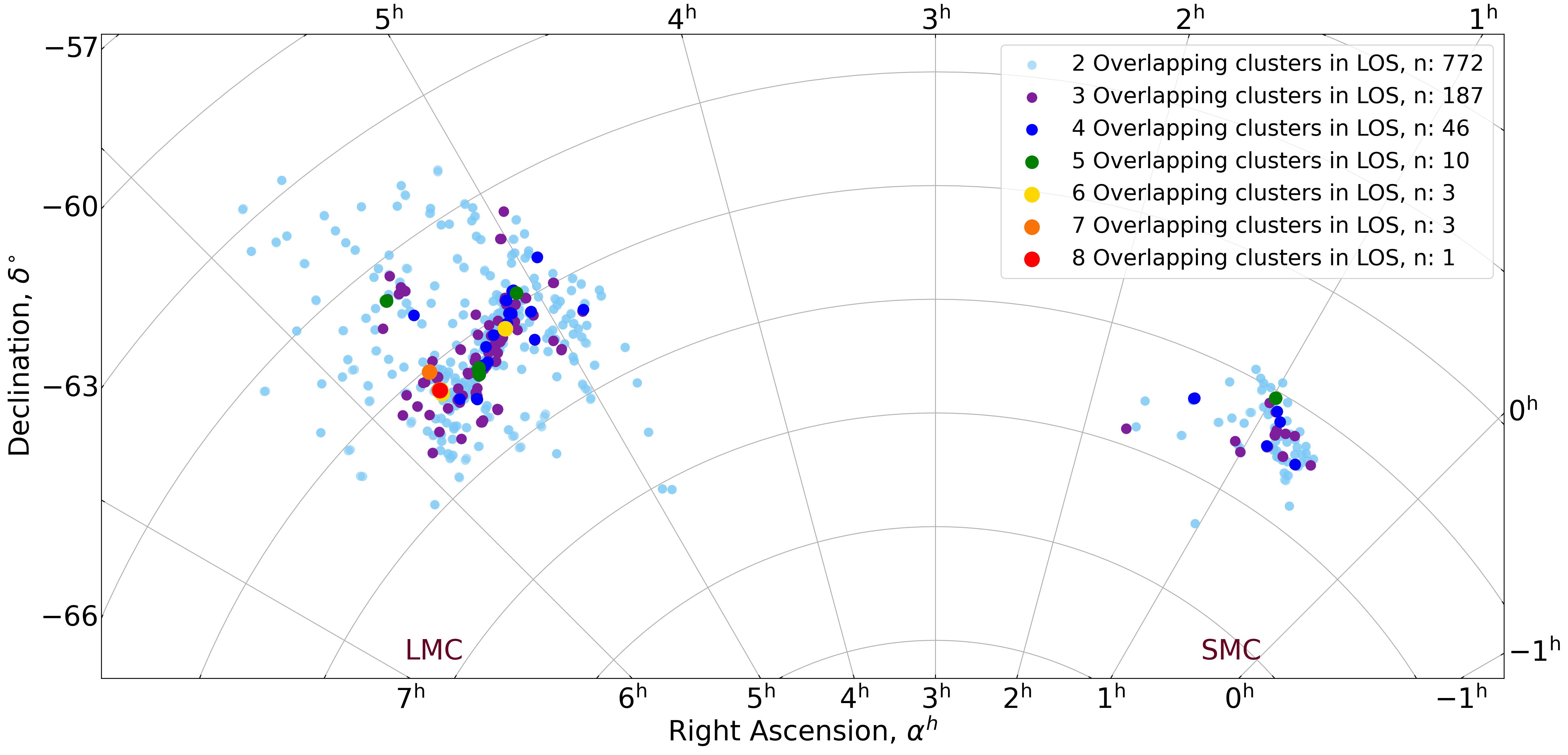}
            \caption[]%
            {{\small }}    
            \label{fig1_b}
        \end{subfigure}
        \caption[]
        {\small Spatial classification of star clusters in the MCs based on the LOS. a) The isolated clusters in the LOS (black dot) and overlapping clusters in the LOS (red dot) of the MCs in the sky plane. The number of clusters in each group is shown in the legend. b) The classification is based on the number (ranging from 2 to 8) of overlapping clusters in the LOS of the MCs color-coded as shown in the legend.} 
        \label{fig1_main}
    \end{figure*}

We identified neighboring star clusters based on the angular separation between their centers. As a result, we classified star clusters as isolated and overlapping based on their spatial distribution in the LOS. The classification of star clusters as per the scheme is shown in Figure \ref{fig1_main}. Towards the central regions, many of the clusters appear to overlap in the LOS, as expected. The outer clusters are mostly isolated as they are well separated from each other. Figure \ref{fig1_a} shows the isolated and spatially overlapping star clusters in the LOS. Out of 4041 cluster candidates, we have found 3019 isolated star clusters. The rest, 1022 clusters, were flagged as overlapping star clusters in the LOS. Also, each cluster is categorized based on the number of neighboring clusters overlapping into the cluster region. Figure \ref{fig1_b} shows the classification based on the number of overlapping clusters ranging from 2 to 8. We note a large population of more than 4 overlapping clusters in the bar region of the LMC and a few such systems in the central SMC. The two-cluster overlap appears to have a wider spread but is confined to the inner regions.

The clusters classified based on spatial overlap need to be cleaned from the LOS field star contamination in the cluster regions. To perform the cleaning, we introduced a workspace associated with each cluster. The following subsection deals with the definition and details of the different types of cluster workspaces we used in this study.

\subsection{Cluster workspaces \label{workspace_sect}}

\begin{figure*}
        \centering
        \begin{subfigure}[b]{0.475\textwidth}
            \centering
            \includegraphics[width=\textwidth]{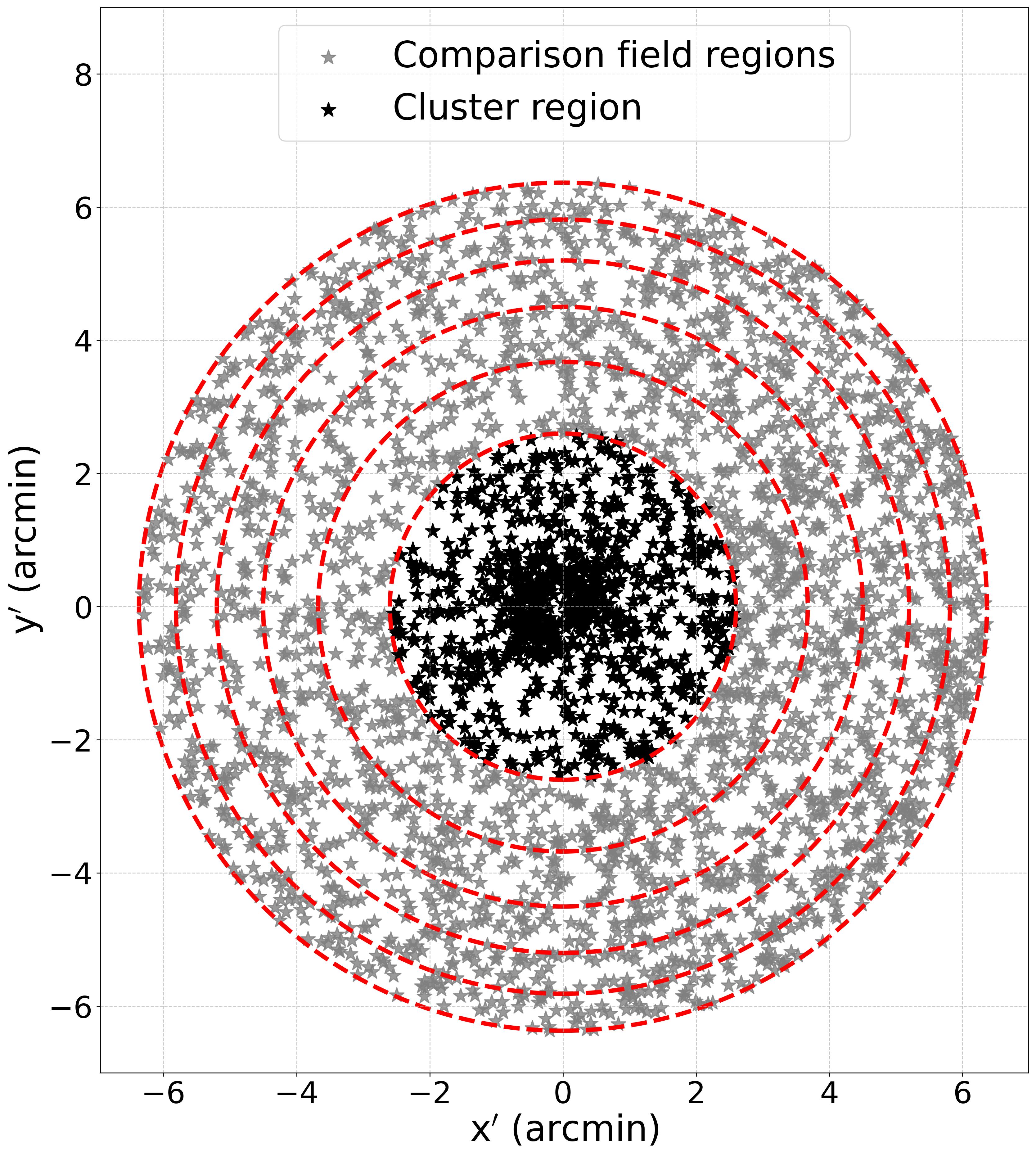}
            \caption[Network2]%
            {{\small}}   
            \label{figisol}
        \end{subfigure}
        \hfill
        \begin{subfigure}[b]{0.475\textwidth}  
            \centering 
            \includegraphics[width=\textwidth]{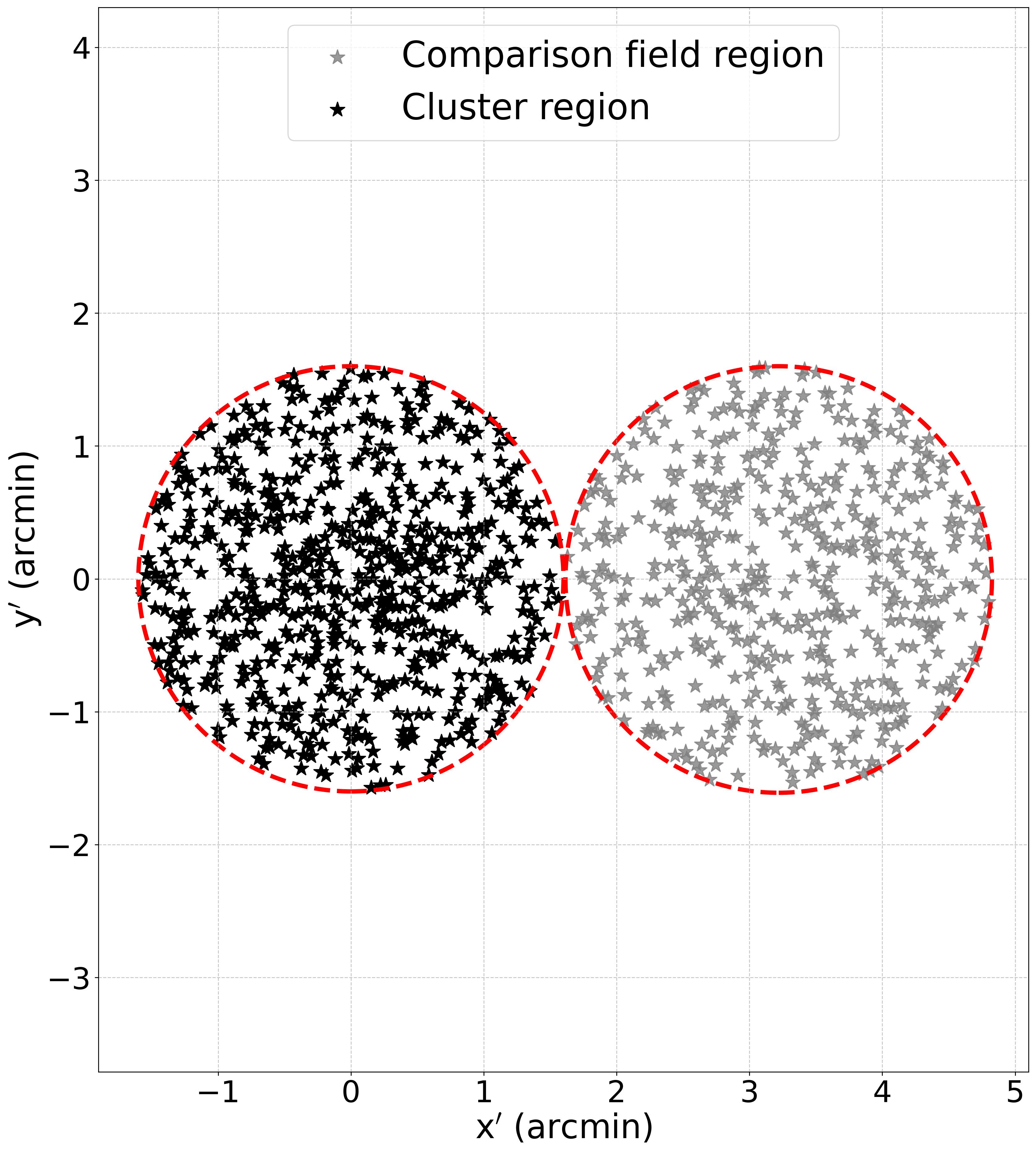}
            \caption[]%
            {{\small}}    
            \label{figcircu}
        \end{subfigure}
        
        \vskip\baselineskip
         \begin{subfigure}[c]{0.475\textwidth}   
            \centering 
            \includegraphics[width=\textwidth]{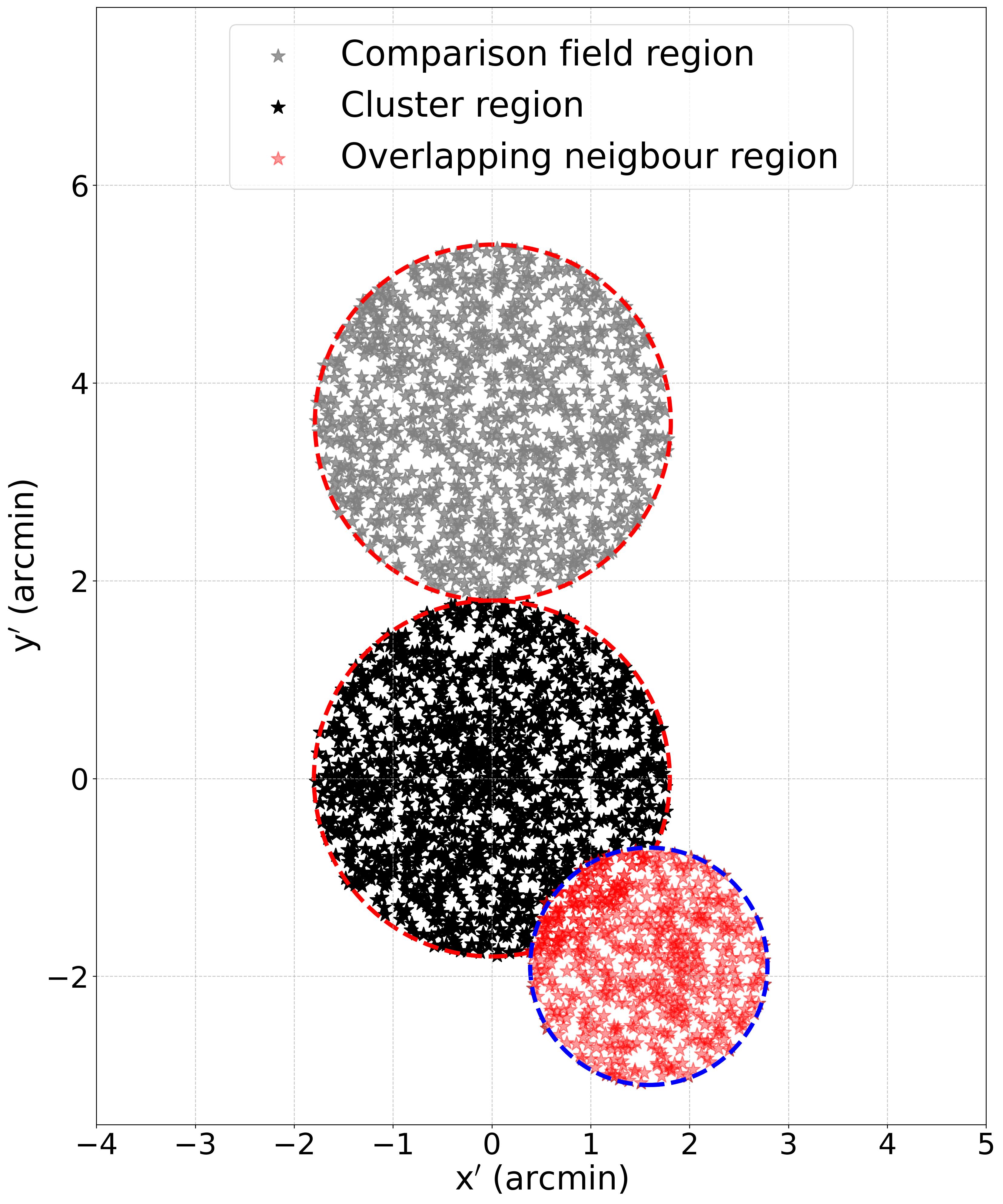}
            \caption[]%
            {{\small}}    
            \label{figmergd}
        \end{subfigure}
        \caption[]
        {\small Different types of cluster workspaces (as defined in subsection \ref{workspace_sect}) for implementing FSD. a) Model workspace of the isolated cluster in the LOS, Group-1 and Group-2 clusters. The cluster region is shown with black stars. The nearby field regions are shown with grey stars, separated by concentric annular circles (dashed red circles). Group-1 clusters use all 5 field regions, and Group-2 clusters use 1 to a max of 4 field regions depending on their availability.  b) Model workspace for a partially isolated cluster in the LOS, Group-3 clusters (black stars: cluster region, grey stars: nearby field region). c) Model workspace for an overlapping pair of clusters in the LOS, Group-4 clusters (black stars: cluster region, red stars: overlapping neighbor cluster region, grey stars: nearby field region).} 
        \label{workspaces}
    \end{figure*}

The workspace for each star cluster in the sky plane is defined such that it comprises a cluster region accompanied by one or more field regions.
The workspace for each cluster depends on the choice of Field Star Decontamination (FSD) algorithm suitable for its location, and it is a function of the cluster size and the presence of neighboring clusters.
The method adopted for FSD depends on the availability of field star area around or near the cluster. Based on the above, we have considered four different types of field regions for FSD. The different types of cluster workspaces are shown in Figure \ref{workspaces}.

\begin{enumerate}
    \item Group-1: These are isolated clusters with sufficient area available around them to define five concentric annular field regions of equal area. About 56\% (2294 out of 4041) of the clusters from our initial selection have spatial freedom to use up to five field regions around them. We chose five annular regions to classify at least 50\% of the clusters from the initial selection as isolated clusters. A graphical depiction of the workspace for isolated clusters is shown in Figure \ref{figisol}. The field annuli are of external radii $\sqrt{n+1}\times r_c$ from the cluster center, where n is the number of annuli. As mentioned this group consists of 2294 clusters.  
    \item Group-2: In several isolated clusters, the outer annular field regions were found to overlap with a nearby cluster. We define Group-2 as partially isolated clusters, where the workspace of a cluster has up to four concentric annular field regions of equal area around the cluster. This group consists of 501 clusters.
    \item Group-3: There were clusters where we could not take the annular field region around them, as even the first annular field region would overlap with a nearby cluster region. In such cases, as shown in Figure \ref{figcircu}, we have taken one circular comparison field near the cluster. Thus, the workspace comprises the cluster region and a nearby field region of the same area. These are defined as Group-3 clusters and 224 clusters fall in this group.
    \item Group-4: A significant number of clusters have one overlapping neighbor in the LOS. Figure \ref{figmergd} shows the workspace for an overlapping pair of clusters as an example. The difference in their workspace from isolated clusters is in the selection of cluster region and the field region. For the clusters with one neighbor, the neighboring cluster boundaries were ensured not to reside within a cut-off radius of 0.1r$_c$ from its center. The choice of the cut-off radius was based on trials and it ensured enough stars from the centrally dense regions of overlapping clusters while performing the parameter estimation. For these clusters, we removed the overlapping region with the neighboring cluster and used a nearby circular field region. This group consists of 772 clusters.
\end{enumerate}

After removing the 250 LOS overlapping clusters based on the choice of cluster workspaces, we are left with 3791 cluster workspace to proceed with the cleaning process of cluster regions. But before applying the FSD method, we adopted specific \textit{Gaia} selection criteria for the cluster workspaces to improve the cleaning phase of cluster regions. These are explained in the next subsection.
    
\subsection{Selection criteria of \textit{Gaia} sources in the workspace}\label{select_gaia_crit}    
The \textit{Gaia} selection criteria for the sources in each cluster workspace are enumerated below.
\begin{enumerate}
    \item Selected sources brighter than 20.5 mag in \textit{$G$}.
    \item Removed the NULL value sources in \textit{$G_{BP}$} and \textit{$G_{RP}$}.
    \item Selected sources with RUWE (Re-normalized Unit Weight Error) < 1.4
    \item Selected sources with $\leq2\sigma$ of proper motion and parallax from the mean proper motion and parallax of all sources. Applied to cluster regions and each available field regions in the workspace.  
\end{enumerate}

The G-band photometric uncertainties are up to $\sim$ 1 mmag for G$\leq$17 and 6 mmag for G$=$ 20. Meanwhile, the G$_{BP}$ and G$_{RP}$> 20 mag have uncertainties greater than 50 mmag. For the reason, we considered sources brighter than 20.5 mag in our study, as we performed the parameter estimation accounting for the error in the magnitude and color of the sources. The \textit{Gaia} data have NULL values for sources not satisfying the photometry selection criteria as mentioned in \cite{gaia_phot_2021A&A...649A...3R}, we removed such data points from each cluster workspace as well. The RUWE value < 1.4 is suggested by \textit{Gaia} studies for the most probable singular sources in their catalog \citep{Lindegreen2018A&A...616A...2L,ruwe2023A&A...674A..34G}.

The proper motion and parallax cut-off of $\leq2\sigma$ we adopted helps in removing the highly deviant proper motion sources from workspace. The criteria were applied to cluster region and available field regions depending on the type of cluster workspace as mentioned in subsection \ref{workspace_sect}. We note that the Gaia parallax is not a good distance estimator for sources located more than a few kpc, since the value lies very close to the parallax zero-point value \citep{luri2021A&A...649A...7G,lind2021A&A...649A...2L}. But parallax is important to remove nearby MW sources that fall in the LOS of MC or MC cluster workspaces.

After applying the stated selection criteria as mentioned above, the cluster workspaces and their associated data sets were fetched onto the local machine utilizing the \textit{Gaia} Asynchronous query. The rest of the field star decontamination from MW and from the MCs itself was carried out with the FSD algorithm we have implemented in our study. The detailed cluster cleaning procedure to remove further field stars is given in the following section.

\subsection{Field Star Decontamination}\label{fsd_stage}

In our study we used a modified version of FSD algorithm used by several authors \citep{pand2007PASJ...59..547P,Choudhury_2015, nayak2016MNRAS.463.1446N} to clean the CMDs. We created $(G, G_{BP}-G_{RP})$ CMDs for the cluster region and the field region(s), for all the groups mentioned above. The CMD of a cluster region will have a combination of cluster and field stars, whereas the CMD of its field region(s) will have only the field stars.
An algorithm was applied to the cluster CMDs to compare and remove field stars. The field stars within the cluster CMD were removed by considering each star in the cluster CMD and removing the closest counterparts in their nearby field CMD(s). We considered [magnitude, color] bins with different sizes, starting with $[\Delta{G}, \Delta(G_{BP}-G_{RP})]$ = [0.02, 0.01] up to a maximum of [0.5, 0.25] about each star in the cluster CMD to search for the closest field star counterpart in the field CMD(s). Those with counterparts in the field CMD(s) are considered as field stars in the cluster region and are removed from the CMD. Stars that remained in the cluster CMD after this statistical cleaning are most likely to be the cluster members and the CMD with such stars can be considered as a cleaned CMD(s).

In the case of Group-1 clusters, the algorithm takes advantage of the multiple field regions. For clusters in this group, we created 5 cleaned CMDs corresponding to using either all annular fields or with a reduced number of annuli up to a minimum of one. For example, 5-FSD CMD is defined as a cleaned CMD where a star from the cluster CMD did not find a counterpart in any one of the five annular field CMDs. Similarly, we created 4-FSD CMDs to 1-FSD CMD as explained above, using a correspondingly less number of field CMDs for a given cluster CMD. The same logic was implemented for Group-2 clusters with n$<$5 (where n, the number of associated field regions available for a given cluster) annular field regions to produce n-FSD CMDs for the same cluster, where n can take a maximum value of 4 and a minimum of value of 1. For Groups-1 \& 2, we retained multiple cleaned CMDs for the same cluster at the end of this iteration. We note that the cleaned CMDs are affected by (1) less number of stars in the case of poor clusters and (2) variation in field star density. Both these result in the loss of features in the CMD and erroneous fits. Therefore, the CMD most suited for the cluster is decided after the parameter estimation with all the cleaned CMDs as detailed in Section \ref{select_crit_CMD}.

Now for the Group-3 clusters where a nearby circular field region by the side of the cluster is identified, we ensured that the field region does not overlap with any other neighboring cluster. In the case of Groups-3 \& 4, only one cleaned CMD is created at the end of this iteration. We note that some parts of the field regions of different star clusters may be common across the Groups, particularly for those located nearby.

The cleaned CMDs after FSD have a few randomly scattered field stars. Therefore, we design the cluster parameter estimation method to adequately take care of the scatter. The implementation of our cluster parameter estimation is explained in the subsequent section.

\section{Estimation of cluster parameters.} \label{sec_3}
At the end stage of FSD we obtained the cleaned cluster CMDs, which are now satisfactory to perform parameter estimation. We automated this stage by setting up prior age and extinction estimates for the clusters, followed by a set of selection criteria for the best-fitted CMDs with prior knowledge, and finally estimated the four cluster parameters with a Bayesian MCMC sampling technique. The detailed procedures are provided below.

\subsection{Setting up prior age and extinction estimates for the clusters}\label{prior_stage}
We used the cleaned CMDs to estimate the maximum likelihood of age by comparing them with theoretical isochrones. We used the PARSEC stellar evolutionary tracks \citep{parsec2012MNRAS.427..127B} to generate the theoretical isochrones for our study. As per the previous studies, prior values of Z$_{\text{LMC}}$ = 0.008 and Z$_{\text{SMC}}$ = 0.004 were adopted as the LMC (N16 sample) and the SMC (N18 sample) metal fractions, respectively. The distance modulus (DM) to the center of the LMC was taken as 18.47 mag \citep{lmc_dit2019Natur.567..200P}. The DM value for a given cluster in the LMC was modified by accounting for an inclination of $i = 23^\circ.26$ and a position angle of the line of nodes, $\Phi = 160^{\circ}.43$ \citep{saroo2022A&A...666A.103S}.  In the case of the SMC, a value of  18.97 mag was assumed for all the clusters. Though there is a large depth in the SMC \citep{depth2009A&A...496..399S}, this quantity cannot be incorporated as a function of any known parameters. Also, a few clusters in the bridge region were given attributes similar to the SMC.

As per the prior assumptions, we estimated the maximum likelihood of age, $\log(t)$ and extinction, $A_G$ for the star clusters. We used an age range of 1 Myr to 10 Gyr ($\log(t) =  6$ to $10$, respectively), with a spacing of $\Delta{\log(t)} = 0.05$ for fitting the isochrone models to the cleaned CMDs. For the interstellar extinction, we adopted a range of $A_G = 0$ to 0.836 mag (corresponding to a range of $A_V = 0$ to 1 mag), with a step size of $\Delta{A_G} = 0.01$ mag. In order to fit the isochrones to the CMDs, we considered stellar evolutionary phases from the main sequence (MS) to the Red Giant Branch (RGB) in the isochrone models, assuming stars in each star cluster stay within the limit of these evolutionary phases. The isochrone fitting to the observed CMD was an iterative process in the parameter space available for each cluster CMD. First, all the theoretical isochrones of different $\log(t)$ and $A_G$ values were pushed to the distant modulus of the LMC and the SMC. Then we implemented a least square fitting method to the observed data points of the cluster CMD and the model isochrone points. The global minima of the least square deviations from about 80 models ($[\log(t)]/\Delta\log(t)$, constant $A_G = 0.45$) were found initially. Later, from the estimated initial fit parameters ($\log(t)_{i},A_{G_i}$), we considered a reduced parameter space of $\log(t)$ range [$\log(t)_{i}-0.6$ to $\log(t)_{i}+0.6$] and $A_G$ range [0 to 0.836 mag] to redo the least square fitting,  we used this iteration to close the gap between the most likely model and the observed data. One more iteration was performed by removing the scattered field stars in the cleaned CMDs. We adopted a 1.5$\sigma$ clip from the mean least square deviations to remove scattered data points. By doing so, convergence to a unique model with reduced error was achieved, with the best-fitting isochrone closely representing the features of the observed CMD.

The ordinary $\chi^2$ minimization does not yield the best-fitting isochrone for the cluster CMD, as multiple combinations of parameters can give the minimum $\chi^2$ values while fitting models to data \citep{D’Antona2018,Souza_2020}. So, we defined an error-weighted modified $\chi^2$ value to the fit for the clusters to define our selection criteria. The \textit{Gaia} G magnitudes have errors of the order of $\sim10^{-3}$ mag, corresponding to significantly less error in observed fluxes, resulting in high $\chi^2$ values. As the error propagation in color (\textit{$G_{BP}-G_{RP}$}) term is larger compared to the magnitude for a given star in the CMD, the error weighting was considered only on the color term, while estimating the $\chi^2$ value of the best fitting isochrone model. We used three modified $\chi^2$ options where the observed flux errors (corresponding to \textit{$G_{BP}$}, \textit{$G_{RP}$} magnitudes) of the stars were at least 2\%, 4\%, or 10\% of the observed fluxes ($F_{i,o}$). After testing the fits with a smaller sample of clusters, we fixed the observed error in flux to be at least 4\% and estimated the modified $\chi^2$ values.
For a CMD with data points of observed color ($c_{i,o}$) and model color ($c_{i,m}$), the modified chi-square ($\chi^2_{f}$) is given by,
\begin{align}
\centering
    \chi^2_{f}=& \frac{1}{N-2} \sum_{i=1}^{N} \left\{\frac{(c_{i,o}-c_{i,m})^2}{s^2_{i}}\right\} \label{eq2}\\
    \textrm{where} \nonumber\\ 
    s_{i,o}= &\textrm{$\sigma_{ci} \forall (\sigma_{fi}\geq 0.04f_{i}), 0.04$ $\forall$ $(\sigma_{fi}\leq 0.04f_{i})$,} \nonumber
\end{align}

\begin{figure*}
        \centering
        \begin{subfigure}[b]{0.475\textwidth}
            \centering
            \includegraphics[width=\textwidth]{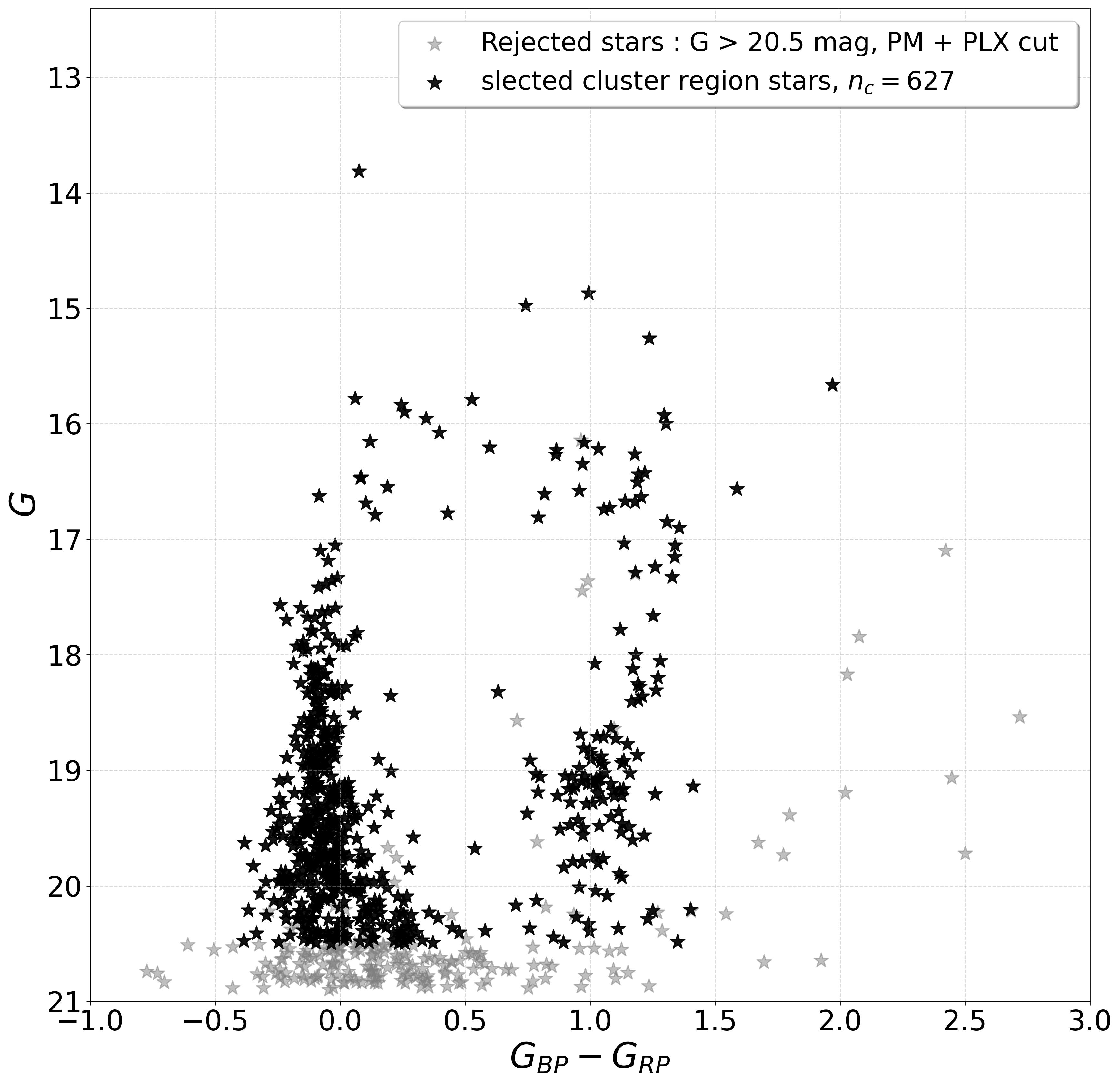}
            \caption[Network2]%
            {{\small NGC458 cluster region CMD}}   
            \label{fig:ngc4581}
        \end{subfigure}
        \hfill
        \begin{subfigure}[b]{0.475\textwidth}   
            \centering 
            \includegraphics[width=\textwidth]{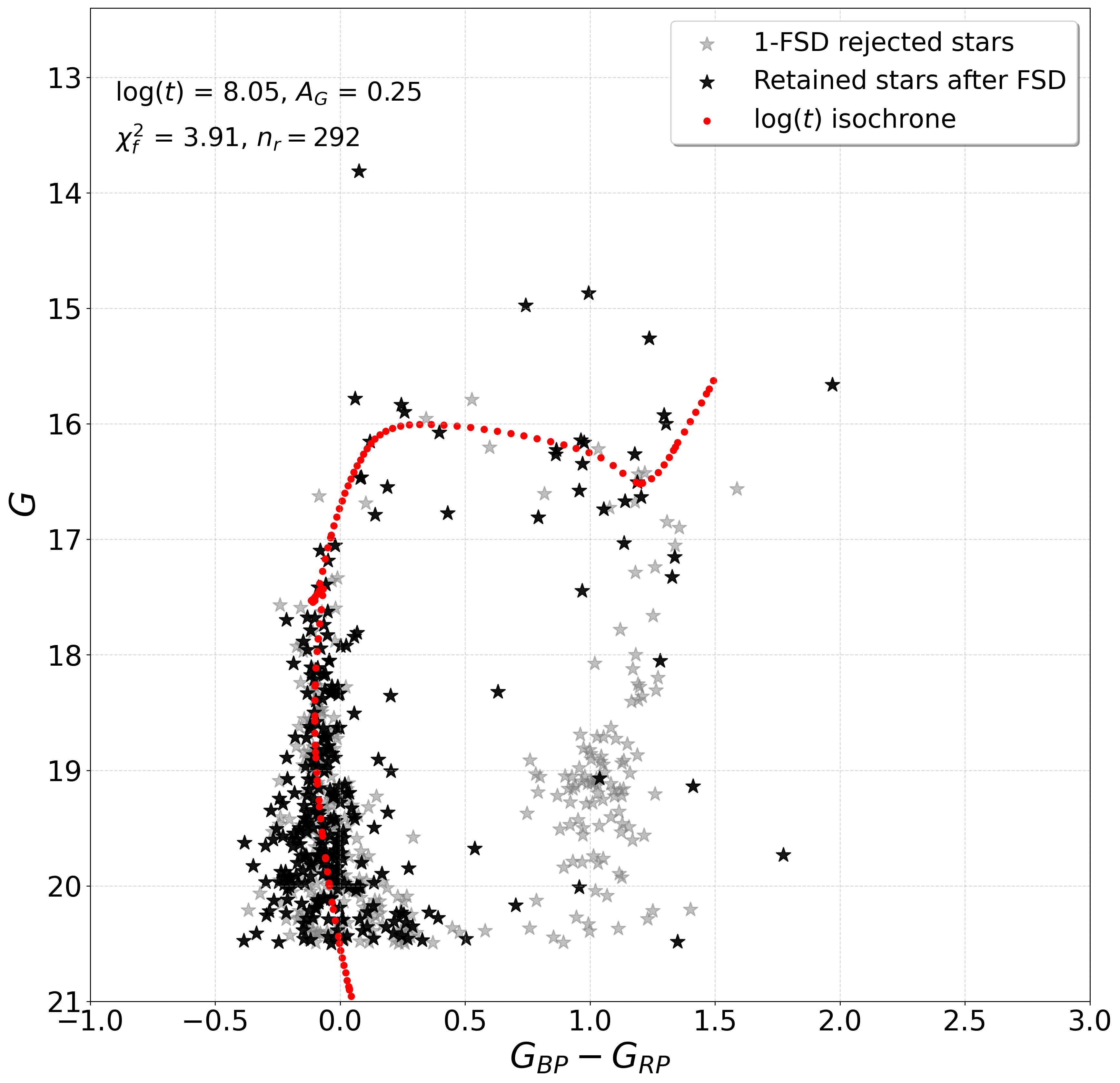}
            \caption[]%
            {{\small Cleaned and fitted CMD after 1-FSD}}    
            \label{fig:ngc4582}
        \end{subfigure}
        \vskip\baselineskip
         \begin{subfigure}[b]{0.475\textwidth}  
            \centering 
            \includegraphics[width=\textwidth]{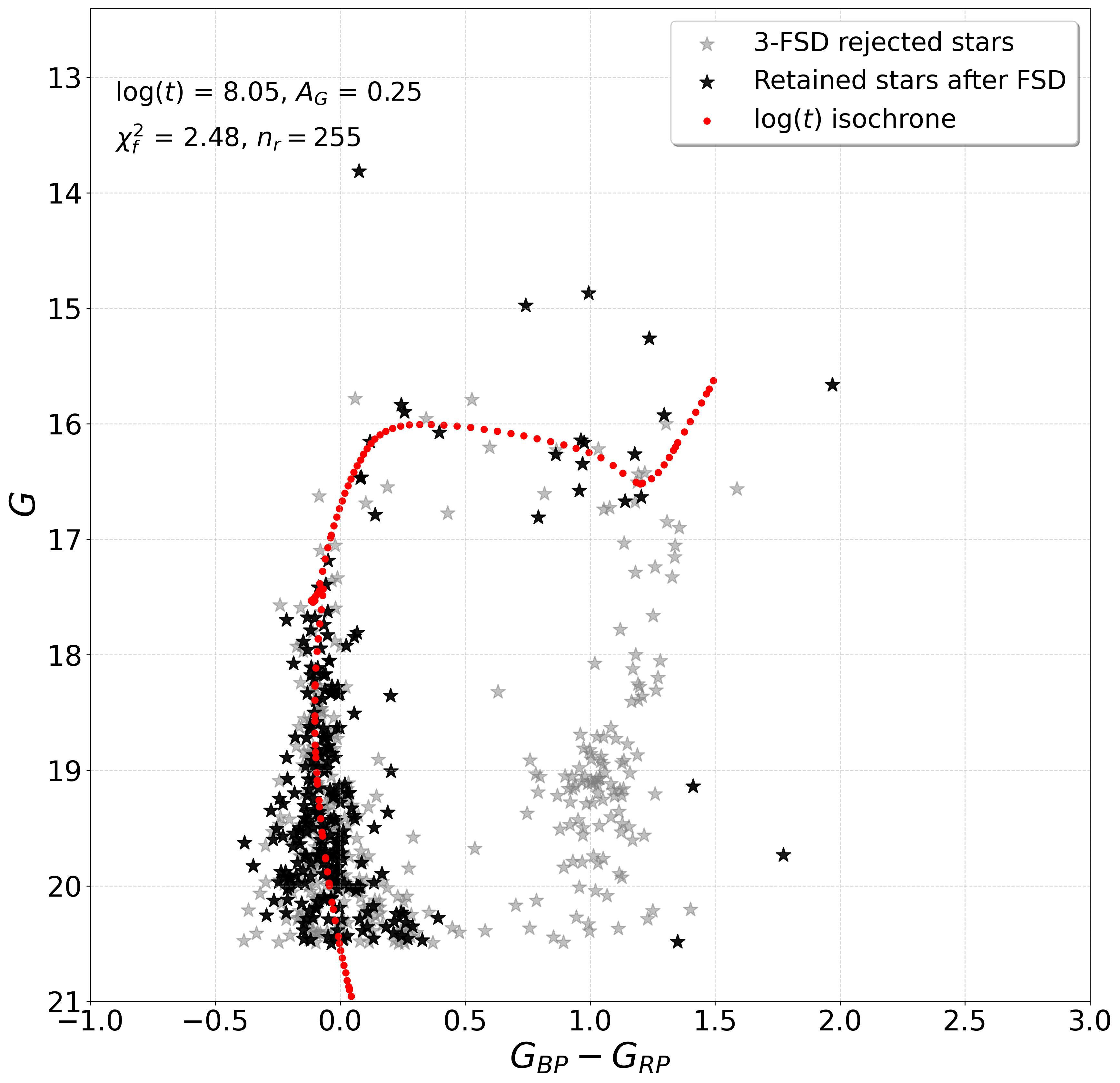}
            \caption[]%
            {{\small Cleaned and fitted CMD after 3-FSD}}    
            \label{fig:ngc4583}
        \end{subfigure}
        \hfill
        \begin{subfigure}[b]{0.475\textwidth}   
            \centering 
            \includegraphics[width=\textwidth]{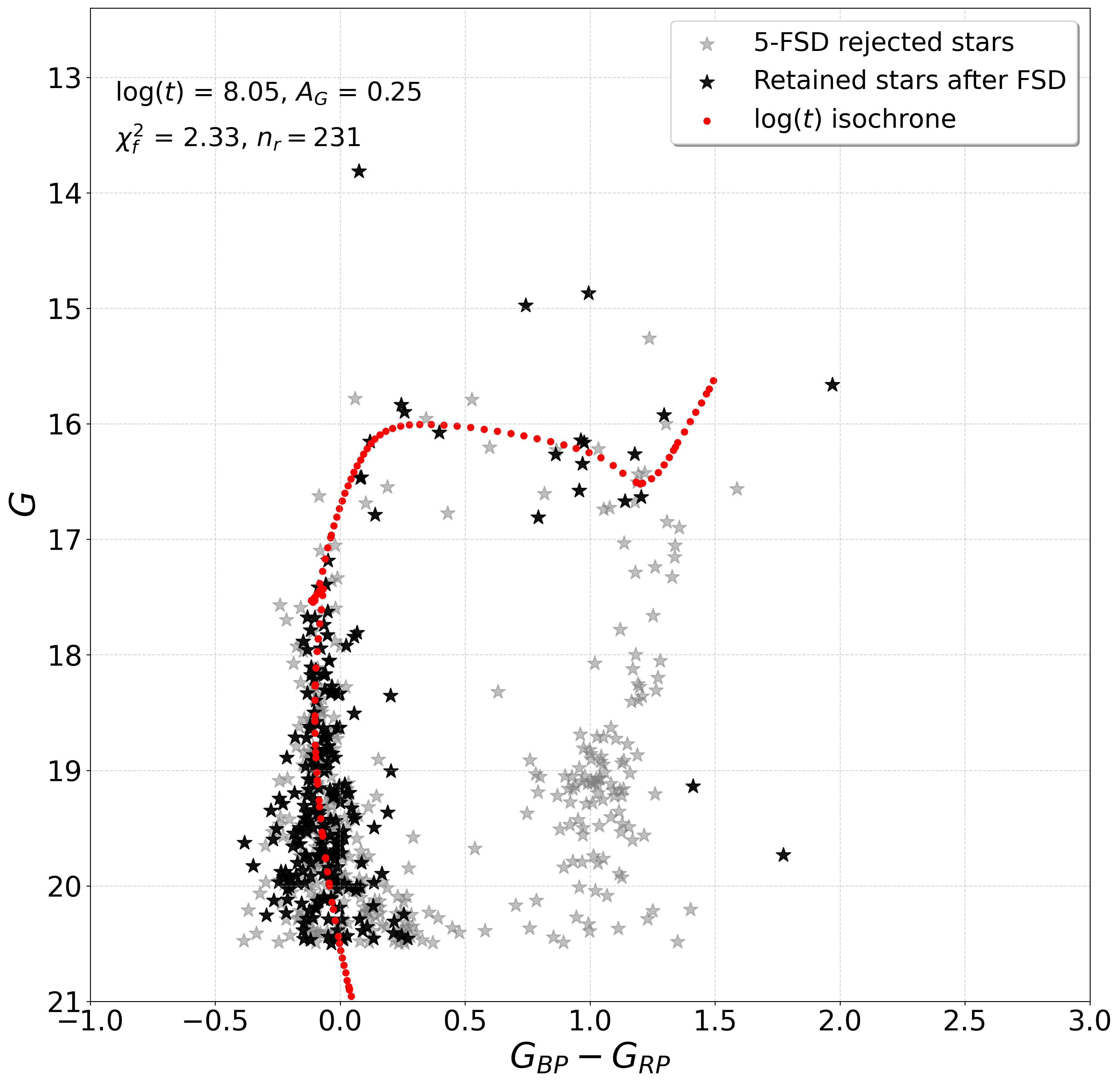}
            \caption[]%
            {{\small Cleaned and fitted CMD after 5-FSD}}    
            \label{fig:ngc4584}
        \end{subfigure}
        \caption[]
        {\small NGC458, a) Cluster CMD with initial selection criteria (proper motion, parallax, and  G magnitude cut-offs). b) Cleaned CMDs (1-FSD, 3-FSD, and 5-FSD) with the parameters ($\log(t)$, A$_G$, and $\chi^{2}$) estimated at DM $=$ 18.977 are provided in each stages of FSD. In higher stages of FSD, the number of retained cluster members decreases because of the field star subtraction. } 
        \label{fig_ngc458}
    \end{figure*}

\begin{figure*}
        \centering
        \begin{subfigure}[b]{0.475\textwidth}
            \centering
            \includegraphics[width=\textwidth]{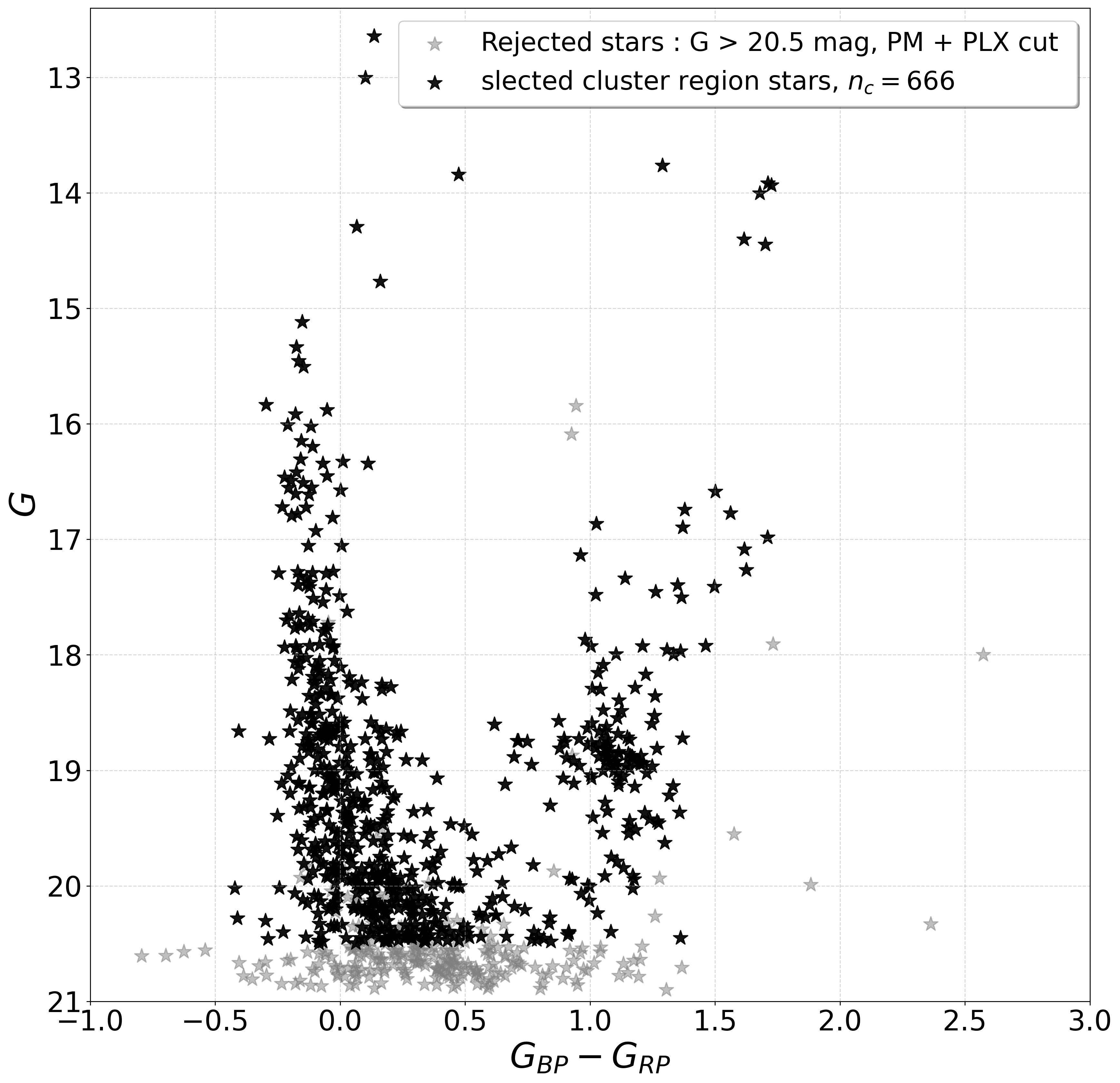}
            \caption[Network2]%
            {{\small NGC1735 cluster region CMD}}    
            \label{fig:merged1}
        \end{subfigure}
        \hfill
        \begin{subfigure}[b]{0.475\textwidth}  
            \centering 
            \includegraphics[width=\textwidth]{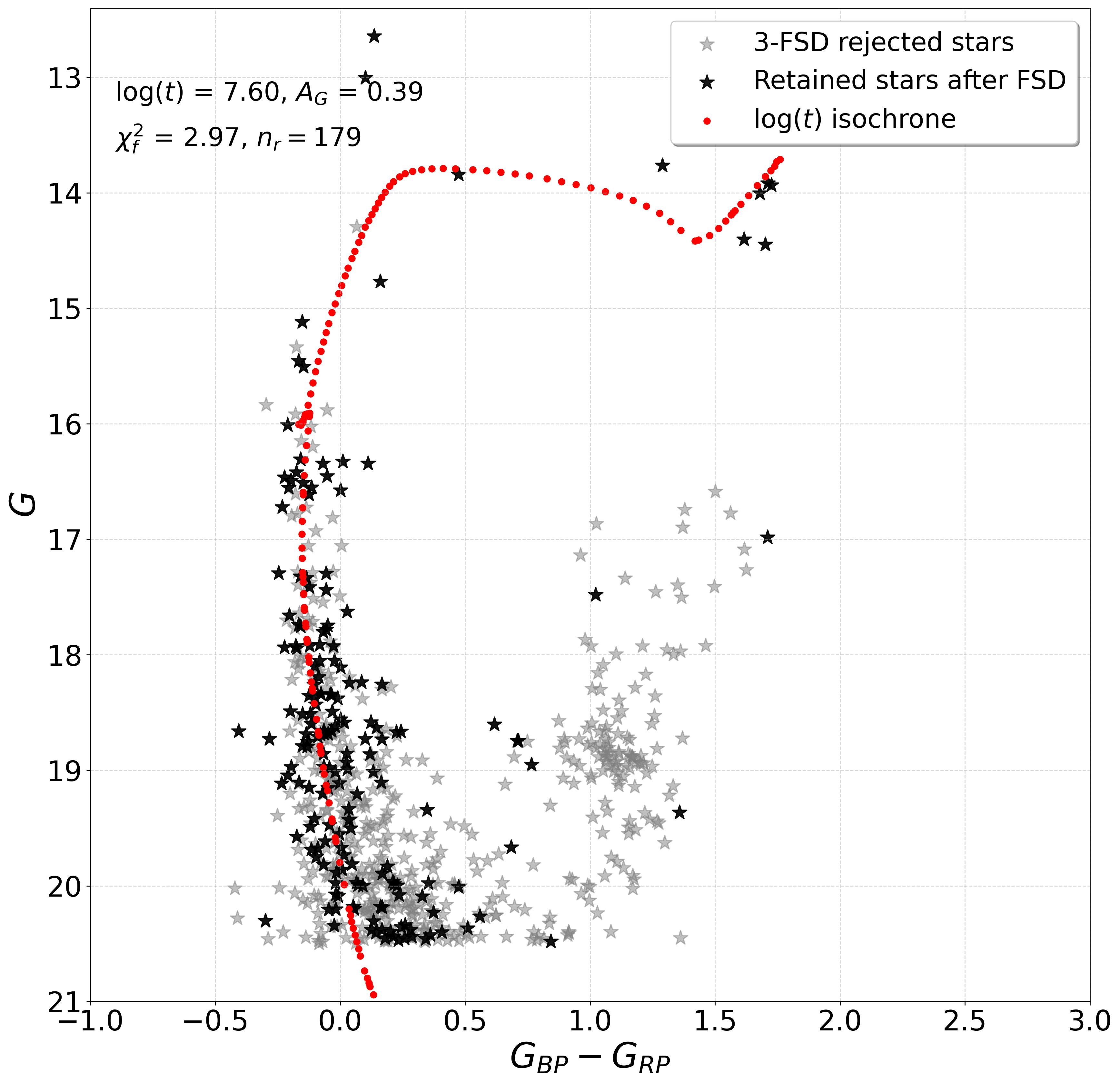}
            \caption[]%
            {{\small Cleaned and fitted CMD after FSD}}    
            \label{fig:NGC1735}
        \end{subfigure}
        \caption[]
        {\small NGC1735: a) Cluster CMD with initial selection criteria (proper motion, parallax, and  G magnitude cut-offs). b) The parameters ( $\log(t)$, A$_G$, and $\chi^{2}$) estimated at DM $=$ 18.53 are provided. The cluster region after FSD shows a significant decrease in field stars.} 
        \label{fig:NGC1735_cleaned}
    \end{figure*}

Hence, we identified the best-fitting isochrones and the corresponding  $\chi^2_f$ values for all the cleaned CMDs. The age and extinction parameters corresponding to the best-fitting isochrone were estimated for all cleaned CMDs. In the case of Groups-3 \& 4, only one cleaned CMD was created and therefore only one set of parameters was estimated per cluster. Two sample clusters and isochrone-fitted CMDs are shown in Figures \ref{fig_ngc458} and \ref{fig:NGC1735_cleaned}. 

In the case of Groups-1 \& 2, where multiple annular field regions are used for FSD, the isochrone fitting along with the parameter estimation was performed on all the cleaned CMDs of a cluster.  We found that star clusters in Groups-1 \& 2 showed variation in the number of retained members in CMDs, depending on the number of annuli used in FSD. Depending on the spatial crowding in nearby field regions, the number of annular regions used for FSD was found to have a considerable effect on the cleaning process, even in the case of dense clusters and particularly for those in the central regions. In the case of isolated poor star clusters, a significant number of cluster members were not retained in the cleaned CMD if all five annular field regions were used. The above observations suggested that the choice of best cleaned CMD for a cluster should be made by considering, (1) the variation in the estimated parameters as a function of the number of annular field regions used in FSD,  (2) over-subtraction of possible cluster members in the CMDs as a function of the number of annular fields used in FSD, (3) $\chi^2_f$ value of the fit, and (4) visual goodness of isochrone fit to the cleaned CMD.  We describe the adopted criteria to choose the best cleaned CMD for clusters belonging to Groups-1 \& 2, in the next section.

\subsection{Selection criteria for the cleaned cluster CMDs.}\label{select_crit_CMD}

In the case of Groups-1 \& 2 clusters, the cluster parameters and the $\chi^2_f$ values differ for the n-FSD cleaned CMDs (n = 1 to 5) of the same cluster. In order to identify and retain the best isochrone-fitted cleaned CMD for a cluster, we implemented the following selection criteria in three steps:
\begin{enumerate}
    \item[(1)] We applied a cut-off of $\chi_f^{2} \leq 15$, based on a visual check of the fitted CMDs of a smaller sample of clusters to remove bad fits.
    \item[(2)] We compared 3-FSD and 5-FSD CMDs for the same cluster and retained those clusters with age separation within $|\Delta \log(t)| \leq 0.1$ (absolute difference between 3-FSD and 5-FSD $\log(t)$), so that the number of field annuli used in the FSD did not influence the estimated age.
    \item[(3)] We used a $|\Delta \chi_f^{2}| \leq 1$ (difference between 3-FSD and 5-FSD $\chi_f^{2}$) to retain clusters that did not show a significant difference in the estimated parameters, such that the number of field annuli used in the FSD did not influence the quality of fit. Since the over-subtraction of cluster members generally happens more in the 5-FSD CMDs than in the 3-FSD CMDs, we selected the 3-FSD CMDs as the best-fit CMDs and the corresponding parameters for these clusters.
\end{enumerate}

 The plots comparing the $\log(t)$ and the $\chi_f^{2}$ in the steps mentioned above for the various FSDs are shown in Figure \ref{fig_FDstages} in the Appendix.

In the case of clusters with $|\Delta \chi_f^{2}| > 1$, as per the criterion in step (3), but satisfying the criterion in step (2), we compared $\chi_f^{2}$ values from 3-FSD and 5-FSD fits, then retained that with the lesser value of $\chi_f^{2}$ and the corresponding parameters. This case suggests the possibility of over-subtraction of members in the 5-FSD.
Clusters with $|\Delta \log(t)| > 0.1$ after step (2) suggest that the estimated parameters are impacted by the choice of FSD, which results from the number of annular regions used. Therefore, as a next step to reduce deviation in the estimated parameters, we compared the fits of the 4-FSD and 2-FSD CMDs as in the above-indicated step (2), followed by the same steps as for the 5-FSD and 3-FSD CMDs. Clusters with $|\Delta \log(t)| > 0.1$  in this step were then taken to the next step of 2-FSD and 1-FSD comparison (see figures in Appendix A for details). All clusters reaching this stage will have either parameter from the 2-FSD or 1-FSD fits.

 The above series of procedures is expected to reduce any impact of the methods of FSD used in the estimation of the cluster parameters. We used two quantities,  the deviation in age as well as the difference in $\chi_f^{2}$, as control parameters to finally choose the number of annuli used in the FSD, instead of arbitrarily deciding the number of annuli for FSD. We believe that this helps to reduce erroneous parameter estimation due to variable field star density, over-subtraction of members in poor clusters, and statistical fluctuation in field stars located in the dense regions of the MCs. In the case of Groups-3 \& 4 clusters with only one CMD after FSD, we applied $\chi_f^{2} \leq 20$ to filter in. Since the clusters in these two groups have only one field star region for FSD and are likely to introduce scatter in the CMDs, a slightly higher cut-off of $\chi_f^{2}$ is justified.

Though the entire process was automated from FSD to isochrone fitting, a visual inspection of the final fitted CMDs was required to check for any spurious fits. Hence we inspected all the cluster CMDs after the completion of parameter estimation and manually rejected a few improperly cleaned/ fitted CMDs (26 clusters).

We note that we did not consider poor clusters with less than 8 stars in the cleaned CMDs, apart from the overlapping clusters, as mentioned earlier. The typical number of our cluster members, retained after FSD, averages to 40. We also did not retain clusters for which we were not able to get good fits (as decided by visual inspection) and/or with $\chi_f^{2} > 20$. A map of the 1082 non-characterized clusters is shown in Figure \ref{fig:nt_ct} in the Appendix.

\begin{figure}
    \centering
       \includegraphics[width=1\linewidth]{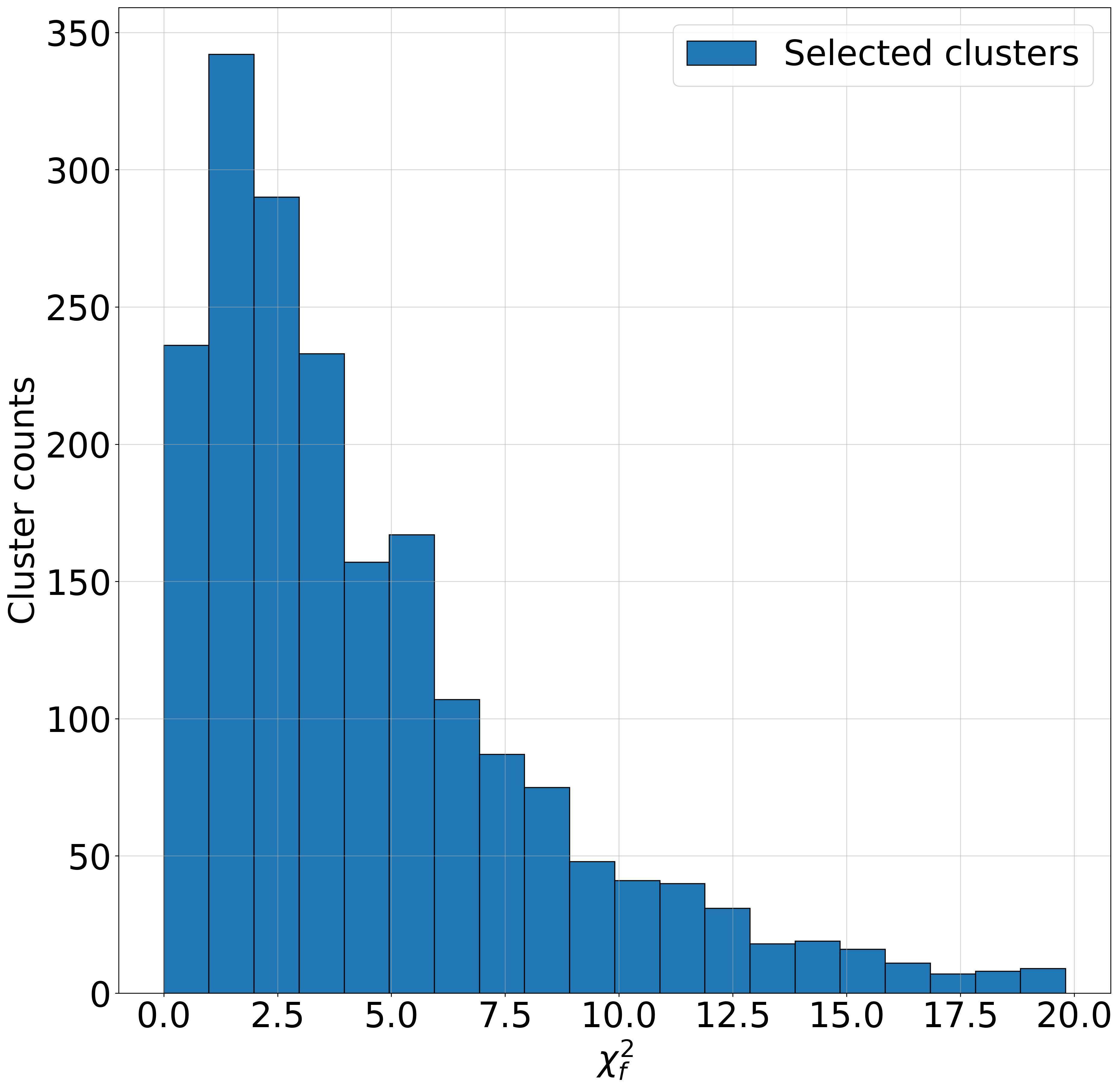}
       \caption{The $\chi_f^{2}$ distribution for the selected 1990 clusters, with a cut-off of $\chi^2_f \leq 20$ to retain the best isochrone-fitted cleaned CMD.}
       \label{fig:main_vgf} 
    \end{figure}

The distribution of $\chi_f^{2}$ for the finally selected CMDs of the retained 1990 clusters is shown in Figure \ref{fig:main_vgf}. The peak of the distribution with more than 300 clusters is at $\sim\chi_f^{2}=1.25$, and about 1000 clusters ($\sim$ 50\%) have $\chi_f^{2}$ value within 1$\sigma$ of the distribution. We conclude that these final fits are therefore satisfactory, based on the $\chi_f^{2}$ values and visual inspection.

 We estimated satisfactory parameters of 1990 clusters, of which 1710 are in the LMC and 280 are in the SMC. Out of the estimated parameters of clusters, 1651 clusters belong to Group-1, 169 clusters belong to Group-2, 117 clusters belong to Group-3, and 53 clusters belong to Group-4. We characterized 960 clusters ($\sim$ 48.2\%) for the first time in comparison to the existing star cluster catalogs of the MCs (\citealt{pu2000AcA....50..337P,pusmc1999AcA....49..157P}; \citealt{chiosi2006yCat..34520179C}; G10; \citealt{palma2016A&A...586A..41P}; N16; N18).  Among these, 847 clusters ($\sim$ 49.5\% of 1710) are in the LMC and 113 ($\sim$ 40.3\% of 280) in the SMC. The following section deals with the MCMC sampling to arrive at the final estimates of age, extinction, distance modulus, and metallicity of clusters from their prior assumptions.

\subsection{MCMC sampling to estimate age, extinction, distance modulus, and metallicity}\label{mcmc_sect}

The estimated age-extinction-metallicity-distance values associated with clusters are sensitive to the choice and combination of model parameter values. Also, in the previous section, age and extinction were estimated for a fixed value of metallicity and distance. We developed an MCMC sampling method to arrive at the final estimates of all of the above four parameters.

In order to perform an MCMC sampling, we defined a range for the parameters, within which the sampling is performed. As 960 clusters ($\sim$ 48\% of 1990) are characterized for the first time, there are no other estimations available to constrain the parameter range. Hence the uniform priors were adopted for the extinction ($A_{G}$) and metallicity during the MCMC sampling. As the SMC is known to be more metal-poor than the LMC, we adopted a Z range of [0.0006 to 0.005] for the SMC, and [0.0015 to 0.01] for the LMC. The assumed Z range is consistent with the prior Z$_{\text{LMC}}$ = 0.008 and Z$_{\text{SMC}}$ = 0.004 that we adopted (subsection \ref{prior_stage}) for the SMC and LMC, respectively. The extinction ($A_G$) was allowed to vary from 0 to 0.836 and the distance modulus (DM) to vary within 0.25 mag from the prior assumption. In the case of age, the sampling was confined within the 15\% change from the maximum likelihood age estimated from the prior information. The goal was to get the expected change in $A_G$, Z, and DM from their prior assumptions and to obtain a robust age estimation.

The sampling likelihood function (LF) is defined to be the average $\chi^2_{f}$ associated with each CMD data point (magnitude, color, magnitude error, color error: $m_{i,o}$, $c_{i,o}$, $sm{_i}$, $sc{_i}$ ) and nearest isochrone model data points (model magnitude, model color: $m_{i,m}$, $c_{i,m}$ ). The $sm{_i}$, $sc{_i}$ represent the scaled error values after taking the flux errors to be at least 4\% of the observed flux values (as explained in subsection \ref{prior_stage}). And the LF is given in the following equation,
\begin{align}
\centering
    LF =& \frac{-1}{N} \sum_{i=1}^{N} \left\{\frac{(m_{i,o}-m_{i,m})^2}{{sm{_i}}^2}+\frac{(c_{i,o}-c_{i,m})^2}{{sc{_i}}^2}\right\} \label{eqlf}
\end{align}
The cleaned CMDs have the presence of randomly scattered field stars (as mentioned in subsection \ref{fsd_stage}). For the reason, within the color-magnitude space for a given cluster, we considered stars within 1$\sigma$ deviation  with respect to the prior estimates ($\log(t)$, $A_{G}$, Z, DM) for that cluster.

The sampling uses a Bayesian approach with an ensemble of multiple walkers (20 walkers) with a combination of stretch and walk proposal moves \citep{gdman2010CAMCS...5...65G} to explore the parameter space. In addition, the standard Metropolis-Hastings algorithm is also used for the proposal moves with a tuned acceptance ratio of $0.225$ to $0.35$ during the iteration phase. The iterations were performed with the mixture of different moves till the convergence was achieved in the posterior distribution of samples. For this study, we developed and implemented our own sampling algorithm, based on the algorithm of \cite{gdman2010CAMCS...5...65G} and adopted for this study. The code is implemented in C language. The code is automated and is capable of handling large number of clusters efficiently.

The sampled posterior distributions of two sample clusters are shown in Figures \ref{fig:mcmc_2gc276} and \ref{fig:mcmc_hw66} in the Appendix.
The MCMC sampling mainly explores the parameter space in $A_G$, DM, and Z to find the range under which our age estimates have confidence. Also, MCMC sampling solves the expected values for $A_G$, DM, and Z with its uncertainties, which cannot be easily obtained using the ordinary $\chi^2$ minimization.

\section{Results}\label{sec_4}

This is the first comprehensive study of star clusters in the MCs using the \textit{Gaia} DR3 data, where the parameters are estimated using an MCMC sampling from the assumed priors. The clusters are parameterized after removing the field stars from the cluster CMD. Considering the uniform data coverage of the \textit{Gaia} data, the exclusion of overlapping clusters in the LOS, and the implementation of the FSD by eliminating foreground stars and kinematically deviant sources, this study provides reliable parameters to explore the star cluster properties of the MCs. As mentioned in subsection \ref{select_crit_CMD}, we estimated parameters for 1990 star clusters. The newly characterized clusters (960 clusters) in this study are mostly located in the outer regions of the MCs. This study, therefore, majorly contributes to the understanding of clusters and their properties in the outer regions of LMC, located more than 4 kpc from the LMC center.

To compare and understand the region-wise difference among clusters in various locations, the cluster coordinates need to be transformed from the skyplane to the galaxy plane. In the case of the LMC, the star cluster coordinates from the sky plane are transformed to the plane containing the LMC disc, using the coordinate transformations given by \cite{vand22001AJ....122.1827V}. Since the morphology of the SMC is not obvious, we keep the SMC cluster coordinates in the Cartesian projection of the sky plane with respect to the center of SMC. The central coordinates for the LMC and the  SMC are adopted from \cite{centerlmc1972VA.....14..163D}. In the following sections, we present the extinction and metallicity maps, CF history, and spatio-temporal map of cluster distribution.

\subsection{Extinction and metallicity maps}\label{ag_met_sec}

\begin{figure*}
        \centering
       \begin{subfigure}[b]{0.475\textwidth}   
            \centering 
            \includegraphics[width=\textwidth]{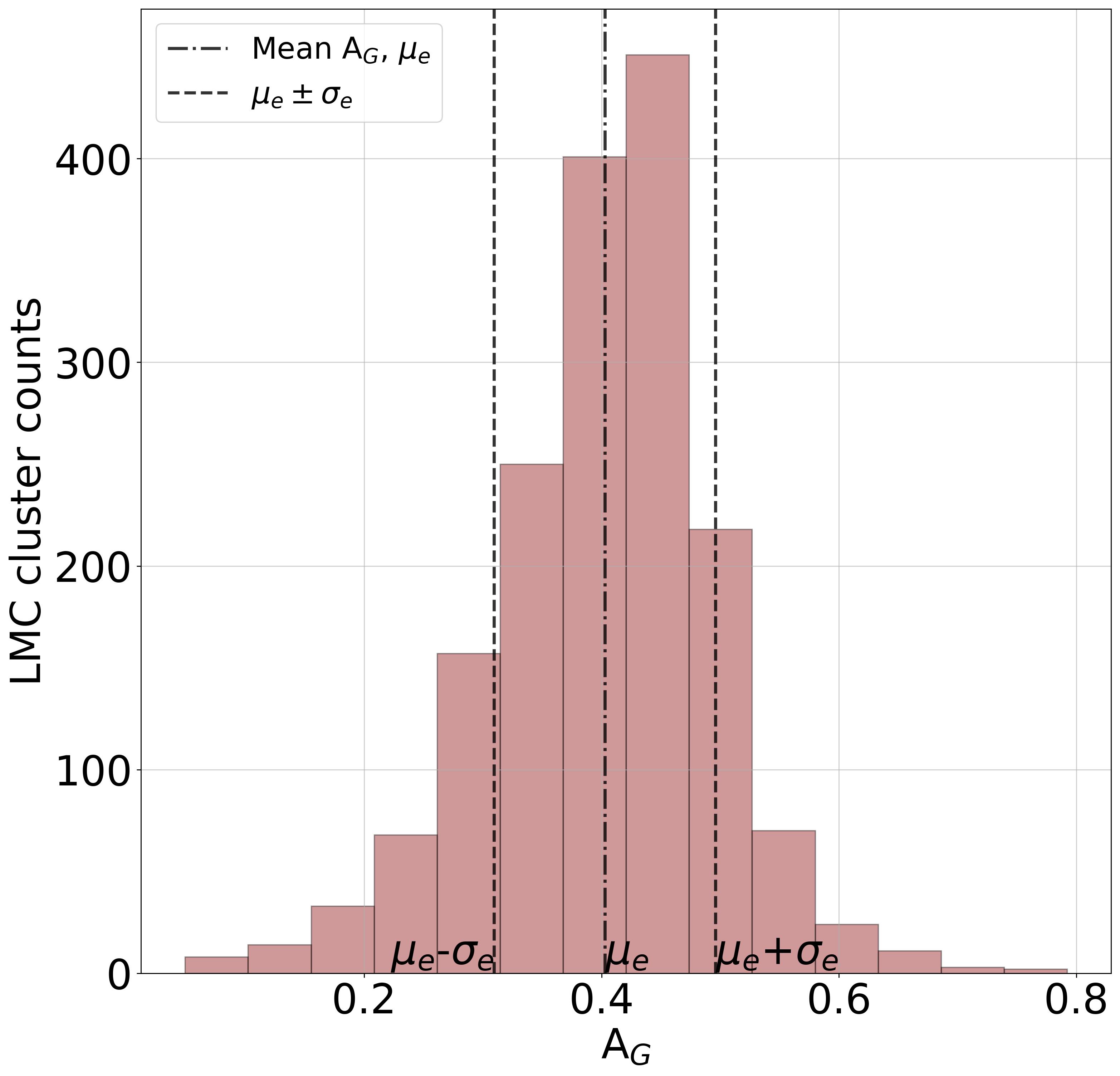}
            \caption[]%
            {{\small  Distribution of extinction ($A_G$) in the LMC.}}    
            \label{fig:LMC_ext_hist}
        \end{subfigure}
        \hspace*{0.1in}
        \begin{subfigure}[b]{0.475\textwidth}   
            \centering 
            \includegraphics[width=\textwidth]{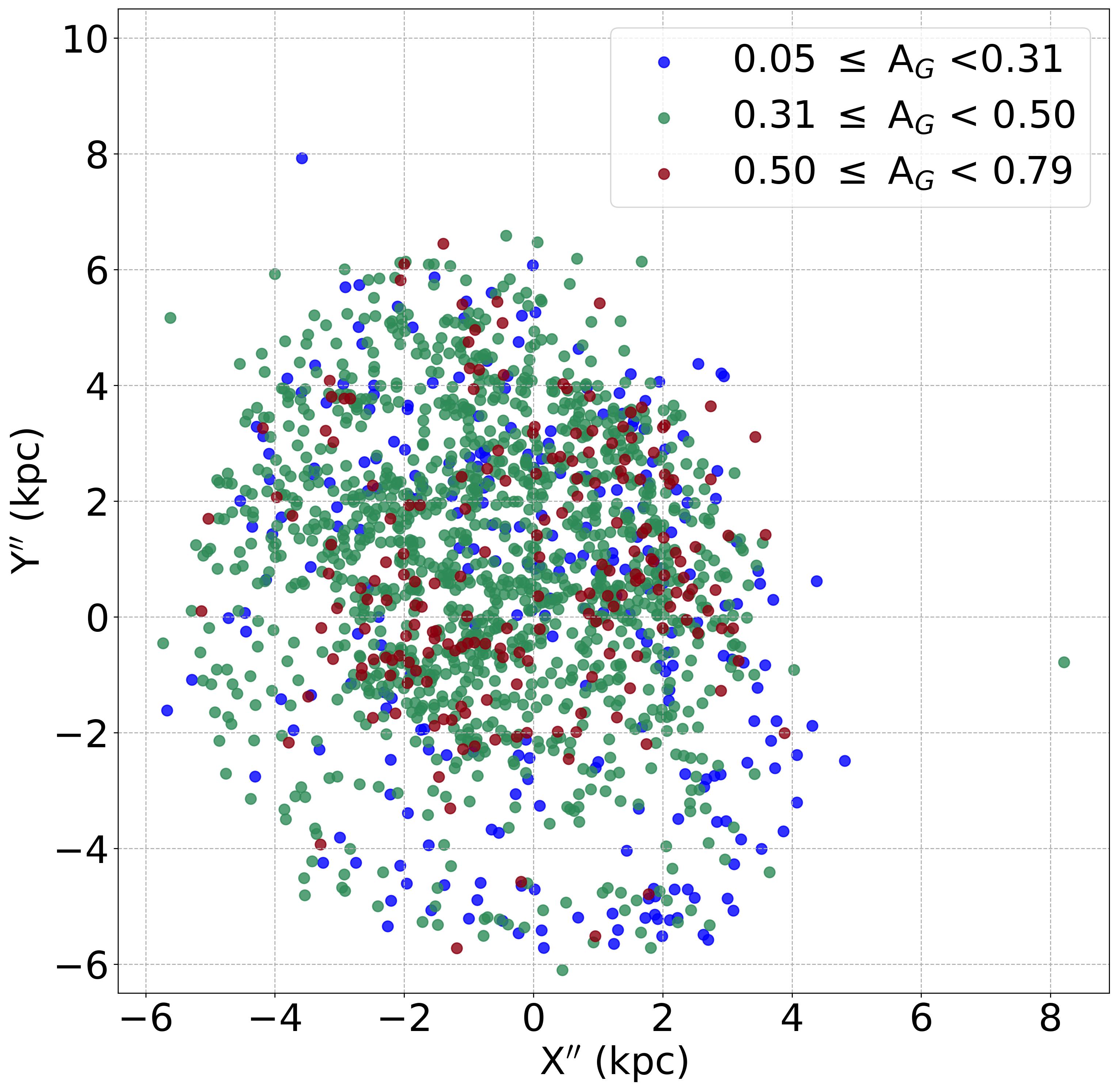}
            \caption[]%
            {{\small Spatial extinction ($A_G$) map of clusters in the LMC plane.}}    
            \label{fig:LMC_ext_map}
        \end{subfigure}
         \vskip\baselineskip
        \begin{subfigure}[b]{0.475\textwidth}   
            \centering 
            \includegraphics[width=\textwidth]{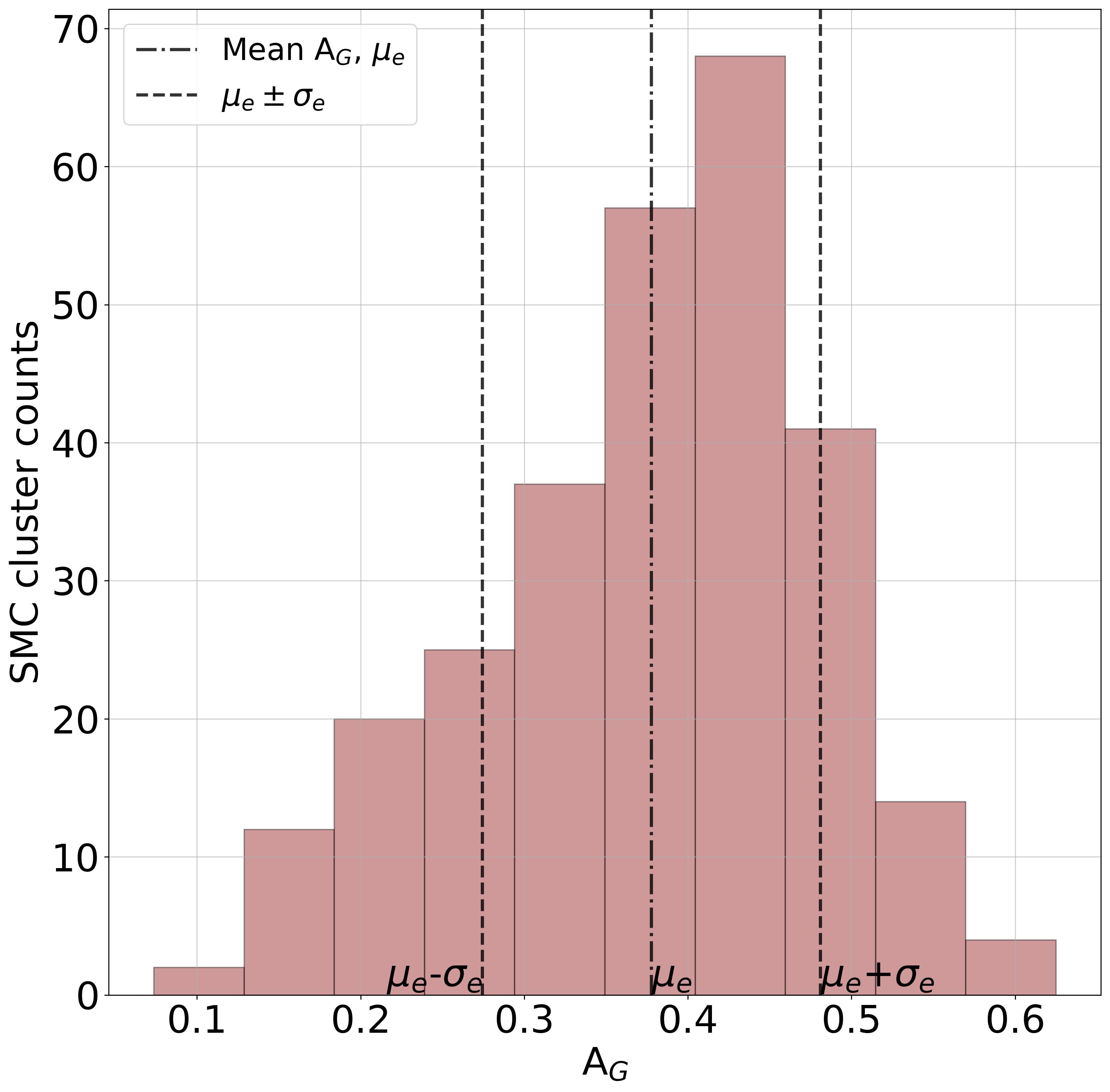}
            \caption[]%
            {{\small Distribution of extinction ($A_G$) in the SMC.}}    
            \label{fig:Extinction_hist_smc}
        \end{subfigure}
        \hspace*{0.1in}
        \begin{subfigure}[b]{0.475\textwidth}   
            \centering 
            \includegraphics[width=\textwidth]{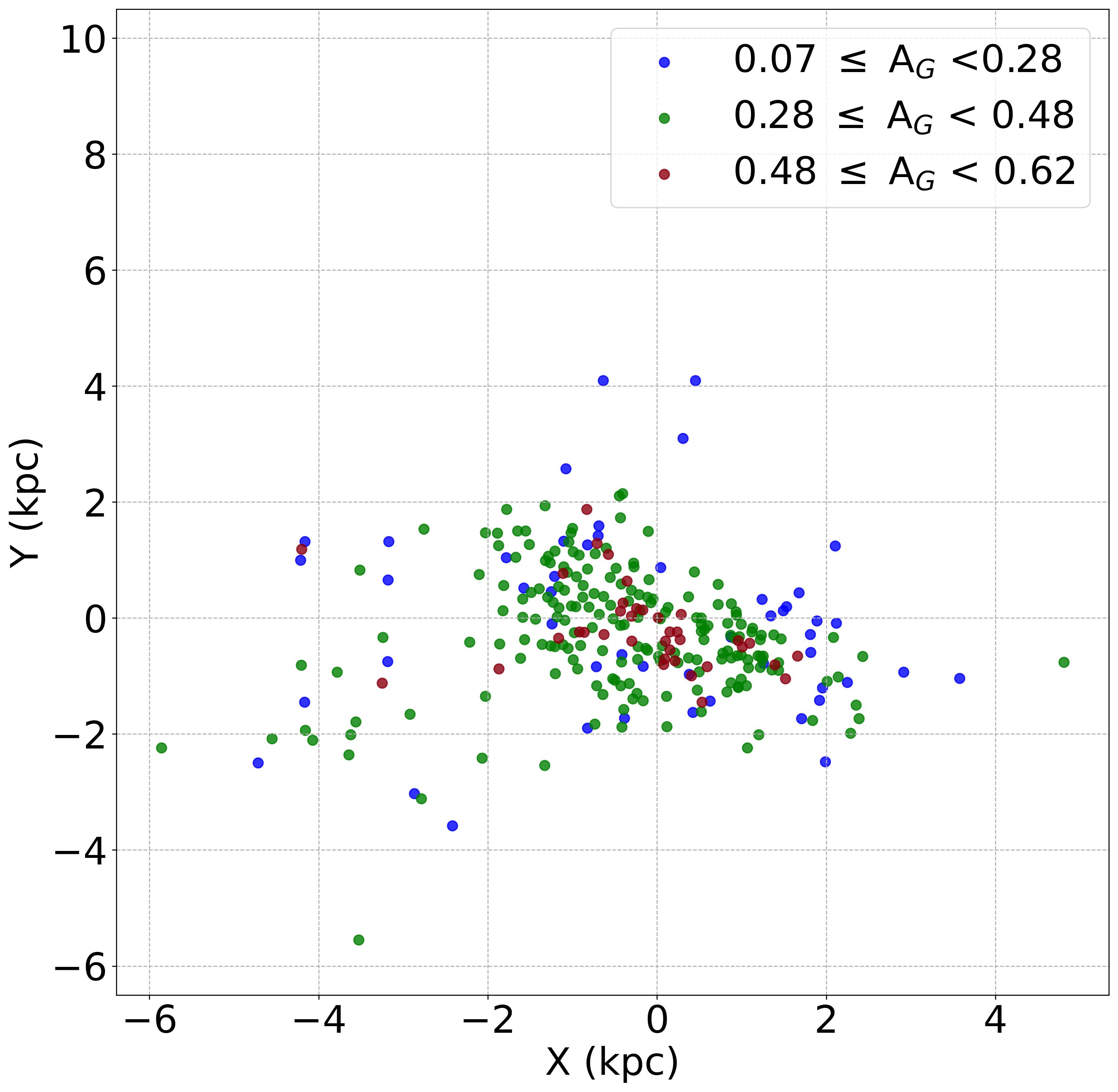}
            \caption[]%
            {{\small Spatial extinction ($A_G$) map of clusters in the SMC (cartesian sky plane).}}    
            \label{fig:SMC_ext_map}
        \end{subfigure}
        \caption[]
        {\small Spatial map and distribution of extinction ($A_G$) from the 1990 star clusters in the MCs. The bin width of $0.25\times$ median error of cluster $A_G$ estimations  (50 percentile) is used here to generate the histograms. The three groups are color-coded based on the mean ($\mu_e$) and standard deviation ($\sigma_e$) of $A_G$ distribution as mentioned in subsection \ref{ag_met_sec}.} 
        \label{fig1_ext_hist_mc}
    \end{figure*}

The extinction and the metallicity parameters for 1990 star clusters are presented here. The extinction histograms for the LMC and the SMC are shown in Figures \ref{fig:LMC_ext_hist} and \ref{fig:Extinction_hist_smc}, with the mean extinction estimated as $\mu_e$ = 0.40$\pm$0.005 mag and 0.38$\pm$0.006 mag respectively. We adopted the constant $R_V$ of 3.41 \citep{rv_lmc2003ApJ...594..279G} for the LMC and 2.72 \citep{rv_dep1994ApJ...422..158O} for the SMC in our reddening estimations. The mean reddening values in $E(B-V)$ are estimated as $0.14\pm0.001$ mag and $0.16\pm0.003$ mag for the LMC and the SMC, respectively. The spatial extinction maps of the LMC and the SMC clusters are shown in Figures \ref{fig:LMC_ext_map} and \ref{fig:SMC_ext_map} respectively, where three groups ($A_G<\mu_e-\sigma_e$, $\mu_e-\sigma_e\leq$$A_G$<$\mu_e+\sigma_e$, $A_G\ge\mu_e+\sigma_e$) are shown with blue, green and red colors, respectively. The central regions of the LMC and the SMC appear to have higher extinction than the outer regions. In the case of the LMC, a significant amount of clusters can be seen towards the northern part, with higher extinction than in the South. The ring of clusters on the outskirts of the LMC has lower extinction than the rest. As this study has covered more clusters in the outer LMC, where the extinction/ reddening is low, the average values for the LMC also tend to be on the lower side.  

\begin{figure*}
        \centering
        \begin{subfigure}[b]{0.475\textwidth}
            \centering
            \includegraphics[width=\textwidth]{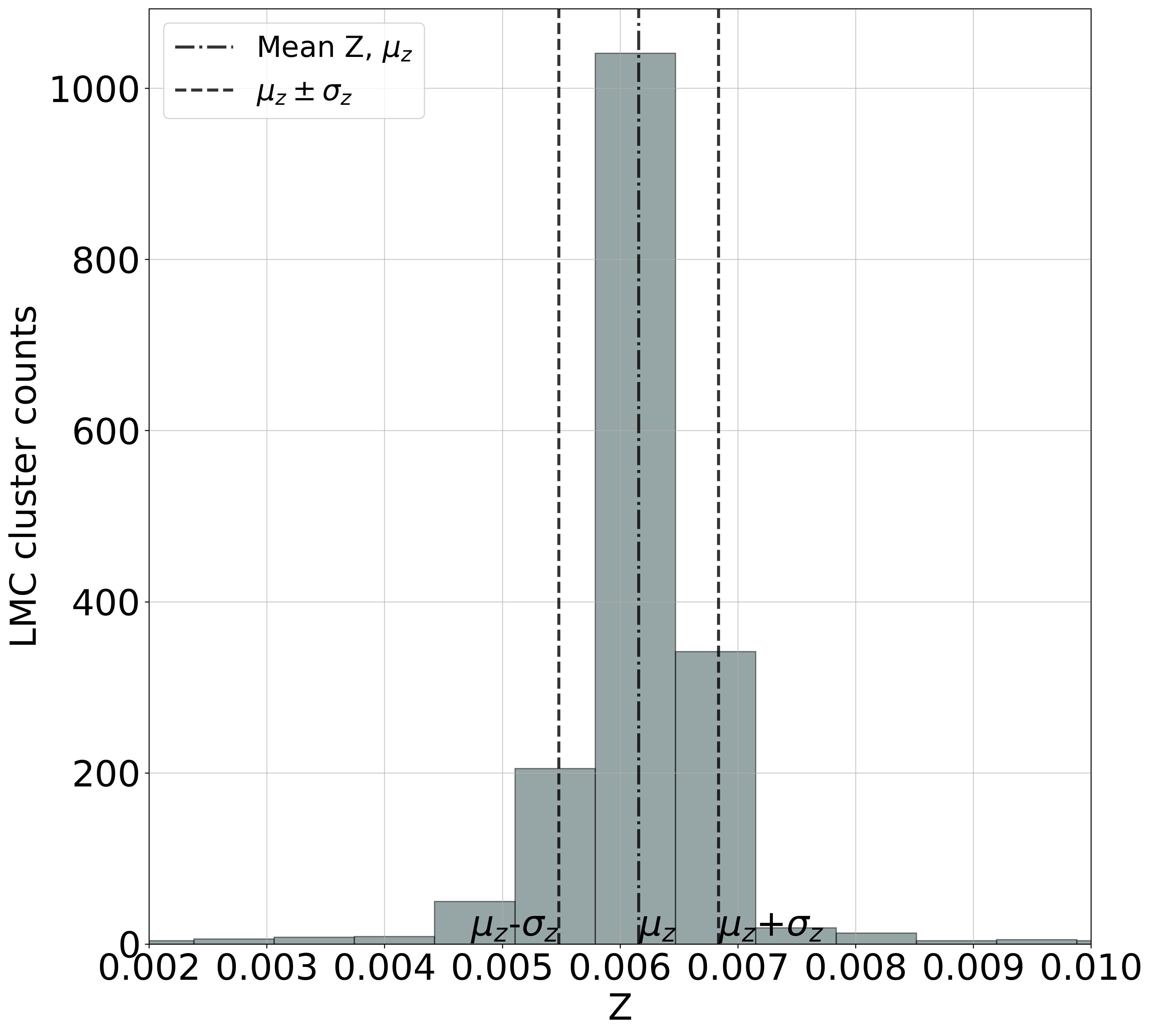}
            \caption[Network2]%
            {{\small Distribution of metallicity (Z) in the LMC.}}   
            \label{met_lmc_hist}
        \end{subfigure}
        \hfill
        \begin{subfigure}[b]{0.475\textwidth}   
            \centering 
            \includegraphics[width=\textwidth]{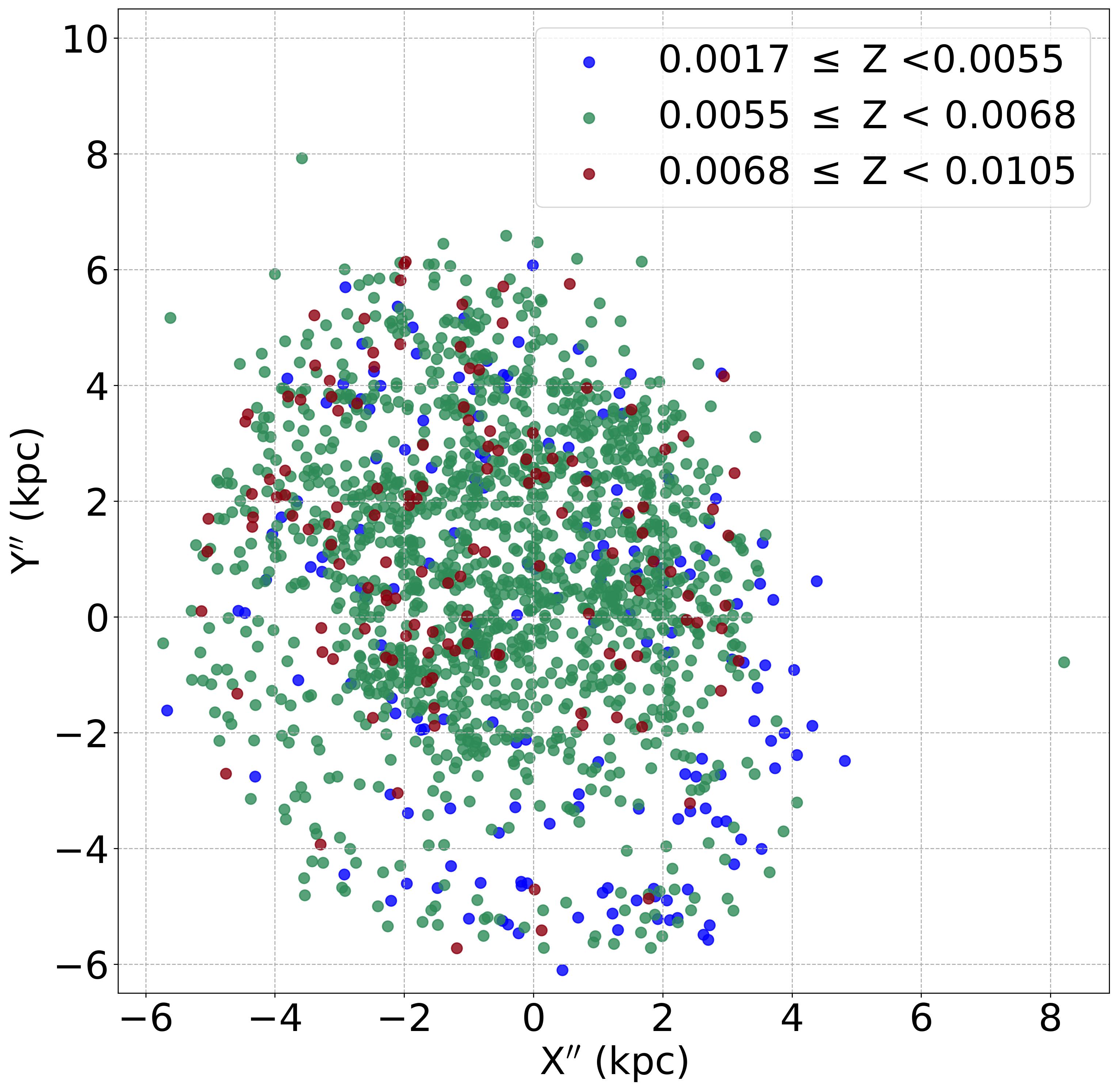}
            \caption[]%
            {{\small Spatial metallicity (Z) map of clusters in the LMC plane.}}    
            \label{met_lmc_map}
        \end{subfigure}
        \vskip\baselineskip
         \begin{subfigure}[b]{0.475\textwidth}  
            \centering 
            \includegraphics[width=\textwidth]{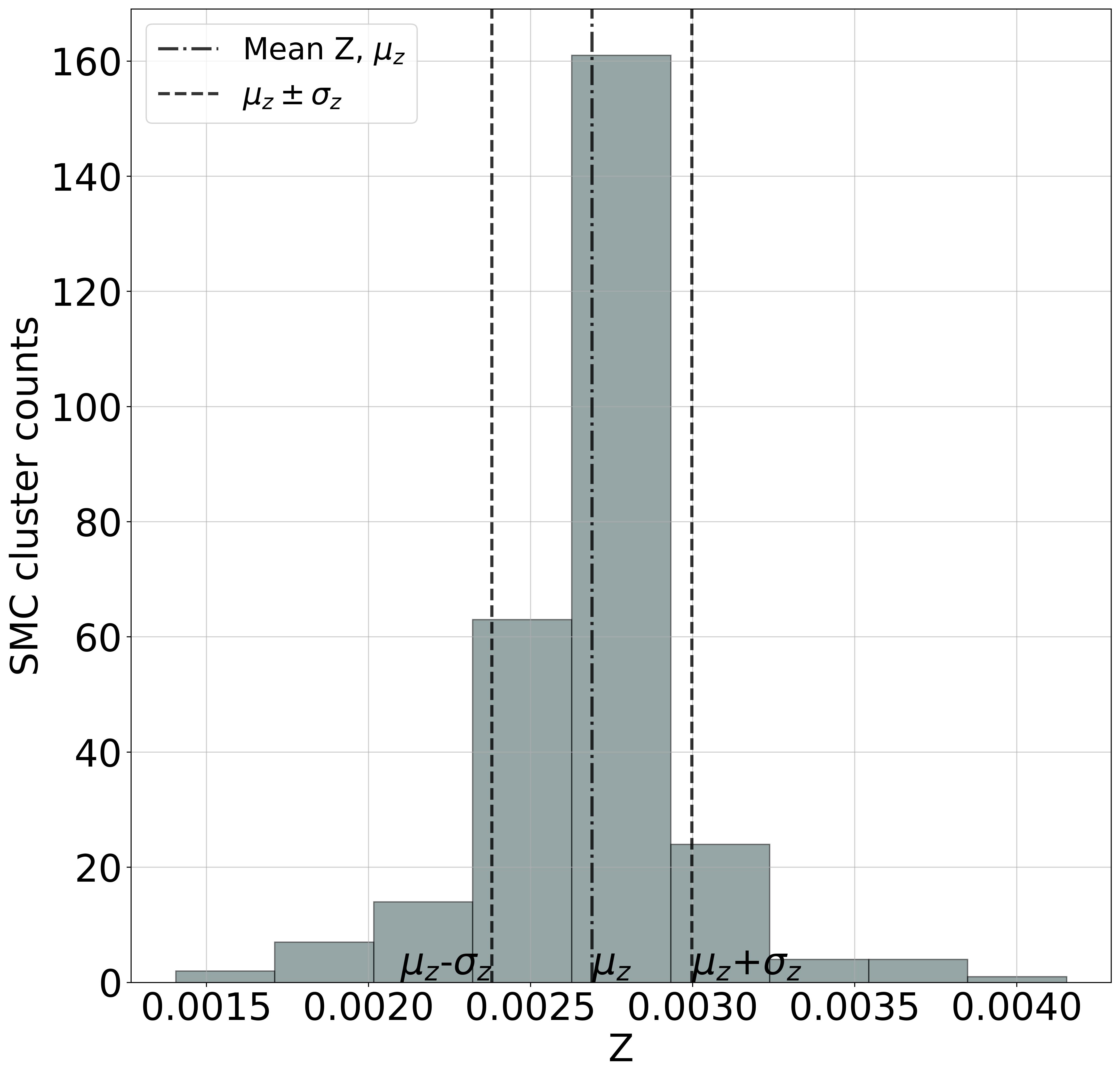}
            \caption[]%
            {{\small Distribution of metallicity (Z) in the SMC.}}    
            \label{met_smc_hist}
        \end{subfigure}
        \hfill
        \begin{subfigure}[b]{0.475\textwidth}   
            \centering 
            \includegraphics[width=\textwidth]{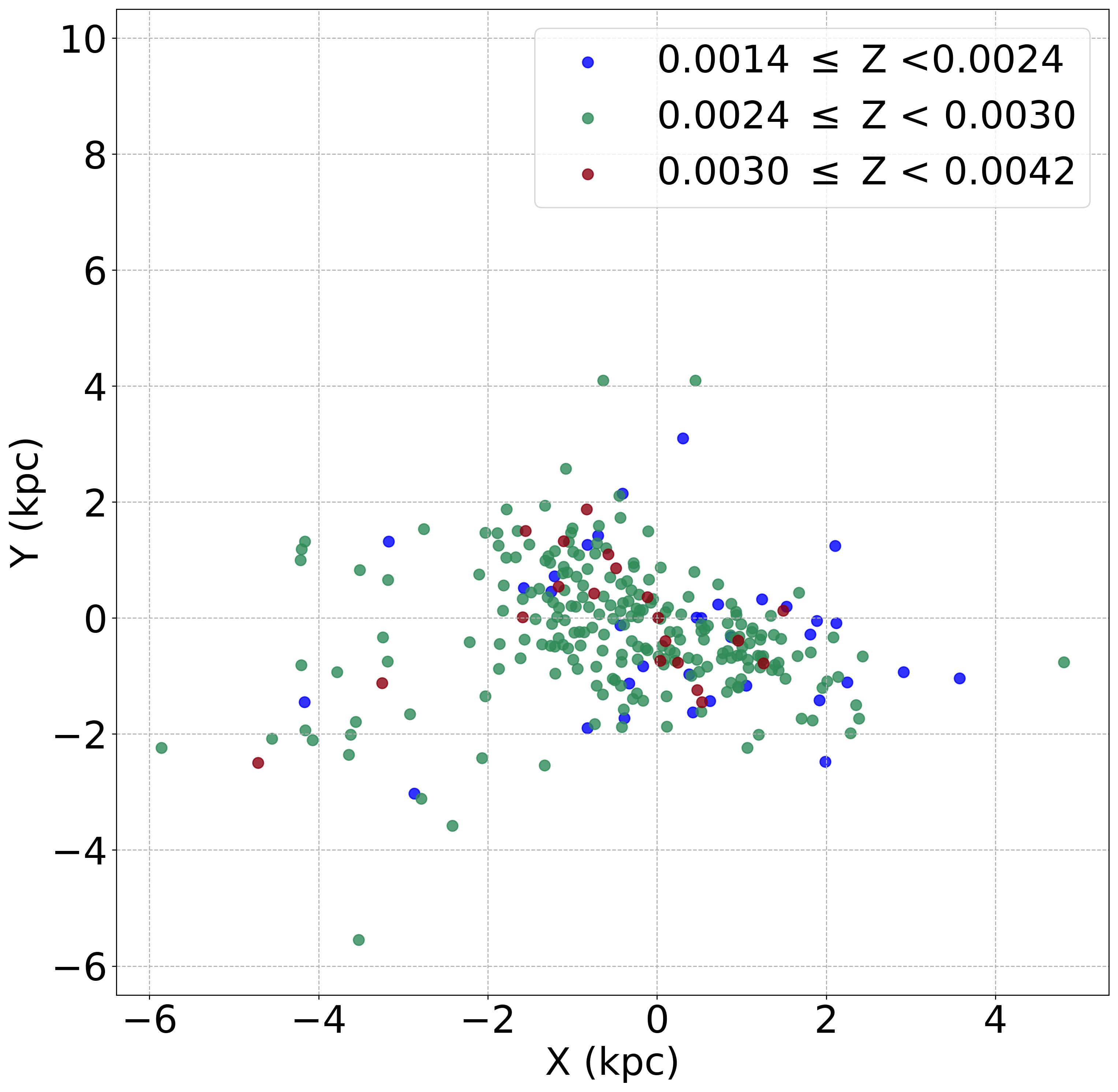}
            \caption[]%
            {{\small Spatial metallicity (Z) map of clusters in the SMC (cartesian sky plane).}}    
            \label{met_smc_map}
        \end{subfigure}
        \caption[]
        {\small Spatial map and metallicity (Z) distribution of 1990 star clusters in the MCs. The bin width of $0.25$ times the median error (50 percentile) in Z estimations are used here to generate the histograms. The three groups are color-coded based on the mean ($\mu_z$) and standard deviation ($\sigma_z$) of Z distribution as mentioned in subsection \ref{ag_met_sec}.} 
        \label{met_mc_plot}
    \end{figure*}

The metal fraction (Z) is also estimated using the uniform priors about the prior knowledge ($Z = 0.008$ for the LMC and $Z = 0.004$ for the SMC) as mentioned in subsection \ref{mcmc_sect}. The distributions of the estimated Z are shown in Figures \ref{met_lmc_hist} and \ref{met_smc_hist} for the LMC and the SMC clusters, respectively. We estimated a mean metal fraction, $\mu_z = 0.006\pm0.00002$ for the clusters in the LMC, and $\mu_z = 0.0027\pm0.00002$ for the clusters in the SMC. The spatial maps of Z for the LMC and the SMC clusters  are shown in Figures \ref{met_lmc_map} and \ref{met_smc_map} respectively, where three groups (Z$<\mu_z-\sigma_z$, $\mu_z-\sigma_z\leq$Z$<\mu_z+\sigma_z$, Z$\ge\mu_z+\sigma_z$) are shown with blue, green and red colors, respectively. In the LMC, the spatial plot shows a relatively more number of metal-rich clusters (shown in red) towards the North and North-East of the galaxy, whereas the southern LMC has more of the metal-poor clusters (shown in blue).  In the case of the SMC, the spatial plot shows metal enrichment towards the central regions compared to the outer regions. Based on this study, we recommend that the choice of Z value for isochrones to fit the CMDs of clusters with age $\le$ 1 - 2 Gyr is 0.004 to 0.008 for the LMC and 0.0016 to 0.004 for the SMC.

The PARSEC models also give the [M/H] values corresponding to the Z values of the isochrones. Based on this, we calculated the mean [M/H] values to be $-0.40\pm0.001$ dex and $-0.76\pm0.003$ dex for the LMC and the SMC, respectively. We note that the mean [M/H] estimated here is for the case of clusters younger than 1-2 Gyr age.

The DM for each star cluster was constrained during MCMC sampling, allowing for a variation of 0.25 mag from the prior values. We note that the posterior values of DM did not show any significant deviation from their prior values. Since the young clusters mostly do not have the RGB population in the cleaned CMDs, the degeneracy of the DM, $A_G$, and $\log (t)$ is present in our estimates. But as mentioned in subsection \ref{mcmc_sect}, the parameter estimation is weighed to find the best age of the cluster and the expected variation of other cluster parameters.

 \subsection{Estimated age distribution and Episodic Cluster Formation in the MCs}\label{mc_spc}
 
\begin{figure*}
        \centering
        \begin{subfigure}[b]{\textwidth}   
            \centering 
            \includegraphics[width=0.85\textwidth]{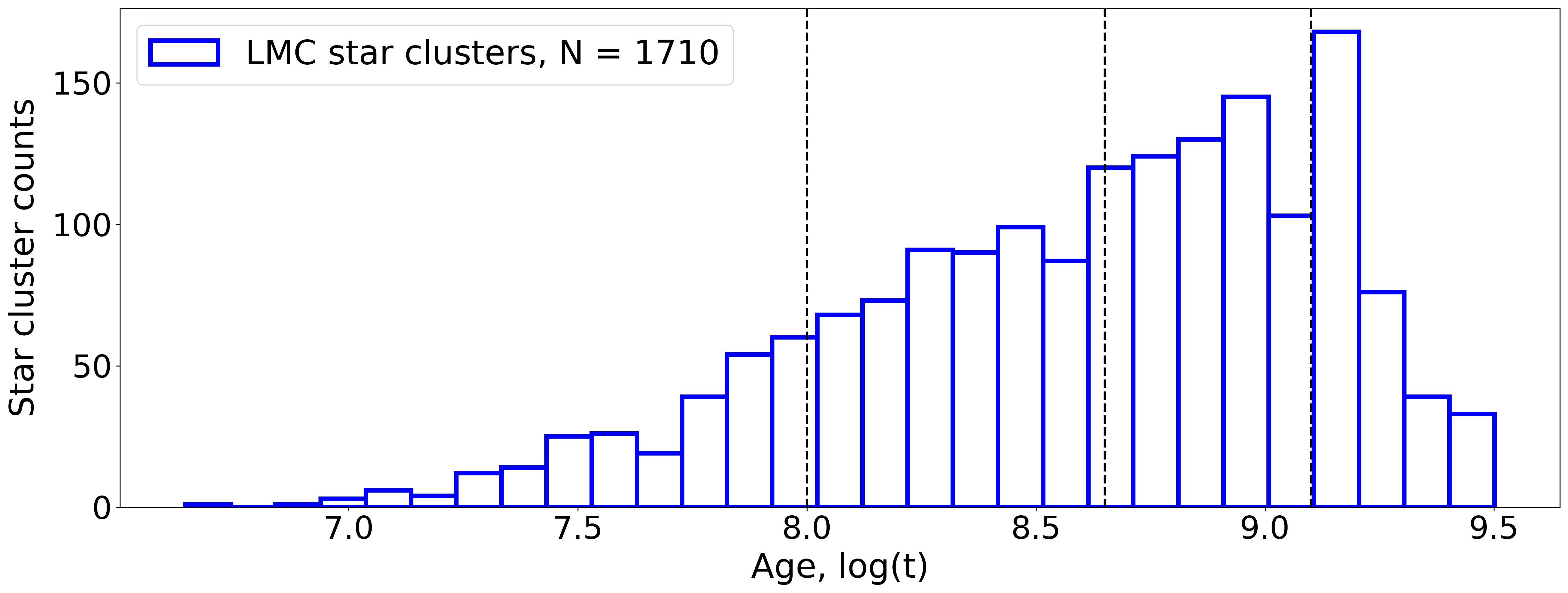}
            \caption[]%
            {{\small Age ($\log(t)$) histogram of clusters in the LMC.}}    
            \label{fig1_age_hist_1}
        \end{subfigure}
        \vskip\baselineskip
         \begin{subfigure}[b]{\textwidth}
            \centering
            \includegraphics[width=0.85\textwidth]{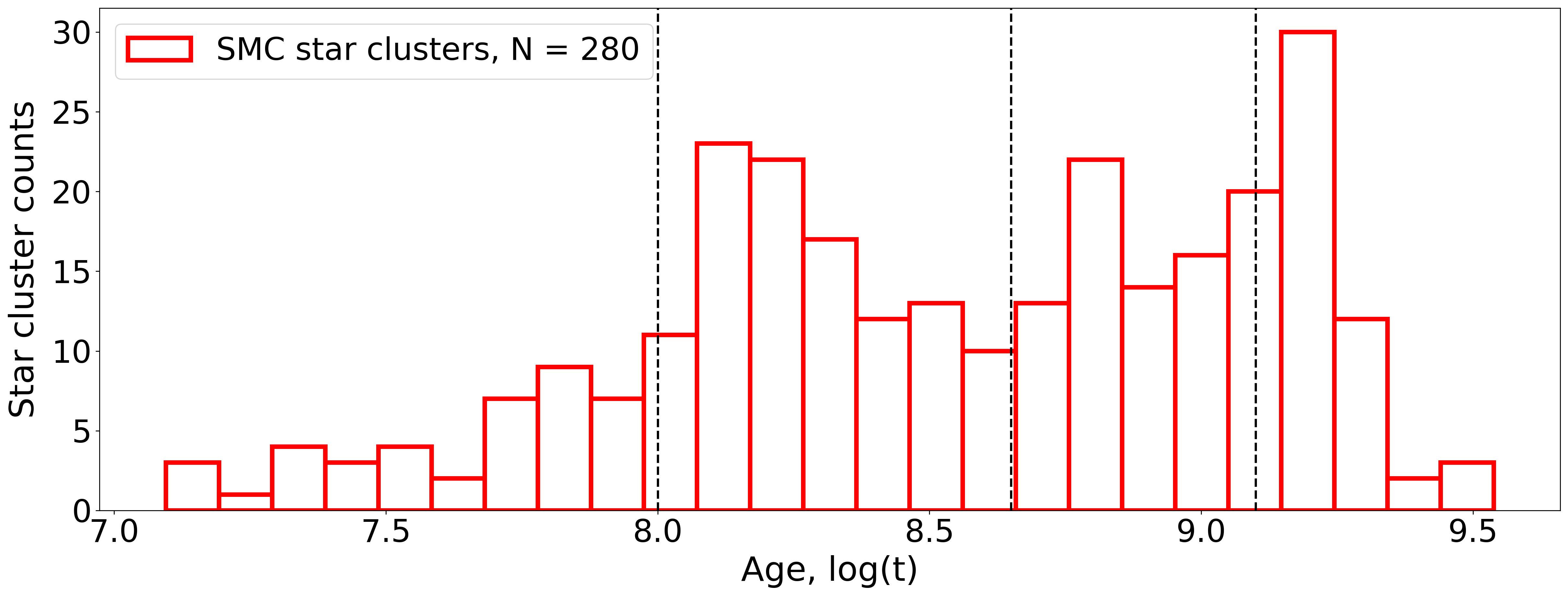}
            \caption[]%
            {{\small Age ($\log(t)$) histogram of clusters in the SMC.}}    
            \label{fig1_age_hist_2}
        \end{subfigure}
         \vskip\baselineskip
        \begin{subfigure}[b]{\textwidth}
            \centering
            \includegraphics[width=0.85\textwidth]{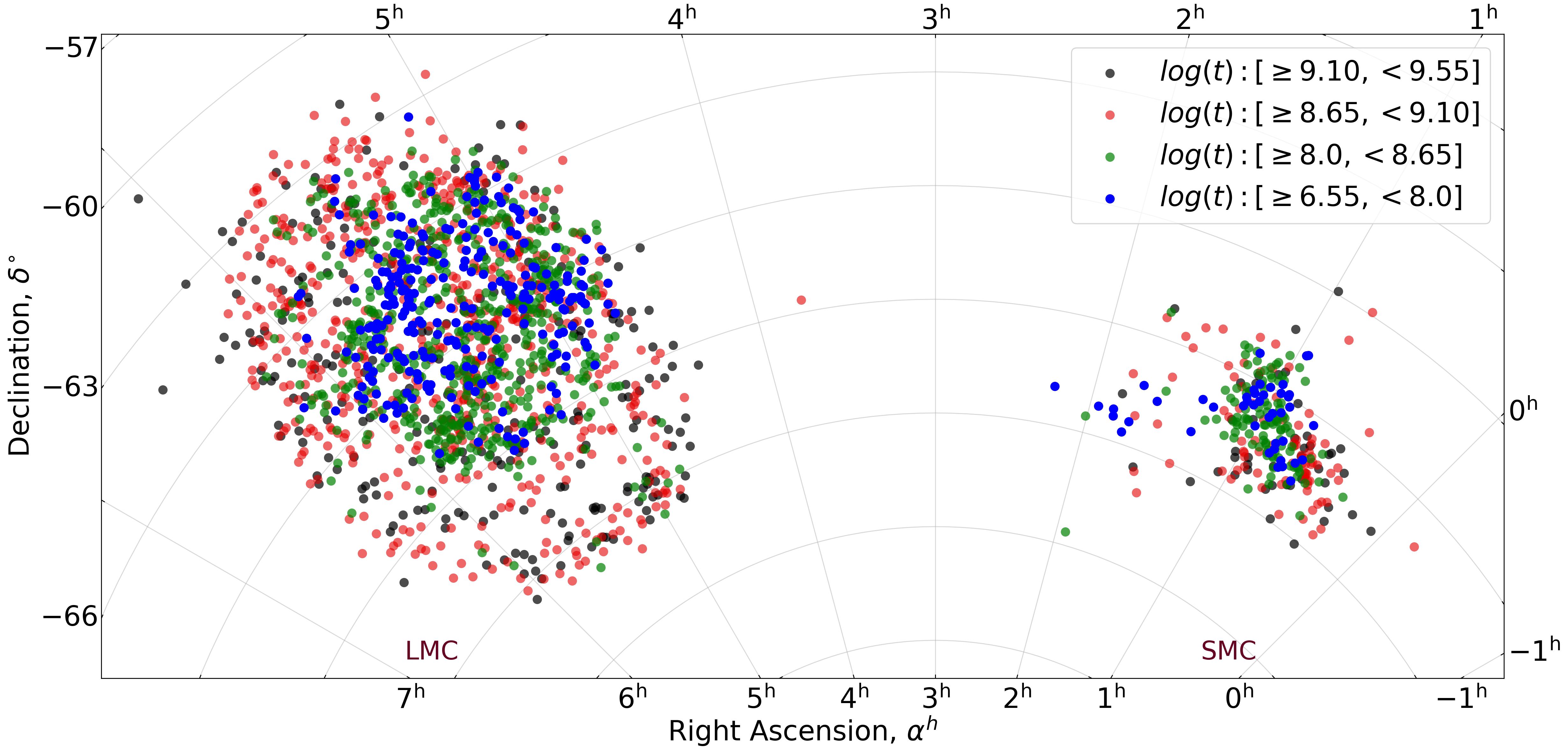}
            \caption[]%
            {{\small Spatial age map (in sky plane ) of clusters in the MCs, color-code as mentioned in the legend.}}    
            \label{fig1_age}
        \end{subfigure}
        \caption[]
        {\small Age distribution and spatial map of star clusters in the MCs. The bin width of 0.1 in $\log(t)$ is used to generate the age histograms, which is $\sim$ thrice the average median error in the $\log(t)$ estimations of the cluster as mentioned in subsection \ref{mc_spc}. The vertical lines in the histograms (a) and (b) show the margins of the age groups, which is grouped and shown in the spatial plot (c).} 
        \label{fig1_age_map_hist}
\end{figure*}

The age distributions for the LMC and the SMC star clusters are shown in Figures \ref{fig1_age_hist_1} and \ref{fig1_age_hist_2}. In general, CF can be continuous or discontinuous in galaxies. When the CF is discontinuous, the CF history is likely to show peaks of cluster formation episodes.  In this study, we aim to detect these episodic cluster formation (ECF) and correlate the time-scales with any external event in the galaxy. The spatial age map of the star clusters in the MCs is shown in Figure \ref{fig1_age} on the sky plane, and they are segregated into four age groups in ranges of $\log(t)$, Age group-1: $[\geq9.10,<9.55]$, Age group-2: $[\geq8.65,<9.10]$, Age group-3: $[\geq8.0,<8.65]$, Age group-4: $[\geq6.55,<8.0]$. The grouping is made to track the age span from old to young clusters in the MCs, and they were visually selected based on the presence of prominent peaks in the ECF of the MCs.

In the LMC, the peaks of ECF are obtained at $\log(t) = 9.17\pm0.036$, $8.93\pm0.036$ corresponding to linear ages of $1.48^{\scaleto{+0.13}{4pt}}_{\scaleto{-0.12}{4pt}}$ Gyr and $851^{\scaleto{+76}{4pt}}_{\scaleto{-70}{4pt}}$ Myr respectively. In the SMC, the peaks of ECF are obtained at  $\log(t) = 9.17\pm0.036$, $8.87\pm0.036$ and $8.17\pm0.036$ corresponding to linear ages of $1.48^{\scaleto{+0.13}{4pt}}_{\scaleto{-0.12}{4pt}}$ Gyr, $741^{\scaleto{+64}{4pt}}_{\scaleto{-60}{4pt}}$ Myr and $149^{\scaleto{+13}{4pt}}_{\scaleto{-12}{4pt}}$ Myr respectively. The first ECF peak coincides in both the Clouds, suggesting that both the galaxies experienced a burst of CF at $\sim$ 1.5 Gyr. The second ECF peak in the LMC is found at $\sim$ 851 Myr, whereas it is found at $\sim$ 741 Myr in the SMC. Though there is a difference of 110 Myr between the second ECF in the LMC and the SMC, this is within 1$\sigma$ error of the estimation. Therefore, we may consider this to be a synchronized CF peak at $\sim$ 800 Myr in the MCs.

The four age groups (as shown in Figure \ref{fig1_age}) help us to locate star clusters in the MCs based on their age. The first synchronized ECF peak of the MCs lies within the Age group-1. These clusters (shown in black) are widely distributed in both the MCs, having a large radial extend. Many can be seen in the outer regions, particularly in the LMC. Age group-2 consists of clusters (shown in red) showing a wider spatial distribution similar to the Age group-1, but with more clusters found towards the central regions of both the MCs. The Age group-2 includes the second synchronized ECF peak in the Clouds. Also, we notice a dip in ECF at $\sim$ 1.26 Gyr between the first and second ECF peaks in both galaxies. There is evidence for an immediate bounce back in CF that resulted in the second synchronized CF, as seen by a significant peak in the LMC, but with a moderate peak in the SMC, though with a difference of 110 Myr. These peaks can be considered synchronized as it is within the margin of error. A decline in CF is detected after the 851 Myr peak in the LMC. In the Age group-3 (shown in green), the CF starts to shrink significantly towards the central regions of both the MCs. The CF shows a slow decline between 600 - 100 Myr in the case of the LMC, noting a small rise and fall in between. In the case of the SMC, the CF is found to increase for ages younger than 300 Myr, which peaks at $\sim 149^{\scaleto{+13}{4pt}}_{\scaleto{-12}{4pt}}$ Myr. Age group-4 (shown in blue) shows a significant shift of clusters to the North-East regions, away from the center of the LMC. Also, in this period, the Magellanic Bridge (MB) structure appears as a trail of young clusters from the South-East region of the SMC. Overall, we note a decline in the number of clusters formed in this age group. 

A distinct radial shrinkage of CF is noticed in the MCs over the last 1 Gyr. Also, the propagation of CF is evident within the age groups. In the next section, we present the directional change in cluster distribution as a function of age within the MCs.

\subsection{Spatio-temporal density map of star clusters}

To trace the spatio-temporal density map of clusters, we produced a 2D-Gaussian kernel density estimation (KDE) within each age group. The cluster coordinates are taken with the conventions as mentioned in Section \ref{sec_4}. Scott's rule was used to determine the optimum bandwidth for the KDE maps among the four age groups of MCs.

\begin{figure*}
        \centering
        \begin{subfigure}[b]{\textwidth}
            \centering
            \includegraphics[width=1\textwidth]{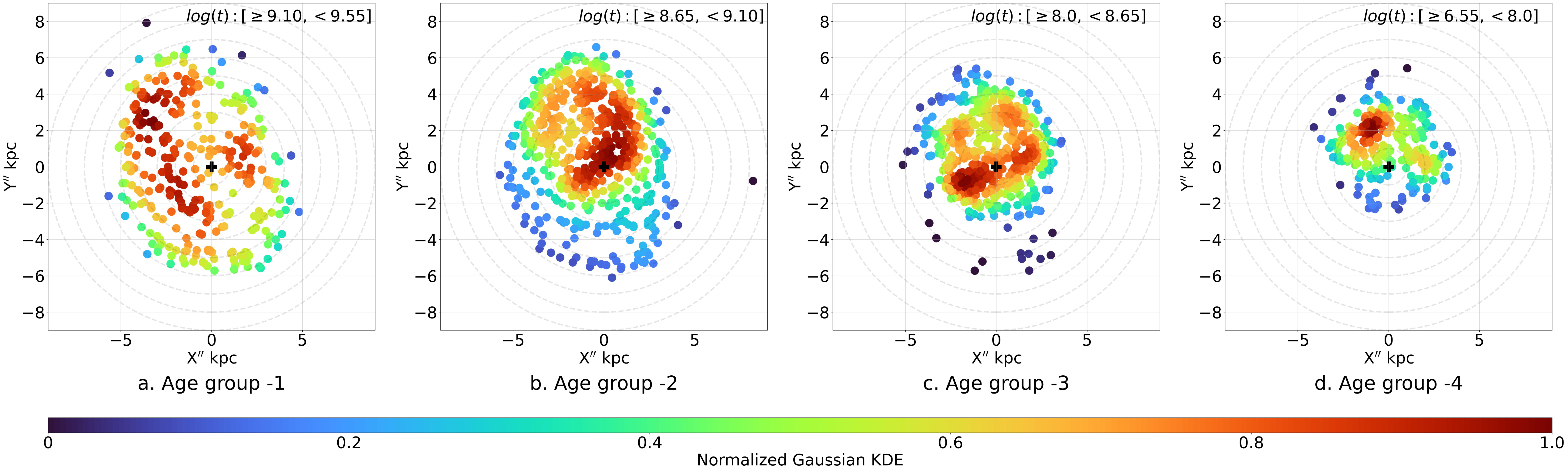}
            \caption[]%
            {{\small}}    
            \label{fig_lmc_prop}
        \end{subfigure}
        \vskip\baselineskip
        \begin{subfigure}[b]{\textwidth}   
            \centering 
            \includegraphics[width=1\textwidth]{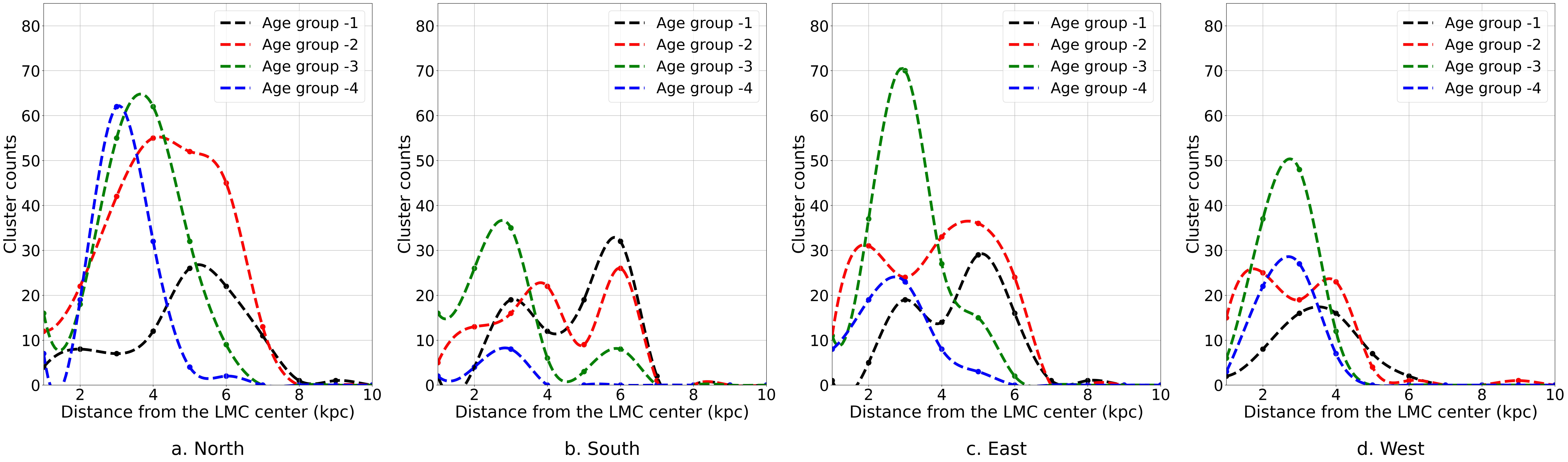}
            \caption[]%
            {{\small}}    
            \label{fig_lmc_curv}
        \end{subfigure}
        \vskip\baselineskip
        \begin{subfigure}[b]{\textwidth}   
            \centering 
            \includegraphics[width=1\textwidth]{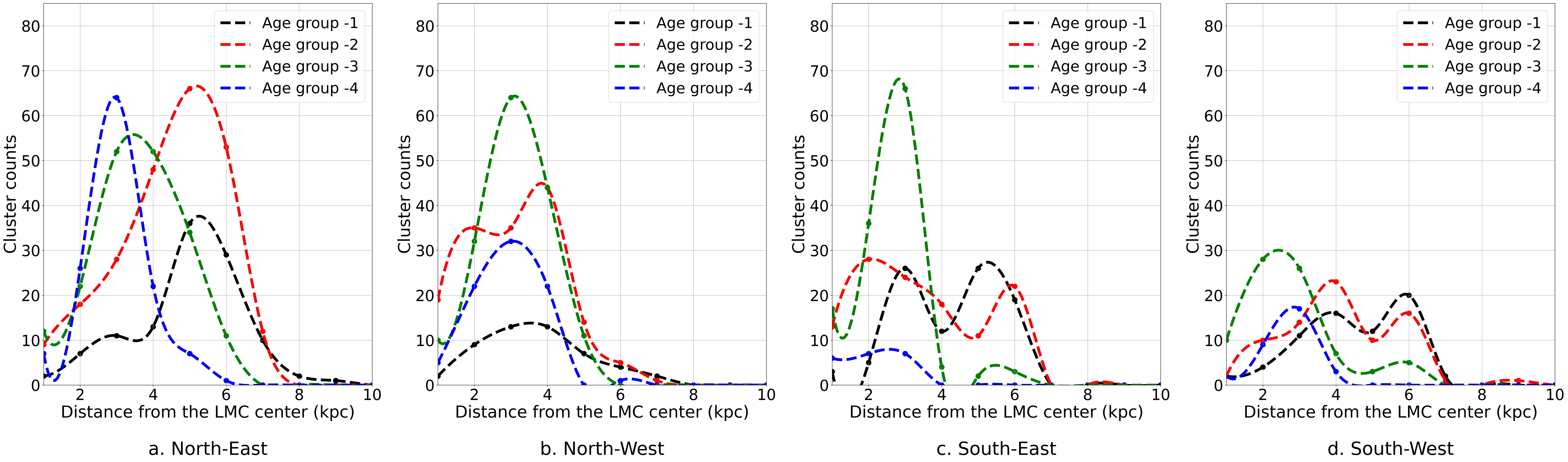}
            \caption[]%
            {{\small}}    
            \label{fig_lmc_curv_plus}
        \end{subfigure}
        \caption[]
        {\small Spatio-temporal density and radial distribution of clusters in the LMC. a) The 2D Gaussian KDE is shown for the clusters within the four age groups (Age group-1, Age group-2, Age group-3, Age group-4 ). b) Radial cluster count profile for the four age groups in the four spatial quadrants (local North, South, East, and West) of the LMC plane. c) Radial cluster count profile for the four age groups in the local North-East, North-West, South-East, South-West directions of the LMC plane. } 
        \label{fig1_lmc_prop_curv}
    \end{figure*}

Figure \ref{fig1_lmc_prop_curv} shows the spatial density distribution of the LMC clusters in the upper panel and their radial distribution in the middle and lower panels. The left-most map in the upper panel suggests a relatively large density of clusters in the eastern region of the LMC within the Age group-1. The density pattern spans from the South to the North, covering the East. The bar pattern is feebly visible, along with a patch of density enhancement in the West. The second map from the left, in the same panel, shows the density distribution of Age group-2, which shows that the cluster density peak shifts toward the North and North-West (NW) regions. We see a bar-like distribution in this age group, which was barely noticeable in the previous age group.  We also notice a significant decrease in cluster density in the southern LMC. We notice that the patch of density enhancement in the West found in the Age group-1 extends to form an arc-like pattern in the NW in this age group. The overall radial extent of clusters in the LMC disk is more or less similar for Age groups-1 and 2.

In the case of the Age group-3 (as shown in the second map from the right, top panel), the density of clusters shifts dominantly towards the central and North-East (NE) regions, whereas most of the clusters in the South are located within $\sim3$ kpc, which is quite inward when compared to the Age group-2. We also note that, in the bar region, the cluster density shifts from the West to the East between the Age groups-2 and 3, such that the younger clusters are more in the eastern part of the bar. The western arm-like pattern still persists to a reduced extend. Further, in the Age group-4 (right-most map of the top panel), the South of the LMC shows the least number of clusters. The highest cluster density is found in the NE part, which started to appear in the Age group-3. In summary, we trace the cluster density in the LMC shifting from various quadrants, such as from the East to NW, then to NE, within the age range studied here. The bar of the LMC is clearly visible only in the Age groups-2 and 3 as a significant density enhancement.

The middle panel of Figure \ref{fig1_lmc_prop_curv} shows the radial distribution of cluster counts in the four quadrants (local North, South, East, and West directions from left to right respectively). We note a wave-like pattern in the cluster count profiles rather than a monotonic decrease from the center to the outer regions. The peaks suggest local density enhancements and are likely tracing the spiral arms in the LMC. The profile shows double peaks in the South and East, whereas the North and West show one prominent peak.  In the northern quadrant, the outer density peak at $\sim$ 4 - 6 kpc grows in the Age group-2, moves inward in the Age group-3 ($\sim$ 4 kpc), and further inward in the Age group-4 ($\sim$ 3kpc). We note a  radial shrinkage of clusters from $\sim$ 8 kpc to $\sim$ 5 kpc in the northern quadrant. In the case of the southern quadrant, we note two peaks in the Age group-1, with a larger strength for the outer one at $\sim$ 6 kpc. In the Age group-2, the outer peak decreases and shrinks in the Age group-3, whereas the inner peak  grows significantly. In the Age group-4, the outer peak disappears and even the inner peak decreases. We notice a large reduction in the outer clusters in the Age groups-3 and 4, with a total shrinkage from $\sim$ 7 kpc to $\sim$ 3 kpc in the southern quadrant.

In the eastern quadrant, the outer peak at $\sim$ 5 kpc grows from the Age group-1 to 2. There appears to be a gradual inward shifting of this peak from $\sim$ 5 kpc to $\sim$ 2 kpc from the Age group-1 to 3. We notice a shrinkage of clusters from $\sim$ 7 kpc to $\sim$ 4 kpc. In the case of the western quadrant, the single peak is found to stay more or less at the same radial distance from the center. Also, we detect only a marginal shrinkage (from $\sim$ 6 kpc to $\sim$ 4 kpc). This is in sharp contrast to the rest of the quadrants. The processes responsible for the inward shifting of younger clusters may be ineffective in this quadrant.

As the LMC is known to move in the NE direction, the KDE plots for the North-East (NE), North-West (NW), South-East (SE), and South-West (SW) quadrants are shown in the bottom panel of \ref{fig_lmc_curv_plus}, to check the effect of the LMC's movement in the MW's halo. We notice a wavy pattern in all four quadrants in all age groups, likely to represent the spiral arms present in the LMC. The NE quadrant shows a peak at $\sim$ 5 kpc in Age group-1, which becomes prominent in the Age group-2. The peak stays more or less the same for the Age groups-2 and 1, but progressively moves inward to at $\sim$ 3 kpc. The radial extends also shrinks from $\sim$ 8 to $\sim$ 6kpc (with a sharp decline at $\sim$ 4 kpc) between the Age groups-2 and 4 in the last 450 Myr. In the NW quadrant, there is one single peak that more-or-less remains at the same radial distance in all age groups. We also do not notice any significant radial shrinkage of clusters in this quadrant, except from $\sim$ 6 - 5 kpc in the last 100 Myr.

In the SE quadrant, the profiles of Age groups-1 and 2 are more or less similar, with two peaks (an outer peak at $\sim$ 5 - 6 kpc and an inner peak at $\sim$ 2-3 kpc). We notice a distinct increase in the number of clusters in the inner peak for the Age group-3 (the peak at $\sim$ 3 kpc), along with an almost disappearance of the outer peak. In the Age group-4, the inner peak also shrinks and appears as a broad profile up to $\sim$ 4 kpc. The radial extend of this quadrant shrinks from $\sim$ 7 - 4 kpc, between the Age groups-2 and 4 in the last 450 Myr. In the case of the SW quadrant, we notice that there is less number of clusters populated in this quadrant with a wavy profile. The Age groups-1 and 2 appear similar with an outer peak at $\sim$ 6 kpc and an inner peak at $\sim$ 4 kpc. In the Age group-3, the outer peak reduces significantly to the extent that it almost disappears, whereas the inner peak is found to move inward to a radial distance of $\sim$ 3 kpc. The radial extend shrinks from  $\sim$ 7 - 4 kpc between the Age groups-2 and 4 (last 450 Myr). In the case of the Age group-4, the inner peak reduces in strength but stays at the same radial distance. We note that except for the NW quadrant, all the other 3 quadrants have experienced considerable radial shrinkage, mostly in the last 450 Myr.

\begin{figure*}
        \centering
        \begin{subfigure}[b]{\textwidth}
            \centering
            \includegraphics[width=1\textwidth]{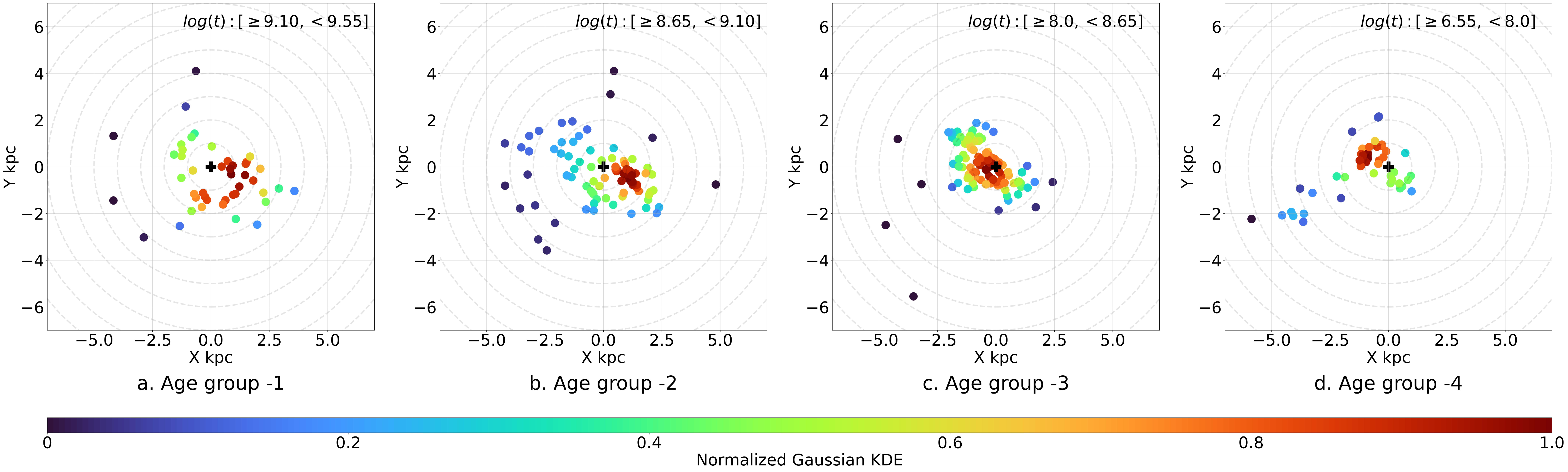}
            \caption[]%
            {{\small}}    
            \label{fig_smc_prop}
        \end{subfigure}
        \vskip\baselineskip
        \begin{subfigure}[b]{\textwidth}   
            \centering 
            \includegraphics[width=1\textwidth]{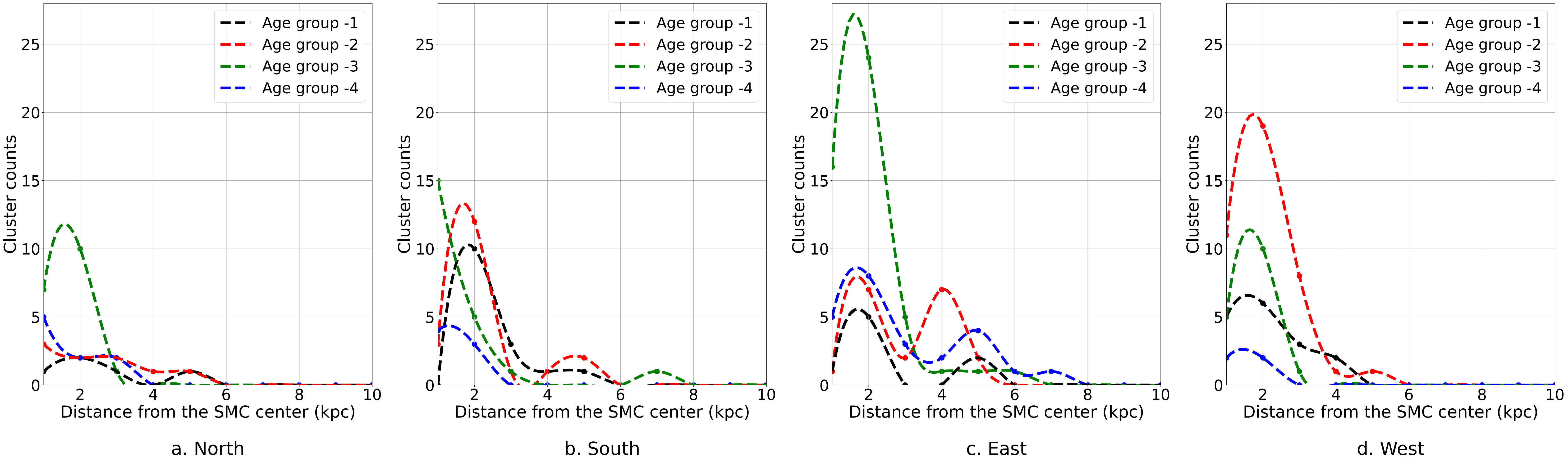}
            \caption[]%
            {{\small}}    
            \label{fig_smc_curv}
        \end{subfigure}
        \vskip\baselineskip
        \begin{subfigure}[b]{\textwidth}   
            \centering 
            \includegraphics[width=1\textwidth]{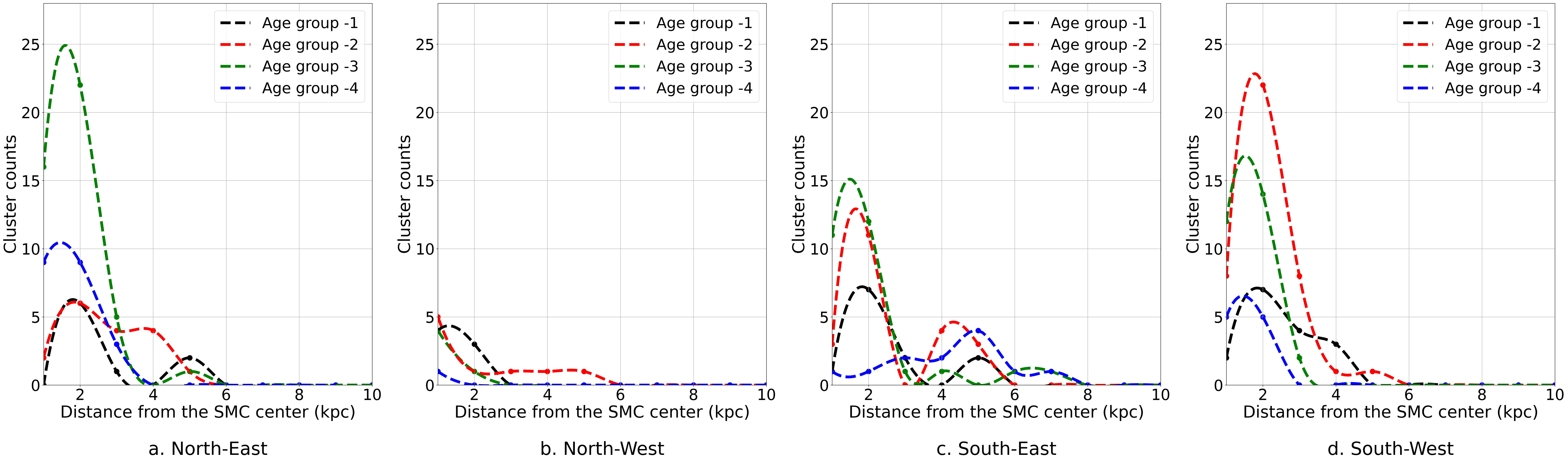}
            \caption[]%
            {{\small}}    
            \label{fig_smc_curv_plus}
        \end{subfigure}
        \caption[]
        {\small Spatio-temporal density and radial distribution of clusters in the SMC. a) The 2D Gaussian KDE is shown for the clusters within the four age groups (Age group-1, Age group-2, Age group-3, Age group-4 ). b) Radial cluster count profile for the four age groups in the four spatial quadrants (local North, South, East, and West) in the sky plane of the SMC. c) Radial cluster count profile for the four age groups in the local North-East, North-West, South-East, South-West directions in the sky plane of the SMC.}
        \label{fig1_smc_prop_curv}
    \end{figure*}

The cluster density is traced in the SMC sky plane as shown in Figure \ref{fig_smc_prop}. In the Age group-1 (shown in the leftmost panel), the cluster density is largest in the South and SW direction at $\sim$ 2 kpc from the SMC center. In the Age group-2 (second panel from the left), the cluster density shifts towards the central and SW regions at $\sim$ 2 kpc. Also, a scatter of old clusters can be noticed at $\sim$ 2 kpc to $\sim$ 5 kpc from the SMC center in the NE and SE regions in the Age groups-1 and 2. The density of clusters further shifts to the NE direction along with an increase in the central region, as evident from the Age group-3 (second panel from the right). In the Age group-4 (right-most panel), we notice a significant density in the NE region close to the center and a tail of clusters in the SE direction towards the LMC, which is the wing. We found that the cluster density in the SMC shifted from SW to NE and SE within the age groups studied here.  

The radial cluster count profiles in the four quadrants of the SMC are traced in the middle panel (local North, South, East, West) and lower panel (local NE, NW, SE, SW) of Figure \ref{fig1_smc_prop_curv}. In the northern quadrant, we notice a significant enhancement in the Age group-3 (100 - 450 Myr), whereas the count profiles of other age groups are more or less similar with less number of clusters. In the southern quadrant, we see a distinct peak in the Age group-1 and 2 at $\sim$ 2 kpc. This peak move towards the center within the Age group-3, followed by a reduction in the cluster counts in the Age group-4. In the eastern quadrant, once again, we note a large enhancement in the Age group-3 (100 - 450 Myr), which is in fact the largest among all quadrants. This quadrant shows a wavy count profile with a similar radial distribution of clusters in the Age groups-1, 2, and 4. The Age group-4 clusters show an extended distribution populating the outer SMC. In the western quadrant, we note an increasing strength of the peak from the Age group-1 to 2, such that the highest peak in the Age group-2 (450 Myr - 1.26 Gyr) is found in this quadrant. The peak reduces progressively in the Age groups-3 and 4. Except in the eastern quadrant, all other quadrants show shrinkage of cluster formation during the age groups considered here.

The NE quadrant shows the biggest peak in the Age group-3 (100-450 Myr), with similar profiles for the other age groups. The NW quadrant has the least number of clusters in all age groups. The SE quadrant shows a growing inner peak through the Age groups-1, 2, and 3. In the Age group-4, the entire profile changes to produce a peak at $\sim$ 5 kpc, populating the SMC wing and the extension to the MB. We also note the presence of the Age group-2 clusters in the SMC wing region. The SW quadrant shows the largest enhancement in the peak for the Age group-2 (450 Myr - 1.26 Gyr), which progressively reduces in the Age groups-3 and 4.

 The NE quadrant has a significantly larger number of clusters than the NW across all age groups, suggesting a very low CF in the NW over the last 3.5 Gyr. The dominant peak in Age group-2 found in the South and SW quadrants suggests an enhancement in CF during 450 Myr - 1.26 Gyr. The South and SW regions of the SMC show an inward radial shrinkage of clusters from $\sim$ 6 kpc to $\sim$ 2 kpc in the last 450 Myr. On the other hand, the dominant peaks in Age group-3 found in the East and NE quadrants suggest an enhancement in CF during 100 - 450 Myr. The East and SE regions show the peak of CF at $\sim$ 5 kpc in Age group-4, which is suggestive of the formation of the wing and the extension to the MB from the SMC in the last 100 Myr.

\section{Discussion}\label{sec_5}

In this study, we have considered clusters with one overlapping neighbor but excluded those with more than one overlapping neighbor in the LOS. Also, this study could not consider poor clusters (particularly in crowded regions), as they have scattered/ poor CMDs after the FSD. We could not characterize about 1000 clusters due to the above reasons. As shown in Figure \ref{fig:nt_ct} in the Appendix, these clusters are found to be located more in the central regions with much less number in the outer regions. Many of them could also be just random density enhancements and not real clusters, as indicated by the poor number of cluster members. For example, N16, N18, and \cite{samya2015AJ....149...52C} have found several such clusters. These clusters are unlikely to significantly alter the overall results derived in this study, particularly for the outer regions of the MCs. The central regions of both Clouds have more such excluded clusters. We expect the real clusters to be not so large in number and therefore may impact the CF statistics only to a small extent. Deeper photometric data are needed to study these poor clusters to increase the number of characterized clusters and to get a complete picture of CF history.

The data selection and the FSD algorithm are designed to effectively eliminate the MW source contamination. To check for any leftover contamination, we cross-matched the cleaned cluster sources from our study with the recent MC-MW source membership probability catalog by \citet[hereafter, J23]{jim2023A&A...669A..91J} and \cite{jimsmcpaper2023A&A...672A..65J}. The authors recommend two probability cut-off (P$_{\text{cut}}$) values for MW source contamination in the MCs (in Section 2.3.3 of J23). They suggest P$_{\text{cut}}$ = 0.01 for completeness in the LMC sample (that is, no LMC objects are missed at the price of an increased MW contamination), and P$_{\text{cut}}$ = 0.52 for the optimal sample with completeness and purity. While the aim of J23 is to make a master catalog of LMC sources with almost zero contamination of MW, our aim is to identify the star clusters located in the central region of MCs as well as in the outskirts and estimate their parameters to understand cluster formation history. As we cannot afford to miss out on LMC stars, we adopted the P$_{\text{cut}}$ = 0.01 criteria to check the leftover MW source contamination. The cross-match showed that only a very small fraction, $\sim$ 1 - 2\% of the final members are MW contaminants and mostly co-located with the cluster sequence in the CMD. We also re-estimated the cluster parameters for a sample of clusters after removing them. The estimated parameters do not change as our optimized parameter estimation technique is not affected by the removal of a very small number of sources.

We estimated the cluster parameters using an automated method, where the best fits are identified based on the $\chi^2_f$ value and MCMC sampling, followed by a visual inspection. The end product is the estimation of 4 parameters (age, extinction, distance, and metallicity) and their errors for 1990 clusters. Among these, we newly parameterized 960 clusters. The age estimation is robust within the expected range of the other three parameters that are constrained in the MCMC sampling. We also note a satisfactory comparison of ages and reddening from this study with those in the literature. In this section, we compare our cluster parameter estimates with the previous literature studies.  Further, we discuss the results presented in Section \ref{sec_4} and highlight a few interesting details on the cluster ages and distribution in the MCs. These are discussed below.

\subsection{Comparison of cluster parameters with the previous estimations}\label{comp_1}

\begin{figure*}
        \centering
        \begin{subfigure}[b]{0.475\textwidth}
            \centering
            \includegraphics[width=\textwidth]{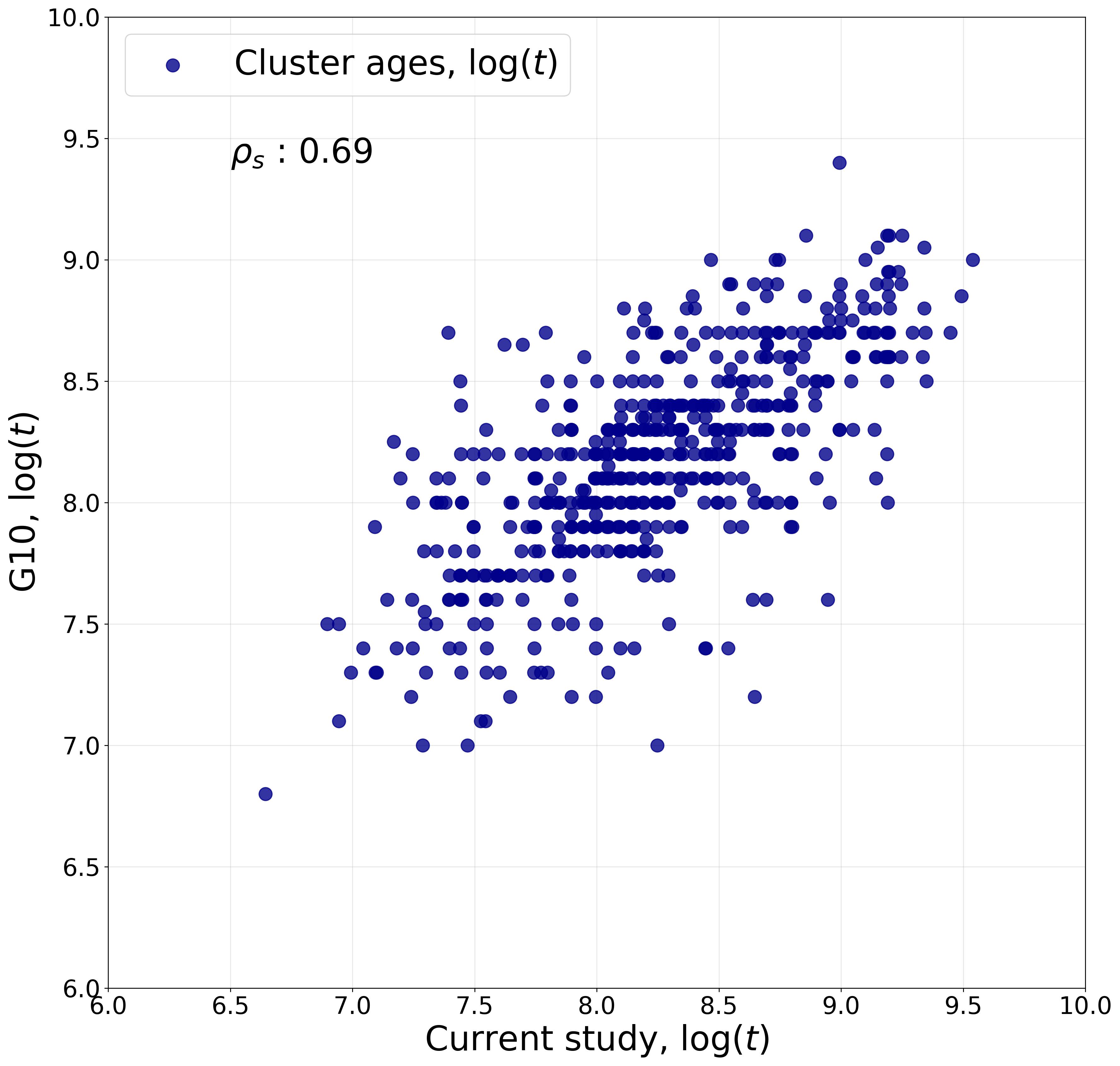}
            \caption[Network2]%
            {{\small }}   
            \label{fig_c1a}
        \end{subfigure}
        \hfill
        \begin{subfigure}[b]{0.475\textwidth}   
            \centering 
            \includegraphics[width=\textwidth]{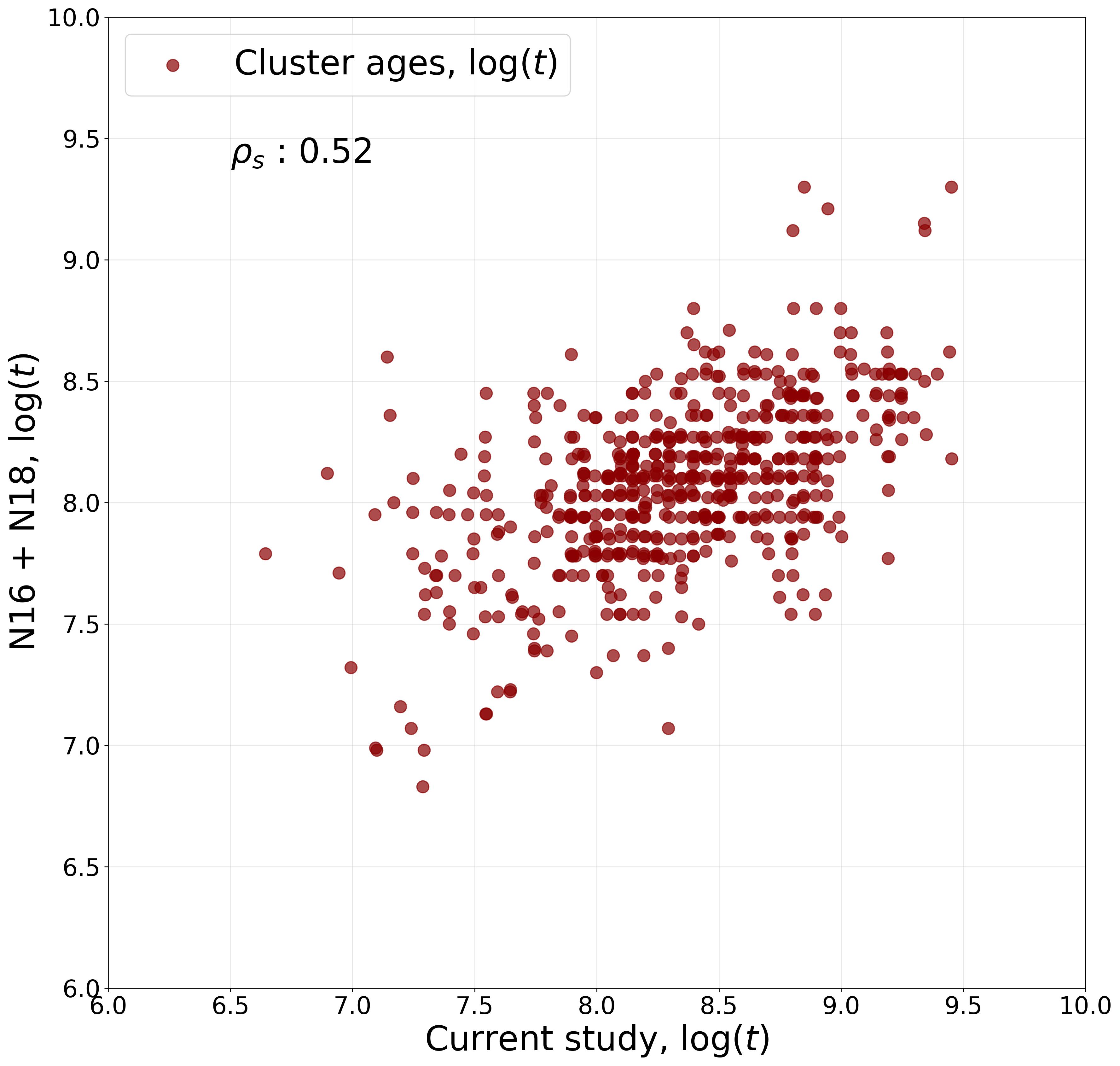}
            \caption[]%
            {{\small }}    
            \label{fig_c2a}
        \end{subfigure}
        \vskip\baselineskip
         \begin{subfigure}[b]{0.475\textwidth}  
            \centering 
            \includegraphics[width=\textwidth]{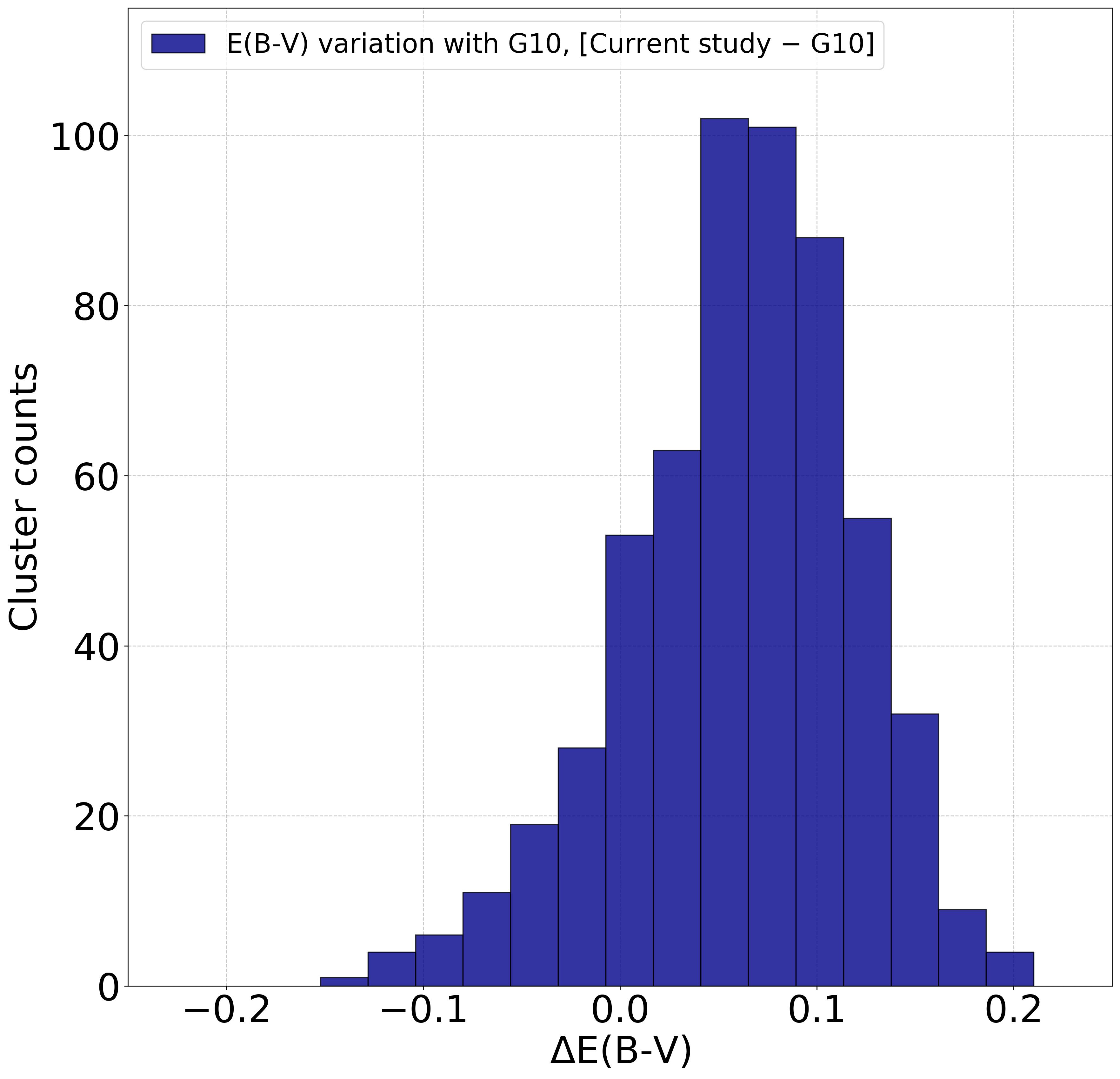}
            \caption[]%
            {{\small }}    
            \label{fig_c1r}
        \end{subfigure}
        \hfill
        \begin{subfigure}[b]{0.475\textwidth}   
            \centering 
            \includegraphics[width=\textwidth]{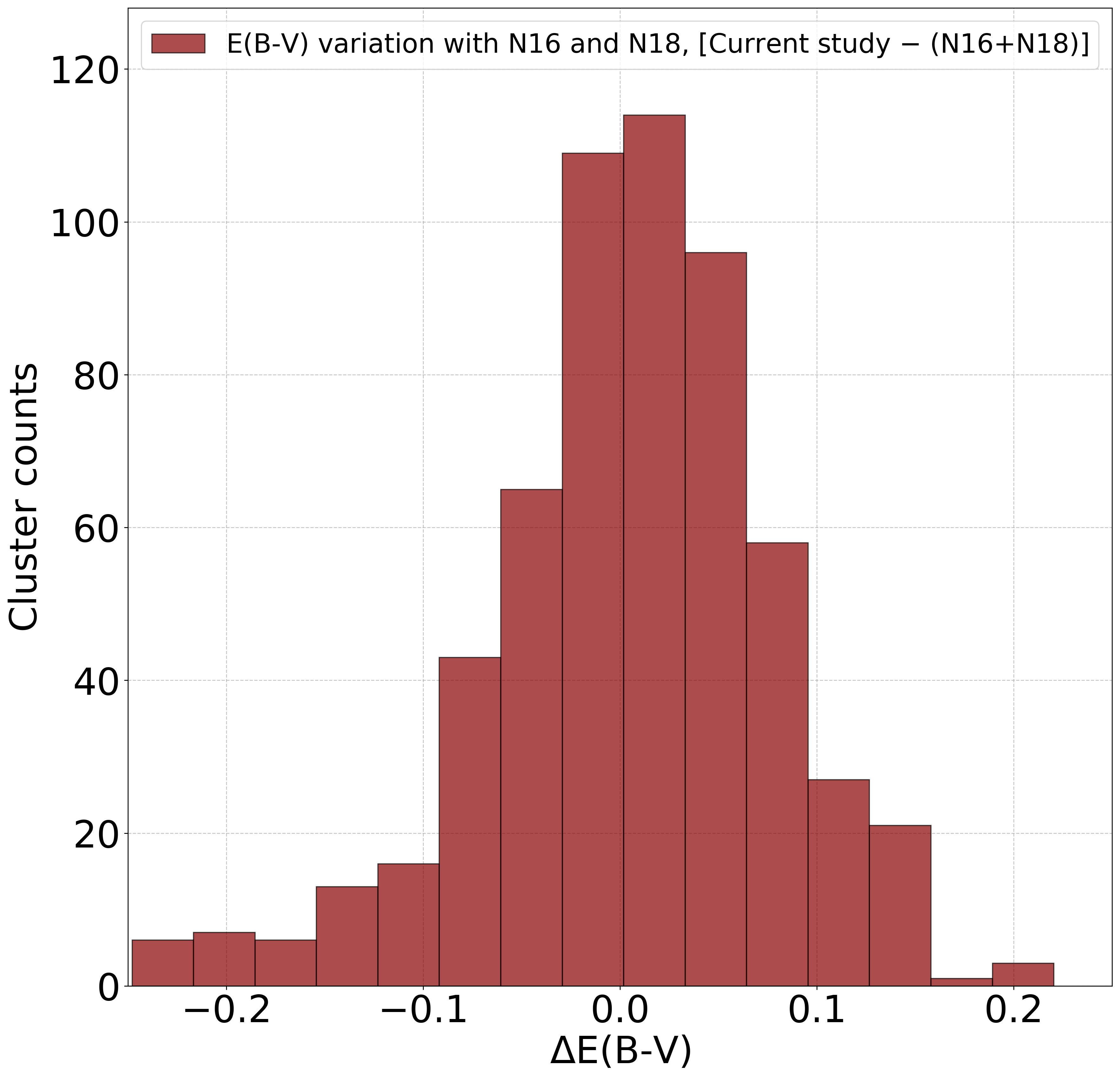}
            \caption[]%
            {{\small }}    
            \label{fig_c2r}
        \end{subfigure}
        \caption[]
        {\small The age-reddening estimates from our current study compared with the literature (as mentioned in Section \ref{comp_1}). The top panels (a and b) compare the age estimates from our study with G10, N16 and N18 (reference catalogs) using Spearman's rank correlation coefficient ($\rho_s$), in which we found positive correlations among the $\log(t)$ estimates as shown in the scatter plots. The reddening variation ($\Delta E(B-V)$) from our study is also compared with the reference catalogs and shown in the bottom panels (c and d).} 
        \label{Compar_plots}
    \end{figure*}

The estimated age-reddening values were compared with two major studies (G10; N16 and N18). We cross-matched 576 clusters from G2010 and 585 clusters from N16 and N18. We used Spearman's rank correlation coefficient, $\rho_{s}$ to compare the age estimations (as in Figures \ref{fig_c1a} and \ref{fig_c2a}). We estimated a positive correlation coefficient of $\rho_{s} = 0.69, 0.52$, suggesting that the age estimations of this study are comparable for the common clusters. The reddening estimations were compared with the above two studies, using histograms of reddening variations, $\Delta E(B-V)$, as shown in Figures \ref{fig_c1r} and \ref{fig_c2r} with constant $R_V$ values as mentioned in subsection \ref{ag_met_sec}. The distribution of $\Delta E(B-V)$ is found to have a 1$\sigma$ width of 0.06 mag, with the peak near 0.05 mag with respect to G10. This difference is statistically insignificant as it is of the order of the error in the current estimation. The distribution of $\Delta E(B-V)$ is found to be $\sim$ 0.0 mag with respect to N16 and  N18. The age and reddening comparisons with G10, N16 and N18 are shown in Figure \ref{Compar_plots}. These suggest that the age-reddening values more or less match the previous estimates. Hence the estimated parameters are similar to the previous studies and are reliable for inferring the cluster properties of the MCs.

There are not many estimations of [M/H], unlike the estimations of [Fe/H]. \cite{rubele2018MNRAS.478.5017R} estimated an [M/H] of $\sim$ 0.65 dex (Z $\sim$ 0.0032) for the SMC stellar population extending up to $\sim$ 1 Gyr, which is very similar to our estimate. We also note that the range in Z or [M/H] is similar in both studies for ages less than $\log(t) < 9.1$. 

\subsection{Synchronized CF peaks and SFH of the MCs}

\begin{figure}
\centering
   \includegraphics[width=1\linewidth]{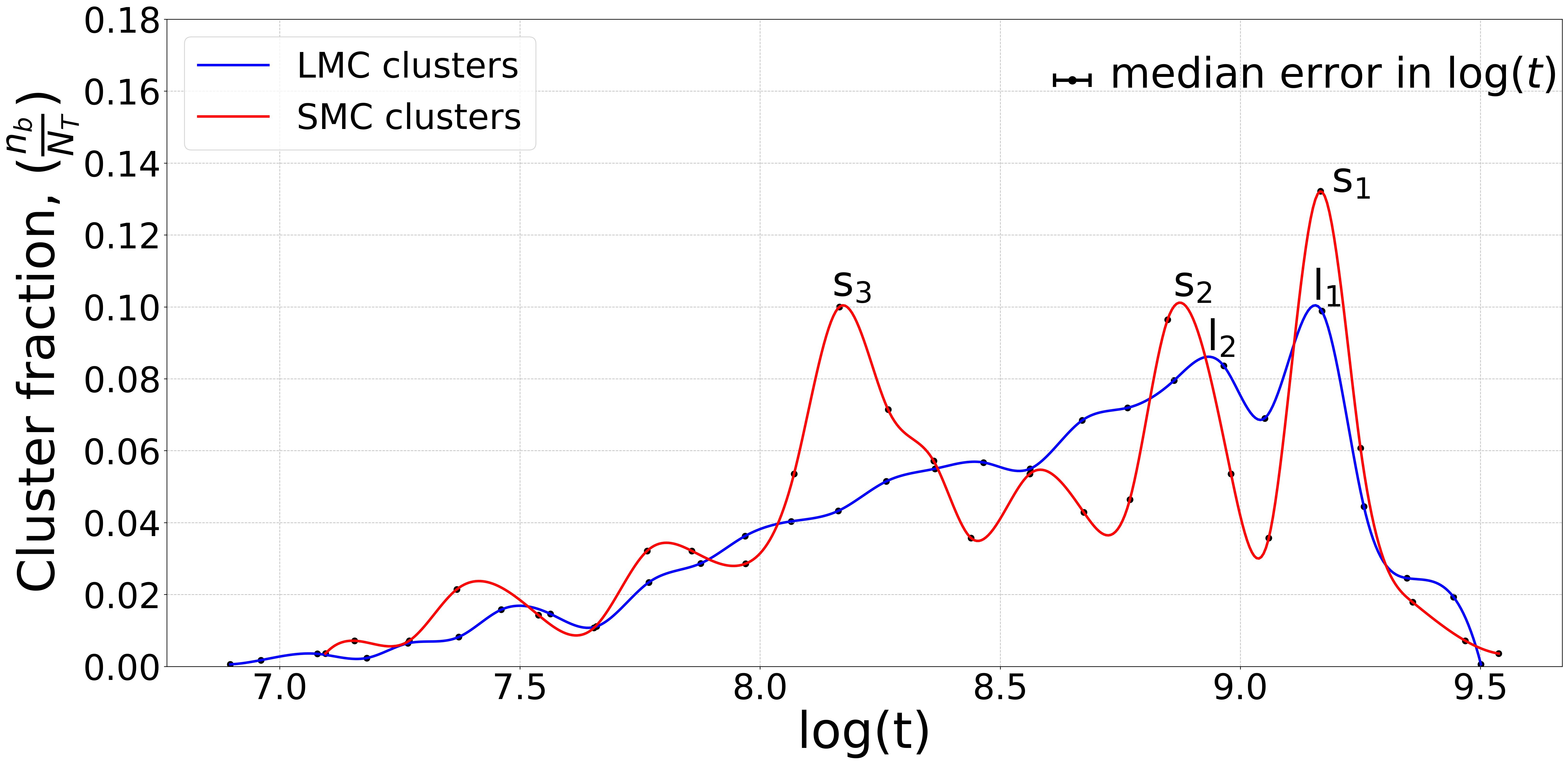}
   \caption{\small The ECF peaks obtained from the clusters we studied in the MCs. The age histograms of the LMC and the SMC from Fig \ref{fig1_age_hist_1} and \ref{fig1_age_hist_2} are normalized ($n_b$/$N_T$) with their total number, $N_T$ ($N_T = $1710 (LMC), 280 (SMC)) and overplotted for comparison. The correlated peaks of CF are marked at $\sim $ 1.48 Gyr ($l_1$ and $s_1$), the other significant CF peaks in the LMC at $\sim$ 851 Myr ($l_2$), and in the SMC at $\sim$ 741 Myr ($s_2$), $\sim$ 149 Myr ($s_3$) are also marked. The median error in $\log(t)$ is also shown.}
   \label{fig:ecf} 
\end{figure}

 To confirm the ages of synchronized CF episodes in the MCs, we generated normalized age distributions for clusters in both MCs, as presented in Figure \ref{fig:ecf}. The age binning was set at $\delta\log(t)=0.10$. The first and most prominent synchronized peak, at approximately 1.48 Gyr ($l_1$ and $s_1$), was observed in both Clouds. Additionally, a second synchronized peak was identified at $\sim$ 851 Myr ($l_2$) in the LMC and $\sim$ 741 Myr ($s_2$) in the SMC, with $l_2$ and $s_2$ showing only a small difference of $\sim$ 110 Myr, well within the margin of error. This is the first time such synchronized peaks in CF are found for the Clouds at 1.5 Gyr and 800 Myr. This is possible due to the spatial coverage of \textit{Gaia} data. Notably, a CF peak at $\sim$ 149 million years ($s_3$) was observed as a significant peak in the SMC, while no corresponding peak was seen in the LMC. The wavy pattern in the profile for the SMC is due to the smaller number of clusters available to estimate the age distribution. A detailed comparison is provided below.

The first peak of synchronized CF is more or less in agreement with the peak of SFH at 2 Gyr obtained by \cite{rube2012A&A...537A.106R}, in the LMC. This peak of CF is likely to be due to the LMC-SMC interaction \citep{rube2012A&A...537A.106R}. The LMC's ring-like clusters in the outskirts are related to the possible LMC-SMC interaction \citep{Choi_2018} with the ages spanning within the Age groups-1 and 2 in our classification. The simulations by \cite{salem_2015ApJ...815...77S} suggested that the outermost parts of the LMC were affected by the ram pressure stripping at $\sim$ 1 Gyr ago. However, we found a global enhancement in CF in the MCs at $\sim$ 800 Myr, so the ram-pressure stripping is likely to have started after this peak. But the second synchronized CF peak noticed in the LMC and the SMC is unlikely to be due to a common event, as no such mutual events are found/ expected at this period. This may be the response of the gaseous disk to the perturbation followed by a stochastic self-propagating star formation \citep{stochastic1978ApJ...223..129G}. The recent study by \cite{mazzi2021MNRAS.508..245M} also found two peaks in the star formation rate (SFR) of the LMC, the first one between 1.6 - 4 Gyr and the second one between 0.63 - 1 Gyr. The two peaks of CF found for the LMC in this study are also consistent with our estimations.

The third peak of CF ($s_3$) in the SMC sets the upper limit of the recent interaction at $\sim$ 300 Myr and the lower limit at $\sim$ 150 Myr, in agreement with previous studies. A smaller peak of CF is noted in the SMC at $\sim$ 400 Myr, which is also noted in SFH by several authors. There are no significantly large peaks of CF in the LMC in the last 600 Myr. We note a marginal enhancement in the cluster numbers between 170 - 400 Myr (Figure \ref{fig1_age_hist_1}) in the LMC. The recent pericentric passage of the SMC about the LMC was reported at $\sim$ 150 Myr ago \citep{lr_culinane_47b51fabdb384e4587e1b62d80927916}, and it is similar to the peak we obtained in our study for the SMC. The CF of the SMC obtained in this study is likely to help constrain the recent interaction history, whereas the LMC (in particular, the outer LMC) does not show any significant impact in CF due to the recent interaction.

The comparison of our Cluster Formation (CF) peak estimates with those of G10 for both the Small Magellanic Cloud (SMC) and Large Magellanic Cloud (LMC) reveals a close alignment, with differences of $\sim$ 100 Myr. Furthermore, the CF peaks identified in studies conducted by N16 and N18 also show consistency with our results. However, our study did not reproduce the younger CF peak of 125 MYr, which was previously reported in both the G10 and N16 studies. This may be because of the missing clusters in the
inner LMC, that are not considered in this study due to crowding and insignificant number of cluster members after the FSD. Also, any disparities in CF peak values could potentially be attributed to changes in the metal fraction values used to estimate the ages of star clusters in previous studies.

A recent study of SFH of the MCs by \cite{synchronised2022MNRAS.513L..40M} suggested that the LMC-SMC had several synchronized star formations in the last  $\sim$ 3.5 Gyr. They detected five star formation peaks in the SMC at $\sim$ 3, $\sim$ 2, $\sim$ 1.1, $\sim$ 0.45 Gyr ago, and an ongoing one. These time scales are in more or less good agreement with the ECF peaks at $\sim$1.5 Gyr (in both the MCs), $\sim$ 800 Myr (in both the MCs), and $\sim$ 149 (SMC) Myr that are found in this study.

\subsection{Age-wise cluster density profiles and its implications}

In our CF density profiles, the LMC exhibits a declining trend in star cluster distribution, particularly in its southern and SW regions. The decrease of star formation in the southern region for ages younger than 2 Gyr and recent active SFH in the North was reported previously by \cite{ruiz_2020A&A...639L...3R} and \cite{indu2011A&A...535A.115I}. These are in good agreement with the results obtained in this study. The outside-in quenching of star formation in the northern part of the LMC was found by \cite{meschin2014MNRAS.438.1067M}, over a similar radial distance of 3.5 - 6.2 kpc, in the last 1 Gyr, which they conclude as due to either consumption of gas or ram-pressure stripping. This is in excellent agreement with the radial shrinkage found in this study.

 The truncation of CF in the outer LMC is dominant for ages younger than Age group-2 clusters, which is at $\sim$ 500 Myr. The truncation could be due to a combination of various effects, such as consumption of gas due to star formation, ram pressure stripping, radial migration of the remnant gas due to loss of angular momentum and tidal stripping. We note that the most severe truncation is in the southern LMC, whereas the motion of the LMC is in the NE direction.

The map of SFR in the LMC as shown in Figure 5 of \cite{mazzi2021MNRAS.508..245M} is in good agreement with the KDE maps presented in our study. The spatial extent as well as the variation in SFRs are well correlated for the corresponding age groups. Their SFR maps also show a significant decline in the NE regions in the last 500 Myr, likely to be due to the ram-pressure effect.

 It is curious to note that the western LMC is the least affected and hardly shows any shrinkage. On the other hand, the extent of the disk in the oldest group is much smaller in the West when compared to the other regions. We also note that this study has comparable East-West coverage with respect to \cite{ruiz_2020A&A...639L...3R}, whereas the North-South coverage is much larger. We note that the outer LMC plays a major role in its global CF for ages older than 1 Gyr. In contrast, the inner LMC plays a major role in deciphering the recent star formation, particularly for ages $\le$ 500 Myr. Therefore, it is important to include the outer LMC while estimating the global SFH of the LMC.

The motion of the LMC implies that any headwind from the MW's circumgalactic medium would compress or impact the NE part of the LMC disk. In this quadrant, HI profile truncates at $\sim$ 6.2 kpc \citep{PUT2003ApJ...586..170P}, in contrast to a much extended stellar profile \citep{vand22001AJ....122.1827V}, ruling out a tidal truncation of HI (as tides would truncate both gas and stars, \citealt{salem_2015ApJ...815...77S}). Therefore, this truncation of HI in the NE quadrant is indeed evidence of ram-pressure stripping \citep{salem_2015ApJ...815...77S}. The shifting of younger stars to the North and NE was noticed by \cite{indu2011A&A...535A.115I}, and they suggested that the recent star formation is impacted by the motion of the LMC in the MW's halo. The radial cluster counts in the North and NE presented in this study (Figures \ref{fig_lmc_curv} and \ref{fig_lmc_curv_plus}) are in good agreement with the above. The radial shift of the CF peak in the last 450 Myr is probably suggestive of the gas compression in the NE due to the LMC's motion.

  Several authors have reported spiral arms in the North of the LMC \citep{devac_1972VA.....14..163D,mackey_10.1093/mnras/stw497}, and the origin of spiral structure is associated with the interaction between the MCs \citep{besla2012MNRAS.421.2109B,besla2016ApJ...825...20B,yozi_bekki10.1093/mnras/stu1132}. The LMC bar shows up as a density enhancement in the Age group-2, and there are more clusters in the East of the bar in the Age group-3 compared to the Age group-2. The SFH of the regions closer to the LMC bar region (South-East Arm, North-West Arm, Blue Arm, North-West Void) are previously studied by \cite{haris_zarits2009AJ....138.1243H}.

  We also note that the wavy pattern in the cluster count profile persists across the age groups and in all the quadrants, suggesting of long-lasting spiral pattern. This is in good agreement with the study of field stars by \cite{ruiz_2020A&A...639L...3R}, who found evidence supporting the long-term stability of the LMC spiral arms, dating the origin of this structure to more than 2 Gyr ago. However, we add that the location of the spiral arms is found to move inwards for younger ages over the last 1 Gyr. The cluster count profiles have demonstrated their ability to trace the probable spiral arms as a function of age in the LMC. We plan to carry out a detailed analysis in the future.

 In the case of the SMC, the CF peak shifted from the SW for Age group-1 to NE for Age group-4. We note that the SMC has far less number of clusters and the results may not be as strong as the LMC due to poor statistics. The study by \cite{smc_ne_sout_2016AcA....66..149J} with the age map of classical cepheids has found the NE part of the SMC to be younger compared to the SW, which is in agreement with our study. Also, the density map (Figure \ref{fig_smc_prop}) for the Age group-3 for SMC, the features identified by \citet[in their Figure 7 for the SMC]{Yousafi10.1093/mnras/stz2400} can be noticed, such as the central bar, the NE and SW extensions (except the wing, that is seen only in the Age group-4). Our study traces the KDE of younger clusters extending from SW to NE for ages younger than 1 Gyr, and it is similar to the KDE map traced by \cite{smckdeyoung2018ApJ...858...31S}. The CF in the SMC is dominated in the South and West during 450 Myr - 3.5 Gyr, and in the North and East during the last 450 Myr. Overall, a general shrinkage of CF is noticed in the SMC. The SMC wing as well as the extension to the MB are traced by clusters younger than 100 Myr. If the wing and MB are formed during the last interaction at 150 Myr, then these clusters might be formed in-situ from the HI gas pulled out of the SMC during the last interaction.

The catalog of 1990 clusters containing the cluster name, coordinates, and the estimated parameters (age, extinction, metallicity, and DM) along with the 16$^{th}$ and 84$^{th}$ percentiles (lower and upper uncertainty bounds) will be made available as an online catalog. Table \ref{tab1} gives a sample of this catalog.

\section{Summary}\label{sec_6}
We summarize the results of this study below:
\begin{enumerate}
    \item We characterized 1990 clusters (1710 in the LMC and 280 in the SMC) using the \textit{Gaia} DR3 data, where 847 clusters in the LMC and 113 clusters in the SMC  are parameterized for the first time (when compared to 7 existing catalogs). The age-extinction-metallicity-DM parameters estimated here are based on an automated fitting of the CMDs after the removal of field stars, which act as prior values. We implemented an MCMC sampling technique to derive the final  parameters along with the errors. The MCMC sampling mainly explores the parameter space in $A_G$, DM, and Z to find the range under which our age estimates have confidence.
    \item The LMC has several clusters that overlap in the LOS. We categorized clusters into isolated, singly overlapping and triple overlapping, etc. This study considers a majority single clusters ($\sim$ 83\%) as well as those with only one overlapping cluster in the LOS. We note that more than 2 overlapping clusters are mostly in the inner regions of the MCs. 
    \item This study could not characterize 1082 clusters that presented either a scattered CMD after FSD or a CMD with less than 8 members after FSD. These clusters are located mostly in the inner regions of the MCs. We note that many of them are likely to be very poor clusters or random density enhancements. The average number of member star retained after FSD is 40, for the 1990 clusters that are characterised.
    \item The mean extinction $A_G$ is estimated to be 0.4$\pm$0.005 mag and 0.38$\pm$0.006 mag for the LMC and the SMC, respectively. We note that the LMC sample has more outer clusters with relatively less extinction. 
    \item We estimated a mean metal fraction, $\mu_z = 0.0060\pm0.00002$ for the clusters in the LMC, and $\mu_z = 0.0027\pm0.00002$ for the clusters in the SMC.  The corresponding mean [M/H] values are estimated as $-0.40\pm0.001$ dex and $-0.76\pm0.003$ dex for the LMC and the SMC, respectively. We recommend that the choice of Z value for isochrones to fit the CMDs of clusters with age $\le$ 1 - 2 Gyr is 0.004 to 0.008 for the LMC and 0.0016 to 0.004 for the SMC. 
    \item For the first time, this study estimated two synchronized peaks in CF in the MCs. The first one is at $\sim$ 1.5 Gyr, likely to be due to their first mutual interaction. The second synchronized CF peak at $\sim$ 740-850 Myr is likely to be of internal origin in the galaxies, as there is no known interaction during this period. 
    \item The cluster count profiles in all the quadrants of the LMC show a wavy pattern with peaks, instead of a monotonic decrease from the center to the outer regions over the age range explored in this study (up to 3.5 Gyr). This is suggestive of the presence of spiral arms as noted by several studies in the literature. In this study, we trace an inward shifting of these peaks (possible spiral arms) to the inner LMC for younger ages.
    \item The cluster count profiles show that there is a significant radial shrinkage in the cluster distribution in the LMC in the last 450 Myr. The KDE maps as well as the cluster count profiles provide evidence for ram-pressure stripping in the NE quadrant due to the movement of the LMC in the MW's halo.
    \item Radial shrinkage of the LMC disk and truncation of outer CF is noted, typically from outer at $\sim$ 8 kpc to inner at $\sim$ 4 kpc. We find a significant shrinkage in CF in the South, NE and SE in the last 450 Myr. The West and NW quadrants, on the other hand, show no significant shrinkage.  
    \item We note a severe truncation of CF in the southern LMC in the last 450 Myr, likely to be due to the combined effect of multiple factors, such as consumption of gas due to star formation, ram pressure stripping, radial migration of the remnant gas due to loss of angular momentum and tidal stripping, etc.
    \item The recent interaction at $\sim$ 150 Myr has impacted the SMC with a CF peak, whereas no such peak is found in the LMC. We note an enhanced CF in the central, NE of the SMC and the northern LMC during this period. 
    \item  The CF in the SMC appears to be enhanced in the South and West during 450 Myr - 3.5 Gyr, whereas the CF gets enhanced in the North and East in the last 450 Myr. The shrinkage of CF in the SMC is noted from $\sim$ 6 kpc to $\sim$ 3 kpc in the NE, NW, and SW of the SMC, mostly in the last 450 Myr.
    \item This study therefore brings out a very detailed view of the CF in the LMC and SMC in the last 3.5 Gyr. The outer LMC clusters have provided a unique mapping of the CF history. This study provides some important insight into the CF episodes and their link with the LMC-SMC-MW interactions.
\end{enumerate}

\section*{Acknowledgements}
We thank the referee for valuable suggestions that helped to improve the manuscript. AS acknowledges support from the SERB Power fellowship. PKN acknowledges TIFR's postdoctoral fellowship and support from the Centro de Astrofisica y Tecnologias Afines (CATA) fellowship via grant Agencia Nacional de Investigacion y Desarrollo (ANID), BASAL FB210003. SS acknowledges support from the Science and Engineering Research Board of India through Ramanujan Fellowship and POWER grant (SPG/2021/002672).
This work made use of the optical data from the European Space Agency (ESA) mission \textit{Gaia} (\url{https://www.cosmos.esa.int/gaia}), which was processed by the \textit{Gaia}
Data Processing and Analysis Consortium (DPAC,
\url{https://www.cosmos.esa.int/web/gaia/dpac/consortium}). The funding for the DPAC has been provided by national institutions, particularly the institutions participating in the \textit{Gaia} Multilateral Agreement.

{\it Software:} ASTROPY \citep{astropy2013A&A...558A..33A}, SCIPY \citep{2020SciPy-NMeth}, MATPLOTLIB \citep{matplot2007CSE.....9...90H}, KALEPY \citep{kalepy}, NUMPY \citep{numpy2020Natur.585..357H} 
\section*{Data Availability}

The data used in this work can be programmatically accessed from the \textit{Gaia} archive (\url{https://www.cosmos.esa.int/web/gaia-users/archive/programmatic-access}).

 \begin{landscape}
 \begin{table}
 
	\centering
	\caption{The sample of the cluster catalog. The cluster name (column 1), location (RA, DEC - columns 2 and 3), and radii (column 4) are from B2008 and S16\&17 catalogs. The parameters ($\log(t)$, $A_G$, Z, DM) for each cluster with their uncertainty bounds (l/u$_{\log(t)}$, l/u$_{A_G}$, l/u$_Z$, l/u$_{DM}$) at 16$^{th}$ (l: lower bound) and 84$^{th}$ (u: upper bound) percentiles are provided in columns 5 to 16. (The full version of the machine-readable catalog is available online.) }
  
	\label{tab1}
	\begin{tabular}{lccccccccccccccr} 
		\hline
		Cluster name & RA & DEC & Radius & $\log(t)$ & $l_{\log(t)}$ & $u_{\log(t)}$ & $A_G$ & $l_{A_G}$&$u_{A_G}$ & Z & $l_Z$ & $u_Z$& DM& $l_{DM}$ & $u_{DM}$\\
      
         & (h m s) & ($^\circ$ $^\prime$ $^{\prime\prime}$) & ($^\prime$) &  & & & (mag) & (mag)& (mag)& &  & & (mag)& (mag) & (mag)\\
		\hline\hline
    NGC458,K69,L96,ESO51SC26&1 14 52.00&-71 33 00.0&2.60&8.046&8.007&8.087&0.349&0.166&0.550&0.00274&0.00148&0.00402&18.984&18.829&19.130\\
NGC1735,SL86,ESO85SC15&4 54 20.00&-67 06 01.0&1.80&7.594&7.556&7.633&0.421&0.274&0.570&0.00636&0.00326&0.00895&18.523&18.380&18.673\\
SL117,KMHK314,GKK-O173&4 56 22.00&-68 58 00.0&1.70&8.045&8.01&8.084&0.469&0.211&0.684&0.00636&0.00362&0.00895&18.652&18.508&18.795\\
OGLE-LMC-CL-0763&6 15 07.95&-66 08 40.9&0.62&8.893&8.857&8.933&0.448&0.231&0.677&0.00664&0.00413&0.00914&17.729&17.595&17.891\\
B155&1 20 22.00&-73 59 55.0&0.55&7.846&7.81&7.882&0.426&0.17&0.652&0.00327&0.00227&0.00426&18.966&18.822&19.115\\
SL18,LW32,HS23,KMHK45&4 42 53.00&-70 07 28.0&1.20&8.789&8.755&8.831&0.434&0.19&0.665&0.00632&0.00374&0.00912&18.915&18.763&19.054\\
KMHK103&4 47 41.00&-70 57 44.0&1.00&8.200&8.159&8.233&0.408&0.18&0.646&0.00582&0.00319&0.0089&18.927&18.777&19.074\\
NGC1774,SL141,ESO85SC26&4 58 06.00&-67 14 32.0&1.70&7.695&7.656&7.732&0.416&0.178&0.649&0.00578&0.00327&0.00869&18.464&18.339&18.62\\
H86-48,SOGLE164&0 38 56.00&-73 24 32.0&0.50&8.247&8.211&8.282&0.460&0.2&0.679&0.00322&0.00221&0.00422&18.983&18.836&19.119\\
IC1655,L90,ESO51SC23&1 11 53.00&-71 19 53.0&1.70&8.094&8.057&8.133&0.353&0.142&0.604&0.00316&0.00214&0.00427&18.933&18.813&19.095\\
OGLE-SMC-CL-0270&0 26 04.88&-74 24 59.5&0.83&8.992&8.957&9.033&0.368&0.158&0.615&0.00308&0.00219&0.00411&18.961&18.816&19.122\\
GKK-O11&5 30 42.00&-72 13 00.0&1.46&8.741&8.705&8.781&0.428&0.195&0.645&0.00632&0.00368&0.00897&18.603&18.473&18.75\\
SL603,LW249,KMHK1151&5 35 43.00&-75 57 32.0&0.80&8.597&8.558&8.635&0.496&0.209&0.706&0.00633&0.00347&0.00895&19.034&18.886&19.188\\
KMHK1149&5 37 44.00&-64 38 41.0&0.65&8.800&8.761&8.835&0.430&0.203&0.673&0.00639&0.00362&0.00896&17.960&17.819&18.11\\
GKK-O2&5 38 42.00&-72 22 00.0&1.46&8.148&8.107&8.185&0.406&0.172&0.641&0.00649&0.00375&0.00901&18.569&18.438&18.717\\

		\hline
	\end{tabular}
\end{table}

\end{landscape}


\bibliographystyle{mnras}
\bibliography{bib_citations}



\appendix

\section{Selection criterion from the different stages of FSD.}   
\begin{figure*}
        \centering
        \begin{subfigure}[b]{0.475\textwidth}
            \centering
            \includegraphics[width=\textwidth]{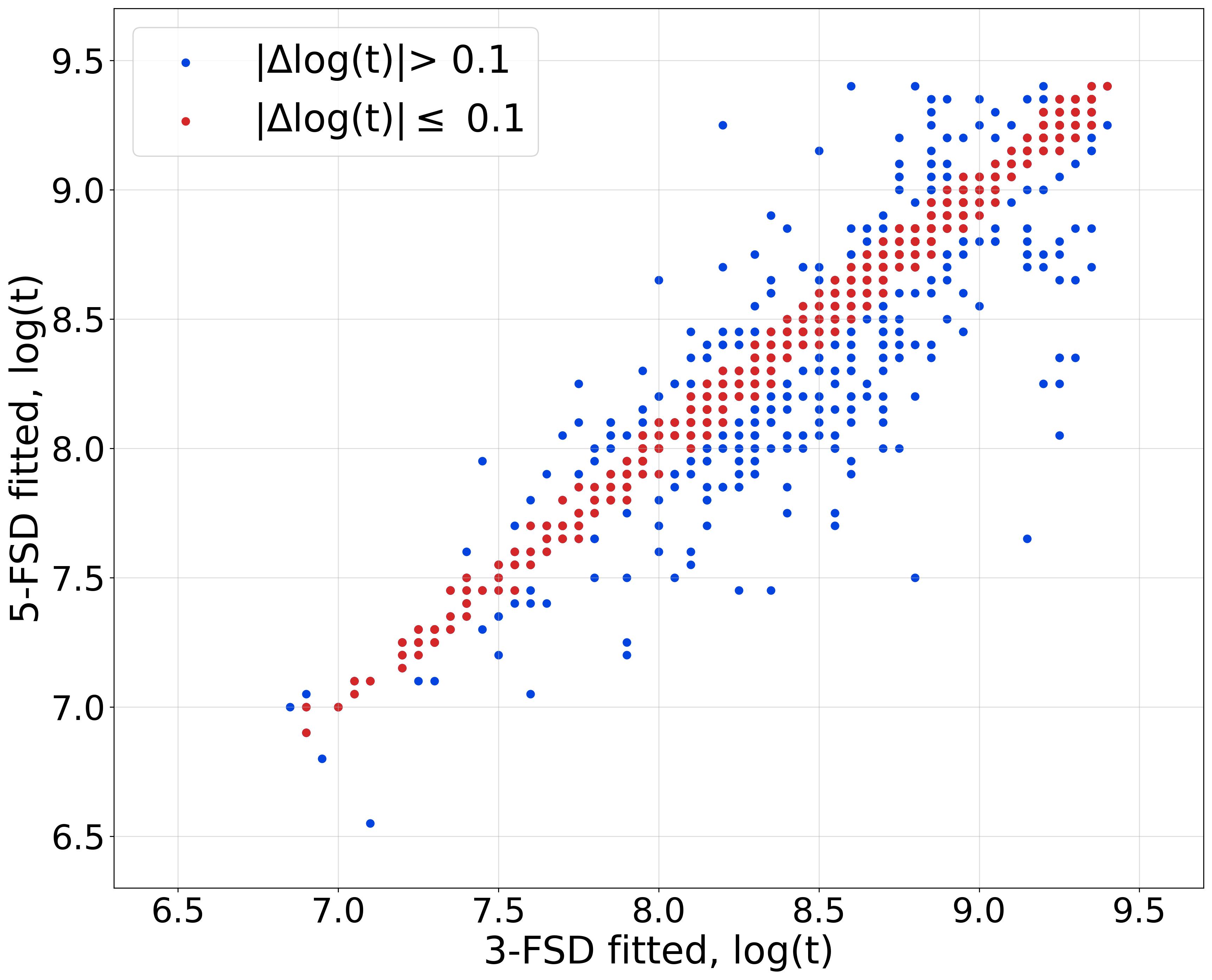}
            \caption[Network2]%
            {{\small }}    
            \label{fig:sc_5_3}
        \end{subfigure}
        \hfill
        \begin{subfigure}[b]{0.475\textwidth}  
            \centering 
            \includegraphics[width=\textwidth]{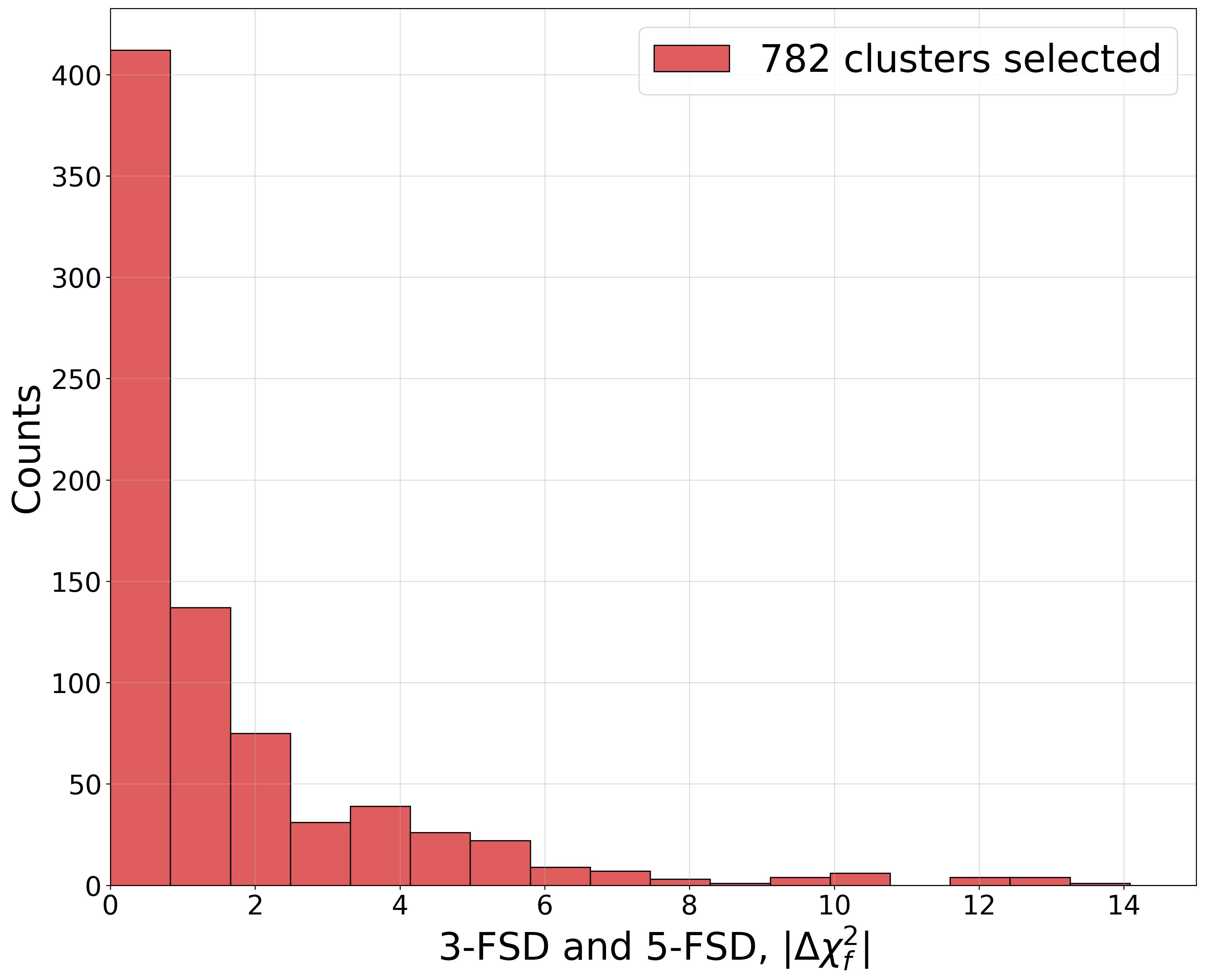}
            \caption[]%
            {{\small }}    
            \label{fig:vgf_5_3}
        \end{subfigure}
        \begin{subfigure}[b]{0.475\textwidth}   
            \centering 
            \includegraphics[width=\textwidth]{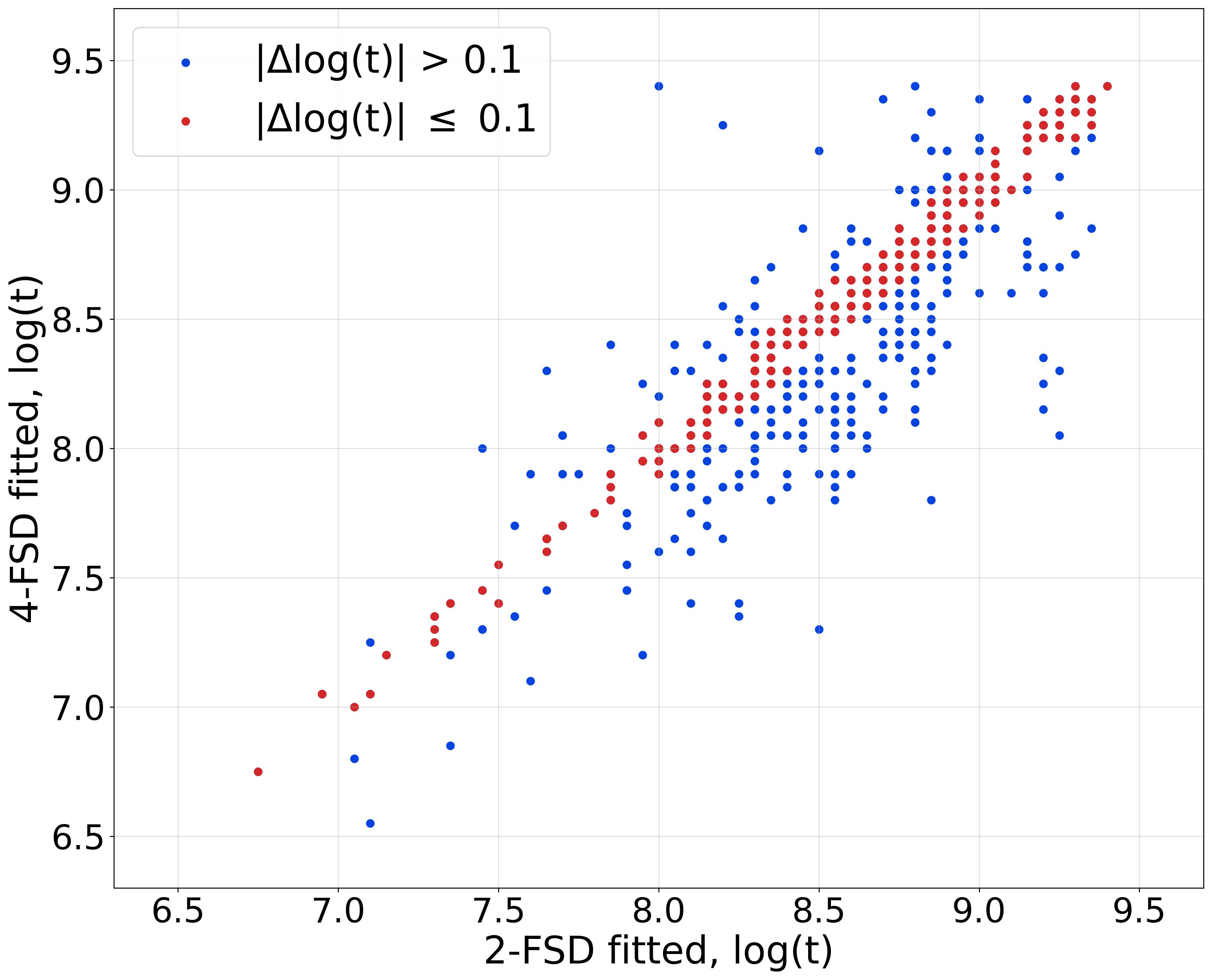}
            \caption[]%
            {{\small }}    
            \label{fig:4_2_sc}
        \end{subfigure}
        \hfill
        \begin{subfigure}[b]{0.475\textwidth}   
            \centering 
            \includegraphics[width=\textwidth]{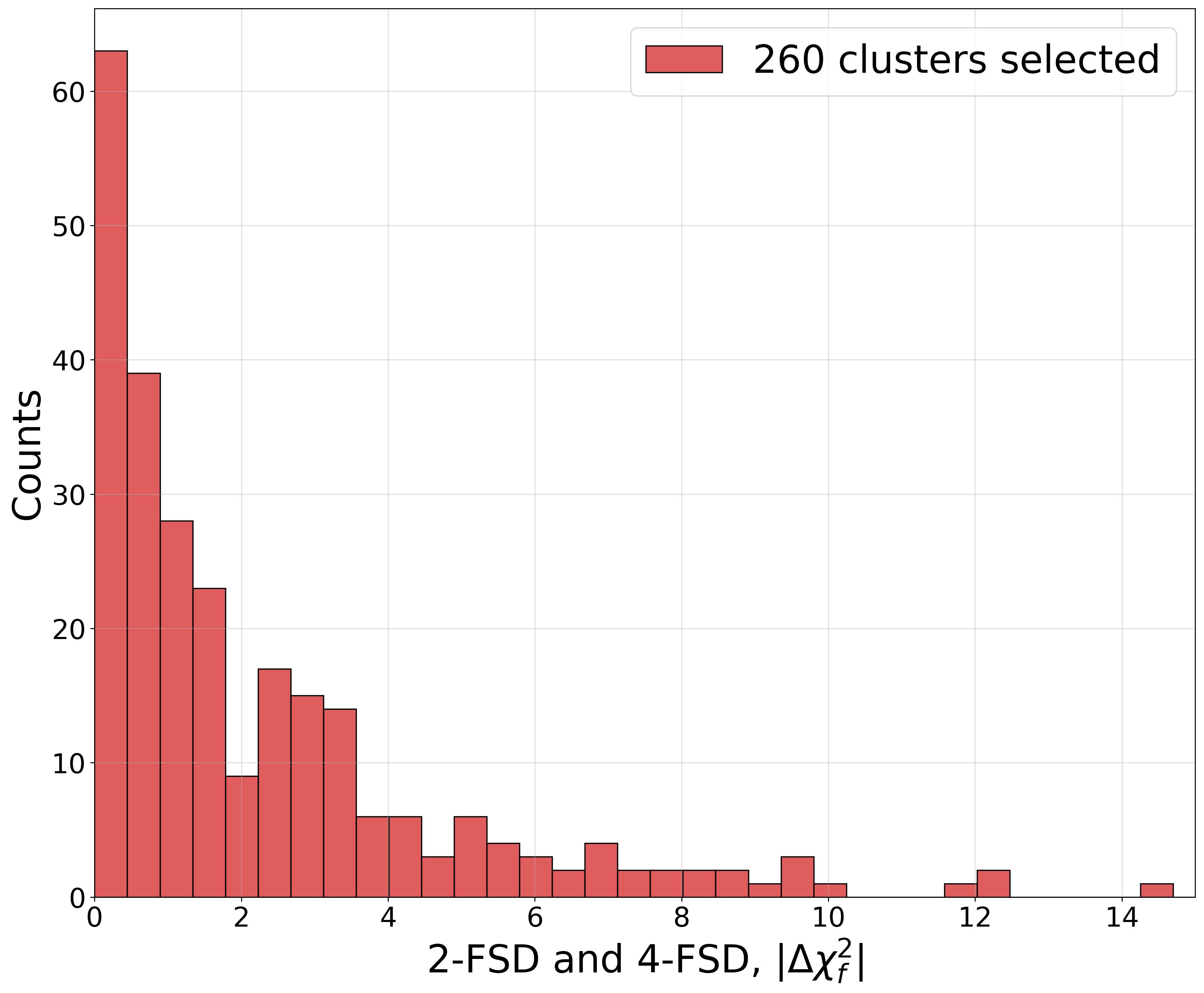}
            \caption[]%
            {{\small }}    
            \label{fig:vgf_4_2}
        \end{subfigure}
        \begin{subfigure}[b]{0.475\textwidth}   
            \centering 
            \includegraphics[width=\textwidth]{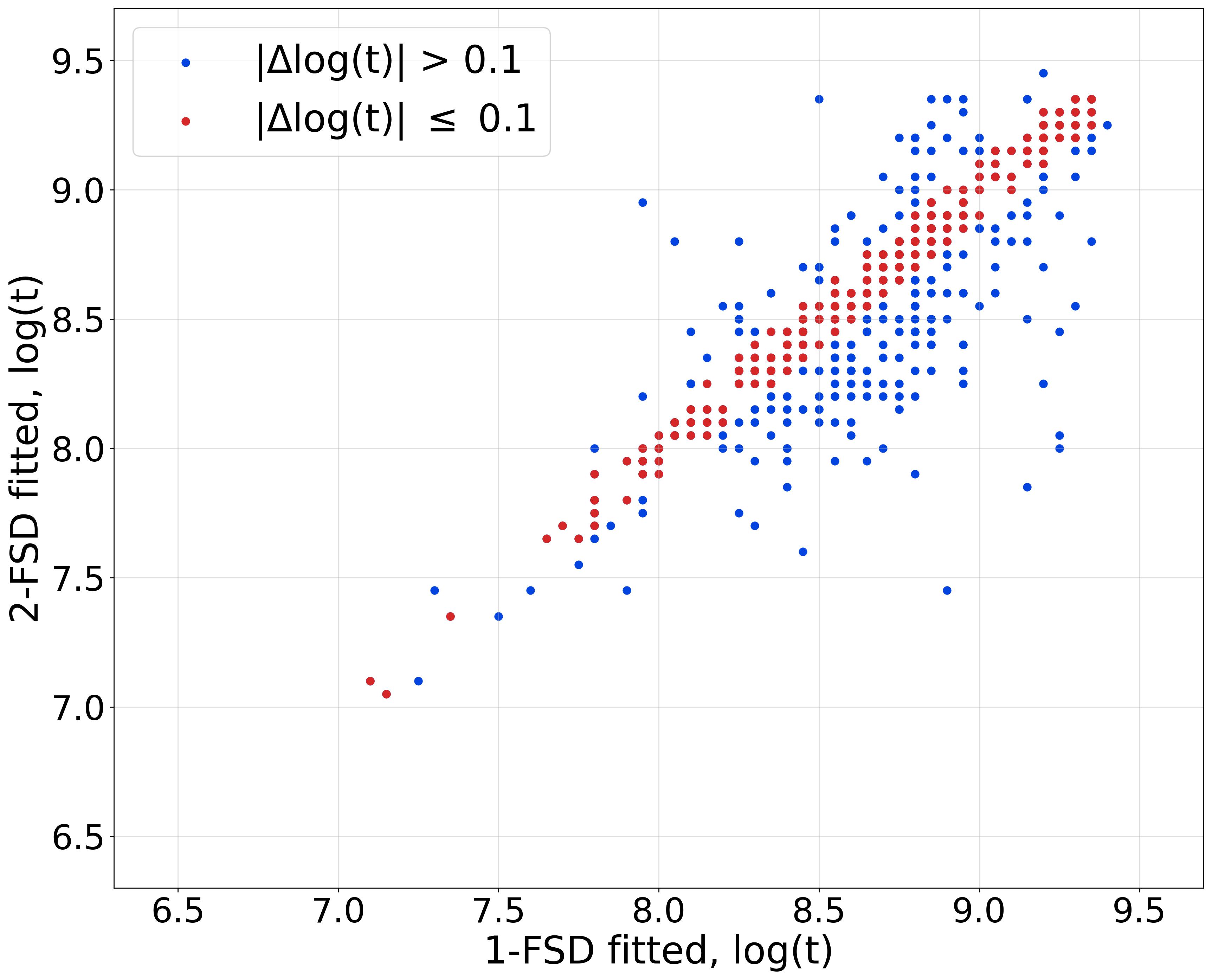}
            \caption[]%
            {{\small }}    
            \label{fig:2_1_sc}
        \end{subfigure}
        \hfill
        \begin{subfigure}[b]{0.475\textwidth}   
            \centering 
            \includegraphics[width=\textwidth]{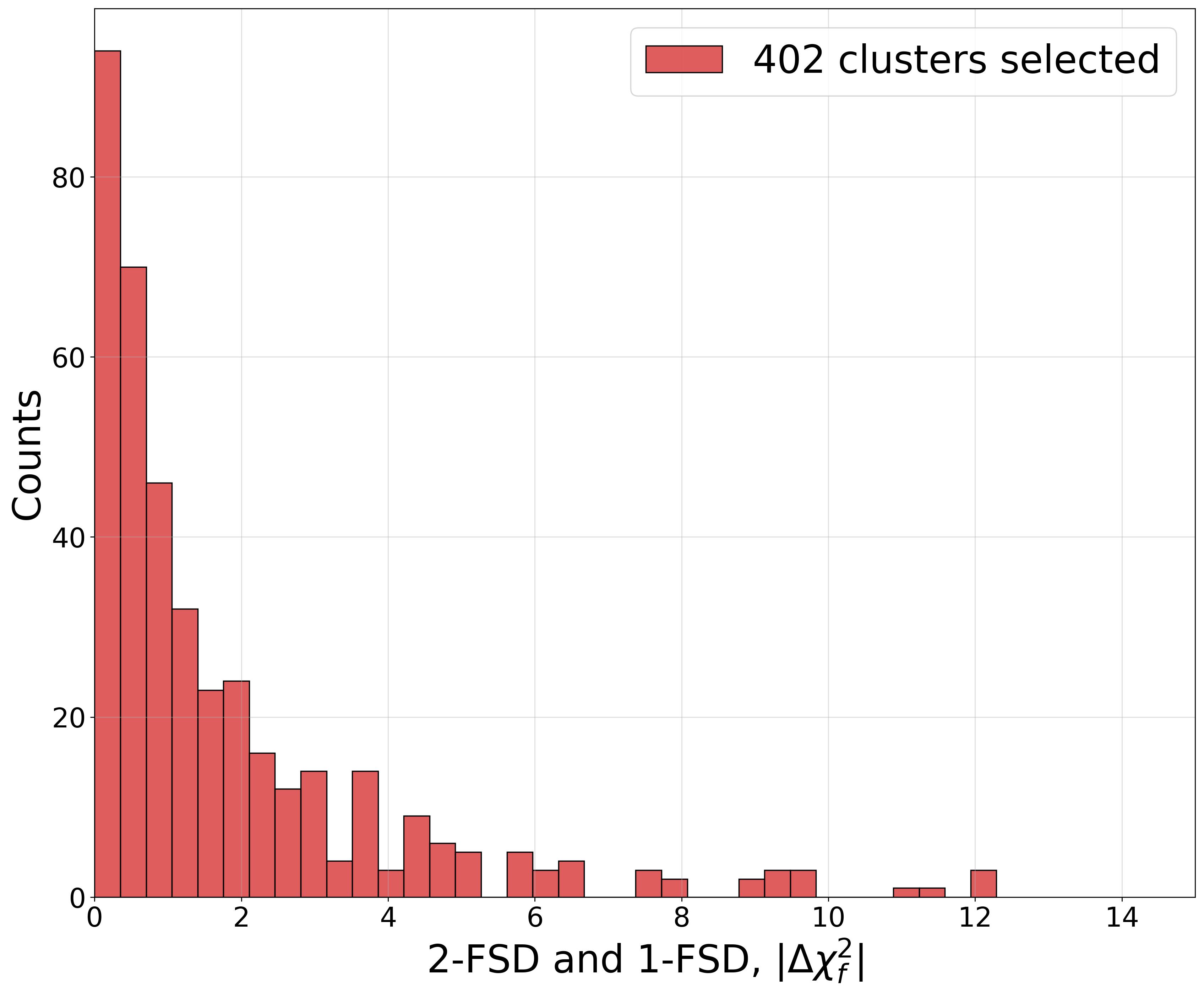}
            \caption[]%
            {{\small }}    
            \label{fig:vgf_2_1}
        \end{subfigure}
        \caption[]
        {\small The selection criterion discussed in subsection \ref{select_crit_CMD} for choosing the best cleaned CMD for clusters having multiple stages of FSD is shown here.} 
        \label{fig_FDstages}
    \end{figure*}

The selection criterion for the cleaned CMDs in the case of Group-1 and Group-2 clusters are detailed in subsection \ref{select_crit_CMD}. Figure \ref{fig_FDstages} graphically depicts the comparison and selection of prior age of the clusters from their n-FSD stages.
\section{Clusters left out from parameterization}
\begin{figure*}
\centering
   \includegraphics[width=1\linewidth]{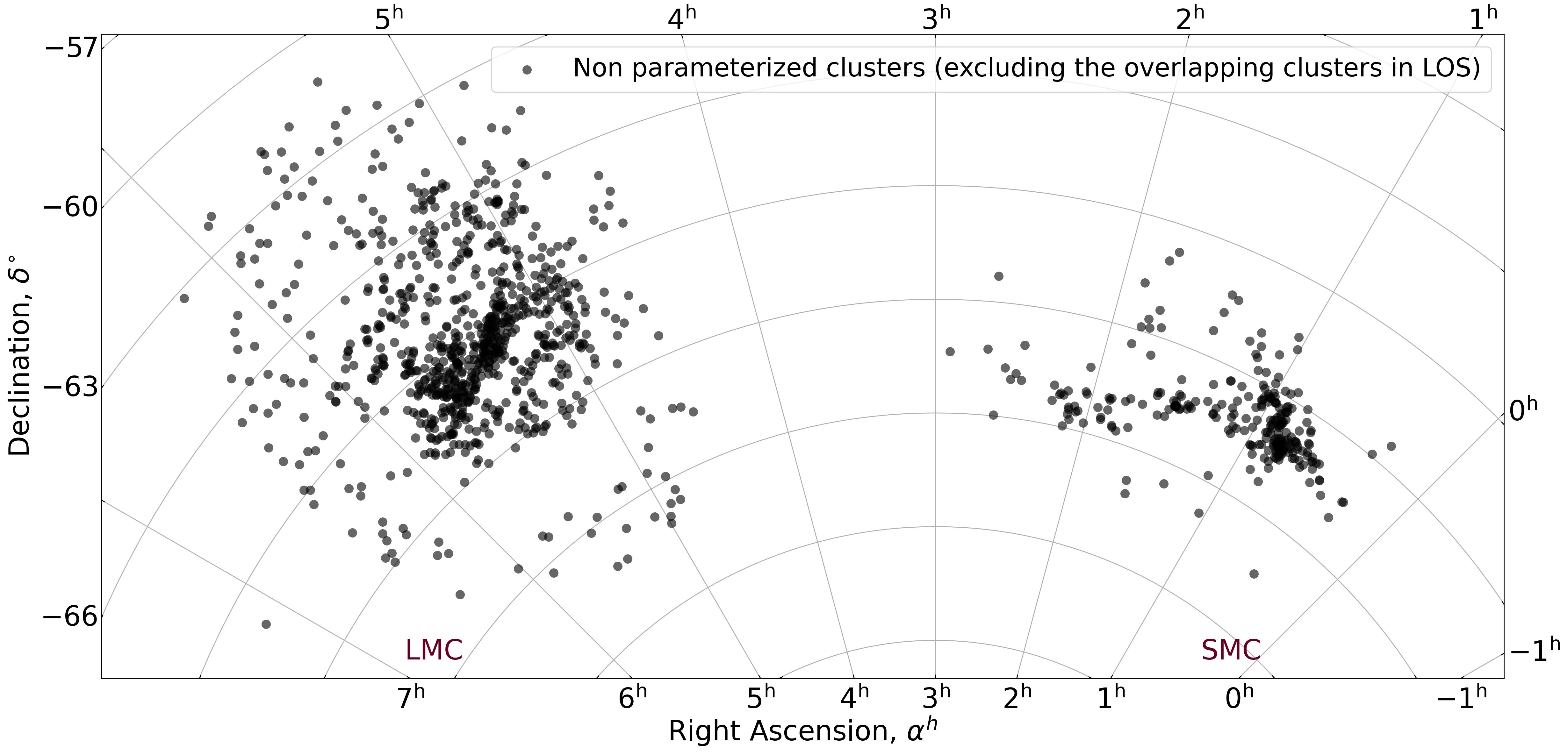}
   \caption{\small The above cluster locations (shown in black) are those we left out from parameterization. It includes 643 isolated clusters and 439 partially overlapping clusters in LOS. They are either poor clusters or residing in crowded regions of the galaxies, resulting in a cleaned CMD that cannot be parameterized. }
   \label{fig:nt_ct} 
\end{figure*}

We were not able to characterize 1082 out of 3791 selected cluster samples in this study. Figure \ref{fig:nt_ct} show the spatial map of their distribution in the skyplane.

\section{Four parameter estimation after MCMC.}
\begin{figure*}
\centering
   \includegraphics[width=1\linewidth]{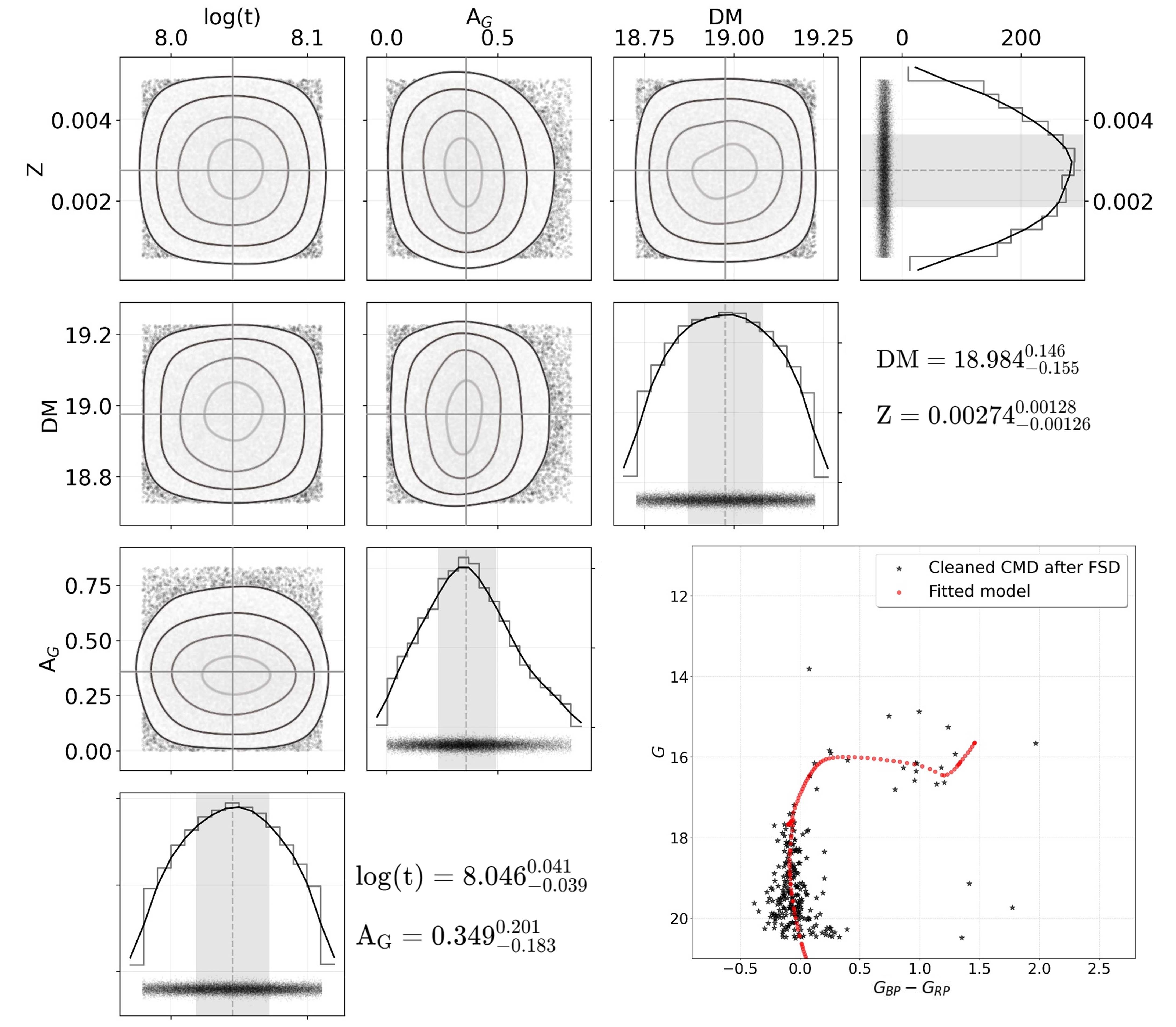}
   \caption{\small The sampled posterior distributions (corner plot) for the four-parameter estimation of cluster NGC458. The shaded regions in the histogram represent the 1$\sigma$ region from the 50 percentile (dashed line in histograms) for $\log(t)$, $A_G$, DM, and Z. The 50 percentile values are chosen as the best parameter for the isochrone fitting. The cleaned CMD of NGC458 is fitted with the isochrone generated using the best-fit parameters.}
   \label{fig:mcmc_2gc276} 
\end{figure*}

\begin{figure*}
\centering
   \includegraphics[width=1\linewidth]{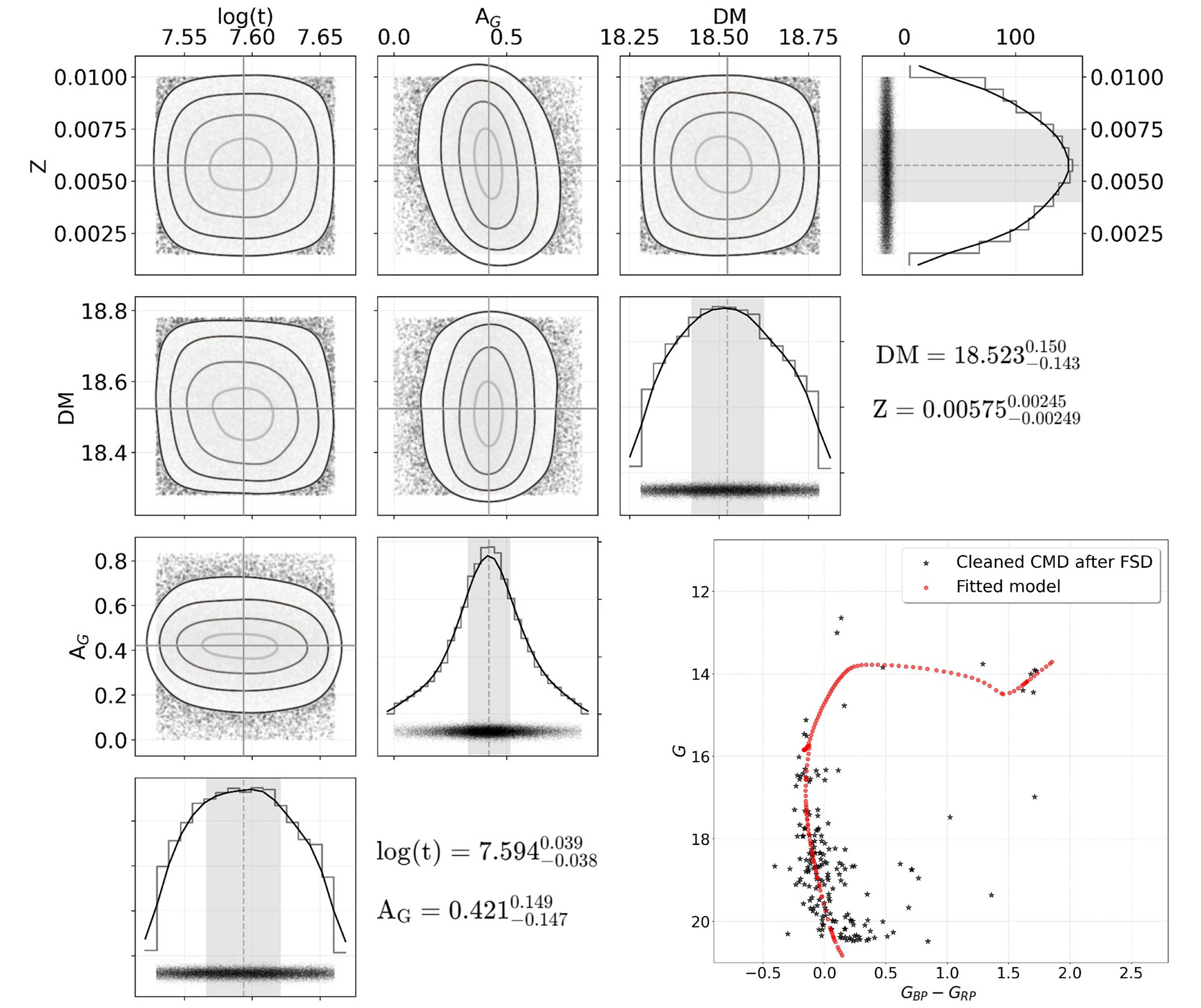}
   \caption{\small The sampled posterior distributions (corner plot) for the four-parameter estimation of cluster NGC1735. The cleaned CMD with the isochrone generated using the best-fit parameters is shown here.}
   \label{fig:mcmc_hw66} 
\end{figure*}

Figures \ref{fig:mcmc_2gc276} and \ref{fig:mcmc_hw66} show the parameter estimations and posterior distributions across four parameters for the clusters NGC458 and NGC1735.


\bsp	
\label{lastpage}
\end{document}